\documentclass[a4paper,twocolumn,10pt]{article}
\usepackage[margin=18mm,columnsep=6mm]{geometry}
\usepackage{authblk}
\pdfoutput=1

\usepackage[T1]{fontenc}
\usepackage{amsmath,amssymb,amsthm,mathtools}
\usepackage{graphicx}
\usepackage{tikz}
\usepackage{pgfplots}
\usepgfplotslibrary{groupplots}
\pgfplotsset{compat=1.18}
\usepackage{float}
\usepackage{placeins}
\usepackage{booktabs}
\usepackage{multirow}
\usepackage{tabularx}
\usepackage[numbers,sort&compress]{natbib}
\usepackage{microtype}
\allowdisplaybreaks[1]
\usepackage{xcolor}
\definecolor{linkblue}{RGB}{20,60,120}
\usepackage[colorlinks=true,linkcolor=linkblue,citecolor=linkblue,urlcolor=linkblue]{hyperref}

\theoremstyle{plain}
\newtheorem{theorem}{Theorem}[section]
\newtheorem{proposition}[theorem]{Proposition}
\newtheorem{lemma}[theorem]{Lemma}
\newtheorem{corollary}[theorem]{Corollary}
\theoremstyle{definition}
\newtheorem{definition}[theorem]{Definition}
\newtheorem{assumption}[theorem]{Assumption}
\newtheorem{example}[theorem]{Example}
\theoremstyle{remark}
\newtheorem{remark}[theorem]{Remark}
\theoremstyle{plain}
\newtheorem*{lemma*}{Lemma}

\newcommand{\calA}{\mathcal A}
\newcommand{\calC}{\mathcal C}
\newcommand{\calE}{\mathcal E}
\newcommand{\calH}{\mathcal H}
\newcommand{\calL}{\mathcal L}
\newcommand{\LBE}{\calL_{\BE}}
\newcommand{\calP}{\mathcal P}

\newcommand{\calW}{\mathcal W}
\newcommand{\BE}{\mathrm{BE}}
\newcommand{\INT}{\mathrm{int}}
\newcommand{\Tr}{\operatorname{Tr}}
\newcommand{\Var}{\operatorname{Var}}
\newcommand{\Cov}{\operatorname{Cov}}
\newcommand{\Lie}{\operatorname{Lie}}
\newcommand{\Span}{\operatorname{span}}
\newcommand{\Alg}{\operatorname{Alg}}
\newcommand{\Herm}{\operatorname{Herm}}
\newcommand{\Skew}{\operatorname{Skew}}
\newcommand{\Comm}{\operatorname{Comm}}
\newcommand{\SU}{\operatorname{SU}}
\newcommand{\U}{\operatorname{U}}
\newcommand{\su}{\mathfrak{su}}
\newcommand{\fu}{\mathfrak{u}}
\newcommand{\so}{\mathfrak{so}}
\newcommand{\fsl}{\mathfrak{sl}}
\newcommand{\fgl}{\mathfrak{gl}}
\newcommand{\id}{\mathbb I}
\newcommand{\ket}[1]{|#1\rangle}
\newcommand{\bra}[1]{\langle #1|}
\newcommand{\ketbra}[2]{|#1\rangle\langle #2|}
\newcommand{\dagg}{^{\dagger}}
\newcommand{\HS}{\mathrm{HS}}
\newcommand{\PREP}{\mathrm{PREP}}
\newcommand{\sd}{\operatorname{sd}}
\newcommand{\gcoeff}{\mathfrak g_{\mathrm{coeff}}}
\newcommand{\Jforms}{\mathfrak J}
\newcommand{\stab}{\operatorname{stab}}
\newcommand{\gint}{\mathfrak g_{\INT}}
\newcommand{\tcalL}{\widetilde{\calL}}
\newcommand{\psucc}{p_{\mathrm{succ}}}
\newcommand{\pacc}{p_{\mathrm{acc}}}
\newcommand{\bpacc}{\bar p_{\mathrm{acc}}}

\title{Barren Plateaus in Quantum Linear Combinations: Interference and Lie-Algebraic Control}
\author[1,3]{Amir Koutahi\thanks{Amir.koutahi@mail.utoronto.ca}}
\author[2]{Pouya Kananian\thanks{pouya.kananian@mail.utoronto.ca}}
\author[2]{Hans-Arno Jacobsen\thanks{jacobsen@eecg.toronto.edu}}
\affil[1]{Division of Engineering Science, University of Toronto}
\affil[2]{Department of Electrical and Computer Engineering, University of Toronto}
\affil[3]{Department of Computer and Mathematical Sciences, University of Toronto Scarborough}
\date{}
\hypersetup{
  pdftitle={Barren Plateaus in Quantum Linear Combinations: Interference and Lie-Algebraic Control},
  pdfauthor={Amir Koutahi; Pouya Kananian; Hans-Arno Jacobsen}
}

\begin{document}
\raggedbottom
\maketitle
\begin{abstract}
A coherent sum of quantum circuits introduces interference that can alter the training signal. We distinguish a single linear combination of unitaries (LCU) from a circuit that repeatedly alternates branch operations and ancilla mixing. For a single-pass LCU, independent branches with zero matrix means cannot erase the differentiated branch's gradient variance: the total variance is at least the branch variance times the fourth power of its coefficient magnitude. This bound holds at any gate position and does not require the circuit unitaries to be Haar-random. Beyond this lower bound, we derive exact loss and gradient formulas that quantify the additional contributions from interference between branches. One of these terms depends on relative coefficient phases and vanishes when the branch ensemble is invariant under multiplication by $i\id$; a four-qubit matchgate example shows that it can be nonzero. For repeated interleaving with full special-unitary ancilla control, we determine the generated Lie algebra from the associative algebra of the branches, including the available central phase directions. This control result predicts concentration only with additional assumptions on circuit sampling. Matchgate examples exhibit polynomial single-pass LCU variances and exponential concentration under joint Haar-random sampling. We also derive coefficient-selection rules and bounds for postselection, measurement error, and approximate moments.
\end{abstract}

\begin{figure*}[t]
\centering
\resizebox{\textwidth}{!}{\begin{tikzpicture}[
  x=1cm,y=1cm,
  font=\small,
  wire/.style={line width=.45pt},
  prep/.style={draw=blue!35!black,fill=blue!3,minimum height=.6cm,inner xsep=4pt},
  branch/.style={draw=teal!75!black,fill=teal!5,minimum height=1.7cm,inner xsep=4pt},
  panel/.style={anchor=west,text=blue!35!black,font=\bfseries\normalsize},
  explanation/.style={anchor=north west,align=left,text width=6.0cm,inner sep=0pt}
]
\path[use as bounding box] (0,0) rectangle (17.1,8.25);

\node[panel] at (.45,7.95) {(a) Single-pass LCU (one SELECT)};
\draw[wire] (1.05,6.8) -- (9.15,6.8);
\draw[wire] (1.05,5.7) -- (9.15,5.7);
\node[anchor=east] at (.9,6.8) {$|\mathrm{ref}\rangle_a$};
\node[anchor=east] at (.9,5.7) {$\rho_s$};
\node[prep] at (2.3,6.8) {$\mathrm{PREP}_R$};
\node[branch,minimum width=1.55cm] at (4.6,6.25) {$\mathrm{SELECT}$};
\node[prep] at (7.0,6.8) {$\mathrm{PREP}_L^\dagger$};
\node[anchor=west] at (9.25,6.8) {$\langle\mathrm{ref}|_a$};
\node[anchor=west] at (9.25,5.7) {$O_s$};
\node[anchor=west] at (1.4,4.8)
  {Projected block: $A/\lambda=\sum_\ell(\alpha_\ell/\lambda)U_\ell$};
\node[explanation] at (10.75,7.35)
  {\textbf{Question}\\
   When does an LCU retain gradient\\variance that decreases at most\\polynomially with qubit count?};
\node[explanation] at (10.75,5.65)
  {\textbf{Required analysis}\\
   Branch-gradient lower bound\\and interference moments.};

\draw[gray!45,line width=.3pt] (.35,4.15) -- (16.85,4.15);

\node[panel] at (.45,3.65) {(b) Repeated branch--mixer layers};
\draw[wire] (1.05,2.55) -- (4.85,2.55);
\draw[wire] (5.65,2.55) -- (9.15,2.55);
\draw[wire] (1.05,1.45) -- (4.85,1.45);
\draw[wire] (5.65,1.45) -- (9.15,1.45);
\node[anchor=east] at (.9,2.55) {$|\mathrm{ref}\rangle_a$};
\node[anchor=east] at (.9,1.45) {$\rho_s$};
\node[prep,minimum width=.7cm] at (1.65,2.55) {$C_0$};
\node[branch,minimum width=.9cm] at (2.9,2.0) {$S_1$};
\node[prep,minimum width=.7cm] at (4.15,2.55) {$C_1$};
\node at (5.25,2.55) {$\cdots$};
\node at (5.25,1.45) {$\cdots$};
\node[branch,minimum width=.9cm] at (6.5,2.0) {$S_T$};
\node[prep,minimum width=.7cm] at (7.75,2.55) {$C_T$};
\node[anchor=west] at (9.25,2.55) {$\langle\mathrm{ref}|_a$};
\node[anchor=west] at (9.25,1.45) {$O_s$};
\node[anchor=west] at (1.4,.55)
  {$S_t$: controlled branches\qquad $C_t$: ancilla mixer};
\node[explanation] at (10.75,3.1)
  {\textbf{Question}\\
   What control becomes reachable, and when does concentration occur?};
\node[explanation] at (10.75,1.75)
  {\textbf{Required analysis}\\
   Joint control algebra,\\then ensemble moments.};
\end{tikzpicture}
}
\caption{Two circuit models. (a) A single-pass LCU prepares an ancilla superposition (PREP), applies each branch unitary conditioned on its label (SELECT), and unprepares the ancilla. Projection onto the reference ancilla state gives the block $A/\lambda$. Its fluctuations are determined by branch and interference moments. (b) Repeated interleaving applies controlled branch layers $S_t$ and ancilla mixers $C_t$ in alternation; its projected block generally contains products of branch operations. The available generators determine reachable control. A variance prediction additionally requires the actual ensemble moments. The detailed definitions appear in Section~\ref{sec:scope}.}
\label{fig:arch}
\end{figure*}

\section{Introduction}\label{sec:intro}

Training a parameterized quantum circuit requires changes in its parameters to produce a measurable change in the loss. In a barren plateau, gradients at random initialization concentrate near zero, with a variance that decreases exponentially as the number of qubits grows. Resolving these gradients can require increasingly costly measurements. Understanding how circuit structure affects this training signal is therefore an important part of designing parameterized quantum models \citep{McClean2018BarrenPlateau,Cerezo2021CostFnBP,Holmes2022Expressibility,Larocca2025Review}.

A linear combination of unitaries (LCU) combines several parameterized circuits coherently \citep{Childs2012LCU}. Each constituent circuit, called a \emph{branch}, contributes with a complex coefficient that sets its amplitude and phase. For $L$ branches acting on the same $n$-qubit system, we write
\begin{equation}
    A(\theta,\alpha)
    =\sum_{\ell=1}^{L}\alpha_\ell U_\ell(\theta_\ell),
    \qquad
    \lambda=\sum_{\ell=1}^{L}|\alpha_\ell|>0.
    \label{eq:A_intro}
\end{equation}
Here $U_\ell(\theta_\ell)$ is the unitary implemented by branch $\ell$, and $\theta_\ell$ contains its trainable parameters. The coefficients allow the model to adjust the relative contributions of the branches. Combining these contributions at the amplitude level makes interference between branches contribute to the loss. The derivatives of these interference terms also contribute to the training signal, so the gradient variance of the combined model depends on more than the variances of its individual branches. We distinguish two circuit models: a \emph{single-pass LCU} implements one sum $A=\sum_\ell\alpha_\ell U_\ell$, while an \emph{interleaved circuit} repeatedly alternates controlled branch operations with gates on an auxiliary register, the \emph{ancilla}, that mix the branch labels. Changing the coefficients of a single-pass LCU changes the amplitudes and phases in that sum. Repeated interleaving can instead produce products of branch operations in the selected block, so that block generally differs from a single LCU.

For an input state $\rho$ and a Hermitian observable $O$, we study the unnormalized loss
\[
    \calL(\theta,\alpha)
    =\Tr\!\left[
        O\,A(\theta,\alpha)\rho A(\theta,\alpha)^\dagger
    \right].
\]
Although $A$ need not be unitary, the LCU construction implements $A/\lambda$ as a selected block of a larger unitary acting on the system and ancilla \citep{Childs2012LCU}. Conditioning on the designated ancilla outcome gives the normalized loss $\calL/\Tr[A\rho A^\dagger]$. Its denominator depends on the circuit parameters, so conclusions about the unnormalized gradient do not automatically extend to the normalized objective.

Two questions guide the paper. When an individual branch's gradient variance decreases at most polynomially with the number of qubits, under what conditions does a single-pass LCU preserve this behavior? What changes when we adjust the coefficients of a single-pass LCU or repeatedly mix the branches in an interleaved circuit? The first question concerns the scaling of the gradient variance inherited from a branch; the second concerns how coefficient tuning changes interference and how repeated mixing changes the available transformations.

Khatri et al.~\cite{Khatri2025PLCU} establish loss-variance bounds and coefficient-averaged formulas for parameterized LCUs, including contributions from interference between branches. We build on their second-moment analysis to examine how complex coefficient phases and differentiation affect those contributions.

For a single-pass LCU, we show that, under the stated independence and zero-mean assumptions, interference cannot reduce the total gradient variance below the differentiated branch's variance multiplied by $|\alpha_\ell|^4$. This bound holds at any gate position. We then evaluate the full loss variance and, under additional assumptions, the gradient variance, including their dependence on relative coefficient phases. These formulas distinguish the contribution inherited from an individual branch from the additional signal supplied by interference, and show why coefficients that increase loss variance need not increase the gradient variance of a chosen parameter.

For repeated interleaving, we first determine what transformations the available gates can generate. The \emph{dynamical Lie algebra} describes the infinitesimal changes in the joint ancilla--system unitary that these gates can produce. These \emph{control directions} include the generators of the individual gates and the commutators obtained by combining and repeating them \citep{Larocca2022DLA}. With full special-unitary ancilla control, described by $\su(L)$ for an ancilla with $L$ branch labels, we characterize this algebra through the branch operators, their products, and the available phase directions.

Variance predictions additionally require a sampling law. The \emph{reachable transformations} are the joint unitaries obtainable by composing the available gates with allowed parameter values and arbitrary repetitions. Random initialization assigns a probability distribution to the circuit parameters and thereby to these unitaries. The reachable set alone does not determine that distribution, and a fixed-depth initialization need not explore the whole set \citep{Ragone2024,Fontana2023Adjoint}. We illustrate this distinction with matchgate circuits: single-pass LCU variances are polynomial in the stated settings, whereas the corresponding interleaved circuits exhibit exponential concentration under joint Haar-random sampling.\footnote{Here, \emph{joint Haar-random sampling} means drawing the complete ancilla--system unitary from the normalized Haar measure on its generated compact group. This measure is invariant under multiplication by any fixed group element. For the loss-variance formulas, an ensemble with the same first two conjugation moments is sufficient. Sampling individual branches or ancilla gates independently does not by itself establish this joint sampling law.}

We also study how to choose the magnitudes and phases of the LCU coefficients to increase the variance of either the loss or its derivative with respect to a selected circuit parameter. In both cases, the variance measures how much that quantity varies across randomly initialized circuits. We then examine how many times the circuit must be run to estimate a gradient accurately. When postselection keeps only runs with the desired ancilla measurement outcome, we also account for the repetitions needed to obtain those accepted outcomes. Finally, we bound the error in our variance predictions when the actual random initialization does not exactly satisfy the sampling assumptions used to derive them.

Section~\ref{sec:variance} treats single-pass LCU loss and gradient moments. Sections~\ref{sec:sandwich} and~\ref{sec:joint_variance} develop the control and variance analyses for interleaved circuits, followed by the matchgate applications in Section~\ref{sec:matchgate}. Section~\ref{sec:audit} discusses coefficient choices and practical predictions. Proofs and supporting results appear in the appendices.

\section{Relation to prior work}\label{sec:related}

Our analysis draws on variance formulas for random unitary circuits and on results for coherent sums of circuits. The former connect control algebras to fluctuations under a specified sampling law.

Let $U_\ell$ and $U_k$ denote two branch circuits, with coefficients $\alpha_\ell$ and $\alpha_k$. For input state $\rho$ and measured observable $O$, their interference contributes
\[
    2\operatorname{Re}\!\left[\alpha_\ell\alpha_k^*m_{\ell k}\right],
    \qquad
    m_{\ell k}=\Tr[OU_\ell\rho U_k^\dagger],
\]
to the loss. To calculate how this contribution varies across random initializations, we evaluate both $\mathbb E[|m_{\ell k}|^2]$ and $\mathbb E[m_{\ell k}^2]$. The first measures the average squared magnitude of the cross-branch term; the second determines how the relative phases of the coefficients affect its variance.

\paragraph{Lie-algebraic and representation-theoretic variance formulas.}
Lie-algebraic methods describe parameterized unitary circuits through the transformations generated by their gates. Larocca et al.~\cite{Larocca2022DLA} connect the control algebra of such a circuit to trainability, while Ragone et al.~\cite{Ragone2024} and Fontana et al.~\cite{Fontana2023Adjoint} establish variance formulas under specified sampling and operator assumptions. These formulas organize the circuit's independently transforming operator families. The algebra-based variance formula requires the state or observable to belong to specified operator spaces; Section~\ref{sec:dla_variance} explains this \emph{membership condition}. More general states and observables can introduce additional contributions. Diaz et al.~\cite{Diaz2023Matchgate} analyze such contributions for matchgate circuits, and Alc\^antara et al.~\cite{Alcantara2026Representation} develop a broader decomposition of the operator space. We use the variance formula based on the generated algebra only where its assumptions hold. For independent Haar-random branches at fixed coefficients, SELECT is Haar-random on the represented product of the branch groups. Theorem~\ref{thm:exact_select} is then a specialization of the general operator-space decomposition of Alc\^antara et al.~\cite{Alcantara2026Representation}, with related representation-theoretic methods used by Diaz et al.~\cite{Diaz2023Matchgate}. Our contribution here is the explicit LCU interpretation of those moments, including relative coefficient phases and the coefficient scale. Our pair-moment decomposition also holds for independent zero-mean branches without Haar-random sampling. To evaluate the group-based moments, we use integration with respect to Haar measure and invariant-form identities \citep{Mele2024,Collins2003,CollinsSniady2006}.

\paragraph{Parameterized LCUs.}
Yao et al.~\cite{Yao2025LCQNN} introduce linear combinations of quantum neural networks, using a trainable ancilla coefficient layer to control several parameterized circuit branches. Their Eq.~(6) measures a weighted average of the branch expectations, with the ancilla left unobserved. Building on this circuit structure, we study the unnormalized loss associated with projecting the ancilla onto a chosen state. This projection allows interference between branches to contribute to the loss, making it sensitive to their relative coefficient phases.

Khatri et al.~\cite{Khatri2025PLCU} derive loss-variance bounds and Dirichlet-averaged formulas for parameterized LCUs, including this interference. In their coefficient average, nonnegative weights summing to one are sampled uniformly from the probability simplex, a distribution called $\mathrm{Dir}(1,\ldots,1)$, independently of the branch parameters. Our fixed-coefficient formulas can be averaged over the same law to make the comparison explicit. Our single-pass LCU analysis builds on their second-moment expansion: Theorem~\ref{thm:KZM_consistency} recovers their identical-group formula when the unconjugated overlap moment vanishes, and Proposition~\ref{cor:floor} uses related independence and positivity reasoning for gradients. We retain and evaluate the phase-sensitive moment $\mathbb E(m^2)$, where $m=\Tr[OU\rho V^\dagger]$ is the overlap between two branches. This is the average of the complex square rather than the squared magnitude; its removal requires an additional condition.

\paragraph{Depth and stacked architectures.}
Repeated interleaving is a different circuit family from a single linear combination. We characterize the Lie algebra generated by alternating branch and coefficient layers. Stacked LCUs compose several LCU maps, whereas quantum singular value transformation (QSVT) alternates a block encoding and its adjoint with phase operations to transform its singular values \citep{Gilyen2019QSVT}. Neither construction alone supplies the unrestricted ancilla control assumed for our fully controlled interleaved circuit. We specify the generator families of that interleaved circuit explicitly and compare its variance bound under joint Haar-random sampling with the fixed-depth stacked-LCU result of Khatri et al.~\cite{Khatri2026Stacked} after deriving the matchgate regimes.

\paragraph{Other non-unitary models and postselection cost.}
Beyond LCU constructions, non-unitary models use engineered dissipation and dynamic circuits to address trainability \citep{Sannia2024Dissipation,Zapusek2025Dissipation,Deshpande2024Dynamic}. LCU constructions also occur in quantum transformers and Boltzmann learning \citep{Khatri2024Quixer,Coopmans2024}. Heredge et al.~\cite{Heredge2025} use LCUs to construct quantum residual connections. Each connection combines two contributions: the state produced by applying a circuit layer and the input state with that layer left unapplied. These contributions form a weighted quantum superposition and can interfere. Repeating this construction across layers preserves contributions that pass through fewer circuit operations, which can help avoid barren plateaus.

Their work also highlights the importance of the probability of obtaining the desired ancilla measurement outcome. Keeping only runs with that outcome is called postselection. If the outcome occurs with probability $\psucc$, collecting one accepted measurement requires $1/\psucc$ circuit runs on average at fixed parameters. A model can therefore retain appreciable gradient variance while requiring many runs to collect accepted measurements. Motivated by this trade-off, Section~\ref{sec:postselection} and Appendix~\ref{app:postselect} examine how many runs are needed to estimate an unnormalized gradient, how postselection changes the loss variance, and the additional repetitions required to obtain accepted measurements.

\section{Preliminaries}\label{sec:prelim}

We describe the measured loss and the transformations available to the circuit in a common matrix language. The notation below connects these descriptions. We then introduce the LCU construction and distinguish its unnormalized signal from the expectation obtained by conditioning on a successful ancilla measurement.

\subsection{Notation}\label{sec:notation}

Unless specified otherwise, matrices are finite-dimensional and Lie-algebra dimensions are real.

We write $M_D(\mathbb C)$ for the space of complex $D\times D$ matrices and $\Herm(D)$ for its real subspace of Hermitian matrices. The identity is $\id$, with a subscript indicating its dimension or underlying space when needed. For a matrix subspace $\calA\subseteq M_D(\mathbb C)$, we use
\[
    \Herm(\calA)=\calA\cap\Herm(D).
\]

The unitary and special-unitary groups are denoted by $\U(D)$ and $\SU(D)$. Their Lie algebras, $\fu(D)$ and $\su(D)$, consist of anti-Hermitian and traceless anti-Hermitian matrices, respectively. The symbols $\Span_{\mathbb R}$ and $\Span_{\mathbb C}$ denote real and complex linear spans.

The Hilbert--Schmidt (HS) inner product and its associated norm are
\[
    \langle A,B\rangle_{\HS}=\Tr(A^\dagger B),
    \qquad
    \|A\|_{\HS}=\sqrt{\Tr(A^\dagger A)}.
\]
We write $\|A\|_\infty$ for the operator norm, equal to the largest singular value, and $\|A\|_1$ for the trace norm, equal to the sum of the singular values.

In a fixed orthonormal basis, $E_{ab}=\ketbra{a}{b}$ is the matrix with a single nonzero entry, equal to one, in row $a$ and column $b$.

\subsection{Lie-algebraic background}\label{sec:lie_background}

Repeated gates can generate transformations unavailable from any individual gate. The dynamical Lie algebra records their infinitesimal directions without listing circuits at every depth. Commutators describe how new directions arise; the decomposition of the resulting algebra will later organize the variance calculation.

A parameterized gate has the form $U(t)=e^{itH}$, where $t\in\mathbb R$ and $H$ is a Hermitian matrix called the \emph{gate Hamiltonian} or \emph{Hermitian gate generator}. The parameter $t$ controls the gate's evolution. To describe the transformations generated by combining such gates, we work with the corresponding anti-Hermitian matrices $iH$. These matrices generate the circuit's dynamical Lie algebra.

A \emph{real Lie algebra of matrices} is a real vector space $\mathfrak g\subseteq M_D(\mathbb C)$ closed under the commutator $[X,Y]=XY-YX$, where $X,Y\in\mathfrak g$. An algebra is \emph{abelian} if all of its matrices commute; otherwise it is nonabelian. Throughout, lowercase fraktur symbols such as $\mathfrak g$ denote real spaces of anti-Hermitian generators.

For $S\subset\fu(D)$, the \emph{Lie closure} $\Lie_{\mathbb R}\{S\}$ is the smallest real Lie subalgebra containing $S$. It can be computed by repeatedly adjoining commutators. Start with $V_0=\Span_{\mathbb R}S$ and set $V_{j+1}=\Span_{\mathbb R}(V_j\cup[V_j,V_j])$, where $[V_j,V_j]=\{[X,Y]:X,Y\in V_j\}$. The sequence stabilizes at the Lie closure after finitely many steps, since its dimension cannot exceed $\dim\fu(D)=D^2$.

\begin{definition}[Dynamical Lie algebra \citep{Larocca2022DLA,DAlessandro2007}]\label{def:dla}
Consider a parameterized circuit $U(\theta)=\prod_{k}e^{i\theta_kH_k}$ with Hermitian gate generators $\{H_k\}_k\subset\Herm(D)$, where $k$ indexes the parameterized gates. Its \emph{dynamical Lie algebra} (DLA) is $\mathfrak g=\Lie_{\mathbb R}\{iH_k\}_k\subseteq\fu(D)$.

The closure includes directions obtained by arbitrary repetition and commutation of the listed generators. It therefore describes the control family, not just the fixed-depth product displayed here.
\end{definition}

For a real Lie algebra $\mathfrak g\subseteq\fu(D)$, write $e^{\mathfrak g}\subseteq\U(D)$ for the finite products $e^{X_1}\cdots e^{X_k}$ with $X_1,\ldots,X_k\in\mathfrak g$. In statements using Haar measure, we assume this generated group is closed: it contains the limits of convergent sequences of its matrices. Because unitary matrices are bounded, closedness makes the group compact and permits normalized Haar-measure averaging. An arbitrary subalgebra of $\fu(D)$ need not generate a closed group.

We will use representations on two different spaces: the group acts on state vectors by $v\mapsto Uv$, and on operators by $X\mapsto UXU^\dagger$. Irreducibility and invariant subspaces always refer to the action and space specified at that point. A \emph{block} refers to a preserved state-vector subspace, in a basis that displays the separate actions on these subspaces.

Averaging random transformations leaves some operators unchanged. To identify them, the \emph{commutant} of a compact group $G\subseteq\U(D)$ collects all matrices that commute with every allowed unitary:
\begin{equation}
    \Comm(G)=\{X\in M_D(\mathbb C):XU=UX\ \text{for all }U\in G\}.
\end{equation}

Next we separate directions that commute with all controls from families that rotate among themselves under the circuit action. Mathematically, a Lie subalgebra of anti-Hermitian matrices is \emph{reductive}: it has the orthogonal direct-sum decomposition
\begin{equation}
    \mathfrak g=\mathfrak z\oplus\bigoplus_{j}\mathfrak g_j.
    \label{eq:reductive}
\end{equation}
The summands in \eqref{eq:reductive} are perpendicular under the Hilbert--Schmidt inner product. The \emph{center} $\mathfrak z=\{Z\in\mathfrak g:[Z,X]=0\ \text{for all }X\in\mathfrak g\}$ contains the directions that commute with every control. The remaining summands are \emph{simple ideals}. An ideal is a subspace preserved by taking commutators with elements of $\mathfrak g$; it is simple if it is nonabelian and has no proper nonzero ideal of its own.

We call these simple ideals the \emph{sectors} of $\mathfrak g$. They are operator-algebra components, distinct from the state-vector subspaces called blocks. Their direct sum is the \emph{semisimple part}, equal to the derived algebra $[\mathfrak g,\mathfrak g]$. Central directions remain fixed under conjugation and give deterministic contributions. Under the membership condition below, all fluctuations come from the simple ideals, so the variance formula sums over sectors.

A control direction contributes to a measured signal only if the input and observable are sensitive to it. Orthogonal projection measures this overlap by retaining the part of an operator in the relevant subspace. For a real Lie algebra $\mathfrak g\subseteq\fu(D)$ and a Hermitian operator $X\in\Herm(D)$, the \emph{projected purity} of $X$ onto $\mathfrak g$ is
\begin{equation}
    \calP_{\mathfrak g}(X)=\|\Pi_{\mathfrak g}(X)\|_{\HS}^2,
    \label{eq:projected_purity}
\end{equation}
where $\Pi_{\mathfrak g}:\Herm(D)\to i\mathfrak g$ is the HS-orthogonal projection onto the real subspace $i\mathfrak g=\{iX:X\in\mathfrak g\}\subseteq\Herm(D)$ of Hermitian operators associated with $\mathfrak g$. Despite its name, this is a squared Hilbert--Schmidt norm and can be applied to observables as well as states; it is not a probability.

For example, if an observable has zero projection onto a sector, that sector cannot contribute to the variance formula even if it contains many control directions. The projected purities retain this dependence on the actual state and measurement. Group averaging $X\mapsto\mathbb E_U[UXU^\dagger]$ is called a \emph{twirl}. Standard facts about this averaging and about passing between real and complex generator spaces are collected in Lemma~\ref{lem:real_form} in the appendix (see also \cite{Mele2024}).

\subsection{The DLA variance formula and the membership hypothesis}\label{sec:dla_variance}

The same control family can have different loss variances under different initialization laws. The following formula applies when the ensemble has the required Haar-measure moments and either the input or the observable lies in the specified operator space. The second condition will explain why single-pass LCU interference needs a separate calculation.

A parameterized circuit $U(\theta)\in\U(D)$ is trained by minimizing losses of the form
\begin{equation}
    f_U(\rho,O)=\Tr\!\big[O\,U(\theta)\rho\,U(\theta)\dagg\big],
    \label{eq:unitary_loss}
\end{equation}
the expectation value of a Hermitian observable $O$ after the circuit acts on an input state $\rho$.

Let $\mathfrak g\subseteq\fu(D)$ be a compact reductive Lie algebra with decomposition \eqref{eq:reductive}, and let $G=e^{\mathfrak g}\subseteq\U(D)$ be the connected compact group it generates. Let $W$ be a random unitary drawn from the Haar measure on $G$, the unique probability measure on $G$ invariant under group multiplication, or, more generally, from an ensemble $\calE$ of unitaries that is an \emph{exact second-moment design} over $G$: an ensemble satisfying
\begin{equation}
\begin{split}
 &\mathbb E_{W\sim\calE}\big[W^{\otimes t}X(W\dagg)^{\otimes t}\big]\\
 &\quad=\mathbb E_{W\sim\mathrm{Haar}(G)}\big[W^{\otimes t}X(W\dagg)^{\otimes t}\big]
\end{split}
\end{equation}
for all $X\in M_{D^t}(\mathbb C)$ and $t\in\{1,2\}$.

The tensor power means $t$ copies of the same sampled unitary, not $t$ independent draws. One copy determines the mean of an expectation value; two copies determine its square. This is why second-moment matching is sufficient for variance. Let $\sigma,X\in\Herm(D)$ satisfy
\begin{equation}
    \sigma\in i\mathfrak g+\mathbb R\,\id_D
    \quad\text{or}\quad
    X\in i\mathfrak g+\mathbb R\,\id_D.
    \label{eq:ragone_hypothesis}
\end{equation}
This is the \emph{membership hypothesis}: at least one operator can be written as a linear combination of Hermitian control directions plus a scalar identity term. The identity term is unchanged by the circuit, so it contributes no variance. Under this condition, the remaining contributions can be calculated sector by sector:
\begin{equation}
    \Var_W\big[\Tr(X\,W\sigma W\dagg)\big]=\sum_{j}\frac{\calP_{\mathfrak g_j}(\sigma)\,\calP_{\mathfrak g_j}(X)}{\dim\mathfrak g_j}
    \label{eq:ragone_formula}
\end{equation}

The numerator measures the overlap of the input and measurement with a sector; the denominator counts the directions over which sampling from Haar measure distributes that contribution. Lemma~\ref{lem:ragone} states the identity \citep{Ragone2024,Fontana2023Adjoint}. For a single unitary circuit, setting $(\sigma,X)=(\rho,O)$ gives its loss variance. Exact second-moment matching gives equality; approximate matching introduces an error controlled by the moment discrepancy. Appendix~\ref{app:design} states explicit moment-error bounds. Without the membership hypothesis \eqref{eq:ragone_hypothesis}, other invariant subspaces and their covariances can contribute, so the displayed formula need not hold \citep{Diaz2023Matchgate}. The hypothesis can be weakened to $\sigma\in i\mathfrak g+\Herm(\Comm(G))$, or the analogous condition on $X$ (Lemma~\ref{lem:gen_membership}); the matchgate analysis of Section~\ref{sec:matchgate} requires this weakened form.

\subsection{Branches, coefficients, and the LCU construction}\label{sec:lcu}

The sum $A$ need not be unitary and therefore cannot generally act as a deterministic gate on the system. The LCU construction implements it as a block of a larger unitary, selected by an ancilla measurement. Following this construction identifies the measured signal and shows where the coefficient weights and phases enter.

We combine $L\in\mathbb N$ parameterized unitary circuits acting on the same $n$-qubit system. Each constituent circuit $U_\ell(\theta_\ell)$ is called a \emph{branch}, and we index the branches by $\ell\in[L]=\{1,\ldots,L\}$. The coefficient $\alpha_\ell$ specifies that branch's amplitude and phase in the linear combination $A$ of \eqref{eq:A_intro}.

Let $\calH_s\cong\mathbb C^d$ denote the system space, with $d=2^n$ for a system of $n$ qubits and computational basis $\{\ket x:x\in\{0,1\}^n\}$, and let $\calH_a\cong\mathbb C^L$ be an ancilla register with orthonormal basis $\{\ket\ell\}_{\ell=1}^L$ indexing the branches. We write $\id_s$ and $\id_a$ for the identities on the system and ancilla spaces, respectively.

We work on this $L$-dimensional ancilla space. If a qubit register is larger, the model assumes that its preparations and mixers preserve the encoded $L$-dimensional subspace; control over unused basis states would define a different circuit model. The unitary $U_\ell(\theta_\ell)\in\U(d)$ has parameter vector $\theta_\ell\in\mathbb R^{M_\ell}$ and Hermitian gate generators $\{H_{\ell,r}\}_{r=1}^{M_\ell}\subset\Herm(d)$, where $r$ indexes the $M_\ell$ parameterized gates of branch $\ell$. Its branch DLA is
\begin{equation}
    \mathfrak g_\ell=\Lie_{\mathbb R}\{iH_{\ell,r}\}_{r=1}^{M_\ell}\subseteq\fu(d).
\end{equation}
The compact connected group $G_\ell=e^{\mathfrak g_\ell}\subseteq\U(d)$ generated by $\mathfrak g_\ell$ is called the \emph{branch group} of branch $\ell$; it is the set of unitaries the branch circuit can reach at sufficient depth.

The model applies the non-unitary operator $A(\theta,\alpha)$ of \eqref{eq:A_intro} to the input state and measures the observable $O$; its \emph{unnormalized loss} is
\begin{equation}
    \calL(\theta,\alpha)=\Tr\!\big[O\,A(\theta,\alpha)\,\rho\,A(\theta,\alpha)\dagg\big],
    \label{eq:N_def}
\end{equation}
the non-unitary counterpart of \eqref{eq:unitary_loss}. The LCU construction realizes this map as follows \citep{Childs2012LCU}.

Write $P_\ell=\ketbra{\ell}{\ell}\in M_L(\mathbb C)$ for the projector onto ancilla basis state $\ket\ell$. Assume $\lambda>0$ throughout; a zero coefficient may be assigned any phase. Decompose each coefficient as $\alpha_\ell=|\alpha_\ell|e^{i\gamma_\ell}$ with phase $\gamma_\ell\in[0,2\pi)$, and define the normalized weights $w_\ell=|\alpha_\ell|/\lambda$, so that $w_\ell\geq0$ and $\sum_\ell w_\ell=1$.

A phase applied to just one branch is a relative phase and can affect interference; we keep the phases $\gamma_\ell$ explicitly.

The overall scale $\lambda$ and the normalized weights play different roles. Changing $\lambda$ rescales the unnormalized loss; changing $w_\ell$ redistributes the branch amplitudes at fixed scale. The ancilla states
\begin{equation}
    \ket{\eta_R}=\sum_{\ell=1}^{L}\sqrt{w_\ell}\,e^{i\gamma_\ell}\ket{\ell},
    \quad
    \ket{\eta_L}=\sum_{\ell=1}^{L}\sqrt{w_\ell}\,\ket{\ell}
\end{equation}
are the preparation and postselection states of the construction. The ancilla circuits that set the coefficients, namely the preparation unitaries $\PREP_R,\PREP_L\in\U(L)$ defined by $\PREP_R\ket{\mathrm{ref}}=\ket{\eta_R}$ and $\PREP_L\ket{\mathrm{ref}}=\ket{\eta_L}$ for the reference ancilla state $\ket{\mathrm{ref}}=\ket1$, constitute the \emph{coefficient layer} of the model, as opposed to the \emph{branch layer} $\{U_\ell\}_{\ell}$. With the SELECT unitary
\begin{equation}
    S(\theta)=\sum_{\ell=1}^{L}P_\ell\otimes U_\ell(\theta_\ell)\in\U(Ld),
\end{equation}
which applies $U_\ell$ to the system controlled on the ancilla being in state $\ket\ell$, the \emph{block-encoding unitary} is
\begin{equation}
    W(\theta,\alpha)=(\PREP_L\dagg\otimes\id_s)\,S(\theta)\,(\PREP_R\otimes\id_s)\in\U(Ld),
    \label{eq:W_def}
\end{equation}
and it satisfies
\begin{equation}
\begin{split}
    (\bra{\mathrm{ref}}\otimes\id_s)\,W\,(\ket{\mathrm{ref}}\otimes\id_s)
    &=(\bra{\eta_L}\otimes\id_s)\,S(\theta)\,(\ket{\eta_R}\otimes\id_s)\\
    &=A/\lambda ;
\end{split}
\end{equation}
the circuit is shown in Figure~\ref{fig:arch}. Running the circuit on $\ketbra{\mathrm{ref}}{\mathrm{ref}}\otimes\rho$ and measuring the ancilla register in the basis $\{\ket\ell\}_{\ell\in[L]}$ yields the outcome corresponding to $\ket{\mathrm{ref}}$ with the \emph{postselection success probability}
\begin{equation}
    \psucc=\Tr\!\big[A\rho A\dagg\big]/\lambda^2\in[0,1],
\end{equation}
and, conditioned on this outcome, the system is left in the state $A\rho A\dagg/\Tr[A\rho A\dagg]$. Postselecting on the reference outcome therefore gives access to the normalized loss $\calL/\Tr[A\rho A\dagg]$; Section~\ref{sec:postselection} analyzes how trainability transfers between the unnormalized and the normalized loss.

For nonnegative coefficients one can choose $\PREP_L=\PREP_R$; fixed phases may be absorbed into SELECT, while trainable phase and mixing gates belong to the coefficient layer. If the coefficient preparations are parameterized circuits with ancilla gate generators $\{K_\xi\}_{\xi=1}^{m_c}\subset\Herm(L)$, where $\xi$ indexes the $m_c$ coefficient-layer gates, their DLA is
\begin{equation}
    \gcoeff=\Lie_{\mathbb R}\{iK_\xi\}_{\xi=1}^{m_c}\subseteq\fu(L);
\end{equation}
for fixed (non-trainable) coefficients, $\gcoeff=\{0\}$. We call a trainable coefficient layer a \emph{mixer} when some generator $K_\xi$ has nonzero off-diagonal entries in the ancilla basis, that is, when the coefficient layer can move amplitude between branches.

\subsection{Lifted states and observables}\label{sec:lifted}

To apply unitary variance formulas to the non-unitary sum $A$, we include the ancilla in the state and measurement. We call these joint-space operators \emph{lifted states and observables}. They express the same measured quantity on a larger space.

There are two convenient ways to write that same expectation. A \emph{frame} here specifies whether the preparation gates are included in the evolving unitary or absorbed into the state and measurement. In the \emph{block-encoding frame}, we keep the complete circuit $W$ and use
\begin{equation}
    \widetilde\rho_{\mathrm{ref}}=P_{\mathrm{ref}}\otimes\rho,
    \qquad
    \widetilde O_{\mathrm{ref}}=P_{\mathrm{ref}}\otimes O,
    \label{eq:lift_BE}
\end{equation}
where $P_{\mathrm{ref}}=\ketbra{\mathrm{ref}}{\mathrm{ref}}$ projects onto the ancilla reference state.

The observable contains this projector because unsuccessful ancilla outcomes contribute zero to the unnormalized measurement. The joint state is normalized, but the selected system block need not be. In the \emph{SELECT frame}, the corresponding operators are
\begin{equation}
    \widetilde\rho_\alpha=\ketbra{\eta_R}{\eta_R}\otimes\rho,
    \qquad
    \widetilde O_\alpha=\ketbra{\eta_L}{\eta_L}\otimes O.
    \label{eq:lift_select}
\end{equation}
The lifted state and observable in \eqref{eq:lift_select} are equivalent to those in \eqref{eq:lift_BE} in the sense that both frames represent the same loss:
\begin{equation}
    \Tr\!\big[\widetilde O_{\mathrm{ref}}\,W\widetilde\rho_{\mathrm{ref}}W\dagg\big]
    =\Tr\!\big[\widetilde O_\alpha\,S\widetilde\rho_\alpha S\dagg\big]
    =\calL/\lambda^2.
    \label{eq:N_lifts}
\end{equation}
The useful frame depends on what is randomized. At fixed coefficients, the preparation gates are deterministic, so the SELECT frame places all coefficient dependence in the state and observable. If preparations vary, the block-encoding frame instead keeps those operators fixed and puts the variation into $W$.

In the SELECT frame, the lifted operators contain terms $\ketbra{\ell}{k}$ with $\ell\ne k$, connecting different branch labels. The fixed-coefficient SELECT algebra has no such connections, so these terms generally violate its membership condition. They encode the interference retained in the direct single-pass LCU calculation. In the block-encoding frame, a different condition still needs checking: random preparations alone do not make $W$ a design.

For a general joint unitary $W$, including a circuit with repeated coefficient and branch layers, define a separate objective
\begin{equation}
    \LBE(W)=\lambda^2\Tr[\widetilde O_{\mathrm{ref}}W\widetilde\rho_{\mathrm{ref}}W\dagg],
    \qquad \lambda>0\ \text{fixed}.
    \label{eq:lifted_loss_general}
\end{equation}
Equation~\eqref{eq:N_lifts} gives $\LBE=\calL$ for the single-pass LCU construction \eqref{eq:W_def}. For a deeper joint circuit, \eqref{eq:lifted_loss_general} remains a projected-block loss, but its block need not equal the original sum $A/\lambda$. We reserve $\calL$ for that original LCU loss and use $\LBE$ for general interleaved-circuit results.

For a general interleaved circuit, $\lambda$ in this definition is only a fixed reporting scale, with no implied coefficient decomposition. Dividing $\LBE$ by $\lambda^2$ gives the directly measured joint expectation; all of its variance formulas are obtained by dividing the displayed variances by $\lambda^4$. We call this the \emph{joint Haar-random regime} when the ensemble of $W$ has first two conjugation moments equal to the Haar-measure moments of its generated group. 

\subsection{Two circuit models and scope}\label{sec:scope}\label{rem:design_scope}\label{rem:indep}

Changing coefficients alters the amplitudes in a sum; adding alternating layers also creates products of branch operations. A variance prediction must specify which architecture is used and how it is sampled. We collect those distinctions and the scope of the results here.

A \emph{single-pass LCU} is precisely the preparation--SELECT--unpreparation circuit in \eqref{eq:W_def}. Its coefficients may be fixed or sampled, but only one SELECT operation appears. An \emph{interleaved circuit} instead has the form
\begin{equation}
 W_T=(C_T\otimes\id_s)S_T\cdots
       (C_1\otimes\id_s)S_1(C_0\otimes\id_s),
 \label{eq:interleaved}
\end{equation}
where $T$ counts branch layers, $C_t$ are allowed ancilla unitaries, and each $S_t$ is an allowed controlled branch layer. Its projected block contains products of branch operations along ancilla paths; it need not equal the original sum $A/\lambda$. The algebra below describes the control family with arbitrarily many such layers and independently available generator directions.

A \emph{joint-space second-moment design} is an ensemble of these unitaries whose first two conjugation moments match those of Haar measure on its generated group, in the sense already defined in Section~\ref{sec:dla_variance}. This is an additional sampling hypothesis, not a consequence of a large generated algebra. Random preparations in a single-pass LCU do not establish it. General results on designs offer methods for proving convergence, but a depth bound must be established for the particular architecture \citep{HarrowLow2009,BrandaoHorodecki2016,Haferkamp2022,Schuster2025}.

The single-pass LCU loss decomposition assumes independent branches with zero first moments; group-based closed forms additionally use specified Haar-measure moments. The gradient floor needs zero means only for the undifferentiated branches and independence from the differentiated branch and its derivative. Shared-parameter architectures and near-identity initializations require checking these hypotheses separately. For example, Quixer uses token unitaries from one common circuit with token-dependent angles; its later QSVT stage supplies the polynomial transform \citep{Khatri2024Quixer}. Its branches are not powers of one shared unitary, and independence does not follow from that construction.

The applications use different measured objectives. Khatri et al.~\cite{Khatri2025PLCU} study $\calL$ with $\lambda=1$, so it also equals $\calL/\lambda^2$. In contrast, Eq.~(6) of Yao et al.~\cite{Yao2025LCQNN} measures $\sum_j p_j\Tr[OU_j\rho U_j^\dagger]$ after leaving the ancilla unobserved. This is an incoherent mixture rather than a projected coherent loss. Their distinct branch parameter blocks give independent branches if sampled independently; separate marginal design assumptions alone do not imply independence. Quixer uses normalized expectation features after its polynomial transform, followed by a classical prediction loss \citep{Khatri2024Quixer}. Its accompanying implementation normalizes the transformed state before measuring those features. They therefore have the form of a normalized postselected expectation for the transformed block, rather than the single-pass LCU $\calL$ or $\calL/\lambda^2$.

For $O\succeq0$, destructive interference can give $\calL=0$ whenever the ansatz permits $A\rho^{1/2}=0$. Such a point is a global minimum of the unnormalized loss but has zero success probability, so the normalized output is undefined there. A gradient floor for $\calL$ therefore does not by itself establish useful learning of the conditional output.

The core single-pass LCU and joint-space formulas concern $\calL$ or the explicitly scaled block objective $\LBE$. A normalized postselected objective also depends on the random success factor. Initialization variance is not a guarantee of optimization success \citep{Anschuetz2022Traps,Arrasmith2022}; neither a variance bound nor an algebra dimension establishes computational advantage \citep{Cerezo2024Sim,Goh2023LieSim}. Section~\ref{sec:audit} explains normalization, finite-moment errors, and measurement costs; Appendices~\ref{app:postselect} and~\ref{app:design} provide the bounds and proofs.

\section{Exact single-pass LCU loss and gradient moments}\label{sec:variance}

For a single-pass LCU, the training signal contains a contribution from each branch and a contribution from interference. We first prove a gradient lower bound under independence and zero-mean assumptions, then evaluate the full loss and gradient variances with more information about the branch moments. Until Section~\ref{sec:random_prep}, coefficients are fixed and all averages concern branch initialization.

For any branch pair, define the overlap
\begin{equation}
 m_{\ell k}=\Tr[OU_\ell\rho U_k^\dagger],\qquad m_\ell=m_{\ell\ell}\in\mathbb R.
\end{equation}
The diagonal overlap is a branch loss. The off-diagonal overlap measures coherence between two branches and satisfies $m_{k\ell}=\overline{m_{\ell k}}$. For a pure input, the two branch outputs are $U_\ell\ket\psi$ and $U_k\ket\psi$, and $m_{\ell k}$ is their matrix element through $O$. This explains why an off-diagonal overlap can be complex even though the final loss is real. With two branches,
\begin{equation}
 \calL=|\alpha_1|^2m_1+|\alpha_2|^2m_2
       +2\operatorname{Re}(\alpha_1\bar\alpha_2m_{12}).
 \label{eq:two_branch_loss}
\end{equation}
Writing $z=\alpha_1\bar\alpha_2m_{12}$, the identity $(2\operatorname{Re}z)^2=2|z|^2+2\operatorname{Re}(z^2)$ explains why the variance can contain both $\mathbb E|m_{12}|^2$ and $\mathbb E(m_{12}^2)$. Only the contribution involving the latter changes with the relative coefficient phase. Before evaluating these moments, we establish the gradient signal that survives irrespective of their values.

\subsection{The single-branch gradient floor}\label{sec:floor}

Can interference erase a branch's measurable gradient signal? Under the conditions below, it cannot reduce the total variance below a fixed multiple of that branch's variance. We call this lower bound the \emph{gradient floor}; it need not account for the entire signal.

The other branches must be independent of the differentiated branch and its derivative, with unitary matrices whose entrywise means are zero. Holding the differentiated branch fixed then gives zero mean for its interference with the others. The resulting squared terms can only add to the variance.

\begin{proposition}[Single-branch gradient floor]\label{cor:floor}
Fix the complex coefficients and let $\theta$ occur only in branch $\ell$. Write $\dot U_\ell=\partial_\theta U_\ell$ and $g_\ell^{\mathrm{branch}}=\partial_\theta\Tr[OU_\ell\rho U_\ell^\dagger]$. Assume that the pair $(U_\ell,\dot U_\ell)$ and the other branch unitaries are mutually independent, that $\mathbb E U_k=0$ for $k\ne\ell$, and that the displayed second moments are finite. Then
\begin{equation}
 \Var(\partial_\theta\calL)
 \geq |\alpha_\ell|^4\Var(g_\ell^{\mathrm{branch}}).
 \label{eq:floor}
\end{equation}
No Haar-random sampling law, zero mean for $U_\ell$, or restriction on the position of the differentiated gate is required.
\end{proposition}

The proof conditions on the differentiated branch and its derivative: it holds them fixed while averaging over the other branches. Zero conditional means remove the covariances. The nonnegative terms that remain are averaged squares, not a claim that interference has a positive sign in every circuit.
\begin{proof}[Proof of Proposition~\ref{cor:floor}]\phantomsection\label{proof:cor:floor}
Set $z_k=\Tr[O\dot U_\ell\rho U_k^\dagger]$ and
$R_k=2\mathrm{Re}(\alpha_\ell\bar\alpha_kz_k)$. Direct differentiation gives
\begin{equation}
 \partial_\theta\calL=|\alpha_\ell|^2g_\ell^{\mathrm{branch}}+\sum_{k\ne\ell}R_k.
\end{equation}
Conditioned on $(U_\ell,\dot U_\ell)$, each $R_k$ has zero mean, and the $R_k$ are independent. Thus they have zero covariance with each other and with $g_\ell^{\mathrm{branch}}$. Consequently
\begin{equation}
 \Var(\partial_\theta\calL)
 =|\alpha_\ell|^4\Var(g_\ell^{\mathrm{branch}})+\sum_{k\ne\ell}\mathbb E[R_k^2],
\end{equation}
which proves the claim by nonnegativity of the squares.
\end{proof}

An inverse-polynomial branch variance and an inverse-polynomial lower bound on $|\alpha_\ell|$ are therefore \emph{sufficient} to retain an inverse-polynomial unnormalized gradient variance. They are not necessary: interference may supply the signal even if the branch term vanishes. For the hardware-scaled objective $\calL/\lambda^2$, the coefficient factor is $w_\ell^4$, equal to $L^{-4}$ at balanced weights. The floor applies to interior parameters as well as edge parameters. Gate placement and Haar-random sampling assumptions enter the closed-form evaluation in Section~\ref{sec:grad_select}.

\subsection{Exact loss moments}\label{sec:exact_select}

To calculate the full loss variance, we must retain both the size and the phase dependence of interference. Independence and zero matrix means remove terms in which a branch appears only once. Two distinct pair moments survive.

\begin{assumption}[Independent branch moments]\label{asm:moments}
The branch unitaries are independent random variables satisfying (a) $\mathbb E[U_\ell]=0$ for every $\ell\in[L]$. We will explicitly indicate when we also impose the optional condition (b) $\mathbb E[U_\ell\otimes U_\ell]=0$ for every branch. Group-based formulas additionally assume Haar-random branches on compact groups $G_\ell\subseteq\U(d)$, or ensembles matching all Haar-measure moments used in those formulas.
\end{assumption}

For Haar-random branches, condition (a) holds when no nonzero state vector is fixed by every element of $G_\ell$. In representation language, its defining action on $\mathbb C^d$ has no trivial subrepresentation. This holds for $\U(d)$, for $\SU(d)$ with $d\geq2$, and for the full matchgate group.

Condition (b) fails when the group preserves a nonzero bilinear form: a matrix $J$ satisfying $UJU^T=J$ for every $U\in G_\ell$. Such forms occur for \emph{self-dual} representations, which are unitarily equivalent to their entrywise complex conjugates, including the defining representations of $\SU(2)$ and the matchgate group. The variance decomposition needs only (a); condition (b) removes its phase-sensitive term.

Zero matrix mean is different from a zero conjugation moment: $\mathbb E U=0$ can hold even though $\mathbb E[U\rho U^\dagger]$ is a nonzero density operator. In the expansion below, the former eliminates terms containing a branch only once; the latter determines the surviving magnitude moments. Haar-measure moments are used only to evaluate the cross moments through the invariant spaces below.

\begin{theorem}[Exact SELECT-frame variance]\label{thm:exact_select}
Under Assumption~\ref{asm:moments}(a), for any fixed coefficients $\alpha_\ell=\lambda w_\ell e^{i\gamma_\ell}$,
\begin{equation}
\begin{split}
    \Var[\calL]
    &=\lambda^4\sum_{\ell=1}^{L}w_\ell^4\,\Var[m_\ell]\\
    &\;+\lambda^4\sum_{\ell\neq k}w_\ell^2w_k^2\Big(\mathbb E\big[|m_{\ell k}|^2\big]\\[-2pt]
    &\hspace{2em}+\mathrm{Re}\big[e^{2i(\gamma_\ell-\gamma_k)}\,\mathbb E(m_{\ell k}^2)\big]\Big).
    \label{eq:exact_select}
\end{split}
\end{equation}
Under Assumption~\ref{asm:moments}(b) the unconjugated moments $\mathbb E(m_{\ell k}^2)$ vanish and the coefficient phases $\gamma_\ell$ drop out entirely; in general the phases enter \emph{only} through the last term.
\end{theorem}
\noindent\hyperref[proof:thm:exact_select]{Proof} in Appendix~\ref{app:select}.

If the ensemble of the branch unitary $U_\ell$ forms a second-moment design over $G_\ell=e^{\mathfrak g_\ell}$ and $\rho$ or $O$ lies in $i\mathfrak g_\ell+\mathbb R\id_s$, the diagonal terms resolve by branch sectors (Lemma~\ref{lem:ragone} applied branchwise, with $\mathfrak g_\ell=\mathfrak z_\ell\oplus\bigoplus_j\mathfrak g_\ell^j$ the reductive decomposition of the branch DLA):
\begin{equation}
    \Var[m_\ell]=\sum_j\frac{\calP_{\mathfrak g_\ell^j}(\rho)\,\calP_{\mathfrak g_\ell^j}(O)}{\dim\mathfrak g_\ell^j}.
    \label{eq:diag_sector}
\end{equation}

\paragraph{The two interference channels.}
The first sum in \eqref{eq:exact_select} contains the individual branch variances, weighted by the fourth powers of their coefficient magnitudes. The second sum contains interference. The conjugated moment $\mathbb E|m_{\ell k}|^2$ measures the squared magnitude of a complex overlap. The unconjugated moment $\mathbb E(m_{\ell k}^2)$ also retains phase information: multiplying the overlap by $e^{i\phi}$ leaves the first moment unchanged and multiplies the second by $e^{2i\phi}$.

To evaluate these averages for Haar-random branches, we look for matrices unchanged by the relevant group action. Squared magnitudes involve conjugation, $Y\mapsto UYU^\dagger$, whose fixed matrices form the commutant. Complex squares instead involve $J\mapsto UJU^T$. Its fixed matrices are the \emph{invariant bilinear forms} defined next. The transpose rather than the adjoint is the essential distinction.

The conjugation average pairs a matrix entry with a complex-conjugated entry. The complex square pairs two unconjugated entries. Their invariant spaces need not agree.

\begin{definition}[Invariant bilinear forms]\label{def:forms}
For a compact $G\subseteq\U(d)$, define
\begin{equation}
 \Jforms(G)=\{J\in M_d(\mathbb C):UJU^T=J\ \text{for all }U\in G\}.
\end{equation}
Its dimension is $\tau(G)=\dim_{\mathbb C}\Jforms(G)$. Each matrix in this space encodes a pairing preserved by two unconjugated copies of the group action.

To relate this space to a moment, arrange a matrix's entries into a vector by $\mathrm{vec}(J)=\sum_{ab}J_{ab}\ket a\otimes\ket b$. The Haar-measure average $\mathbb E_{U\sim G}[U\otimes U]$ is the orthogonal projector onto $\mathrm{vec}\,\Jforms(G)$. Thus $\tau(G)=0$ exactly when Assumption~\ref{asm:moments}(b) holds.

Equivalently, $\tau(G)$ counts ordered mutually dual pairs of irreducible subrepresentations, including multiplicities. Here two unitary representations are dual when one is equivalent to the entrywise complex conjugate of the other. Schur's lemma says that an invariant map between inequivalent irreducible actions is zero, and between equivalent ones is fixed up to a scalar. This explains why only dual pairs are counted. The projector description is what we use for the pair-moment calculation.
\end{definition}

For branch $\ell$, write $\kappa_\ell=\dim_{\mathbb C}\Comm(G_\ell)$ and $\tau_\ell=\tau(G_\ell)$. Both invariant spaces can be found by solving homogeneous linear equations constructed from the generators. Their kernels, meaning the matrices mapped to zero by those equations, are precisely the allowed invariants. The bases themselves enter the exact pairings; their dimensions only count how many terms are available.

The next formula expresses the magnitude moment using invariant matrices from both branches. An HS-orthonormal basis has Hilbert--Schmidt inner products $\delta_{ij}$. The traces below measure how the input and observable overlap pairs of these matrices.

\begin{lemma}[Commutant formula for conjugated cross moments]\label{lem:cross_moments}
For distinct branches $\ell\neq k$ with independent Haar-random unitaries (or the same required moments), let $\{P_i\}_i$ and $\{Q_j\}_j$ be HS-orthonormal bases of $\Comm(G_\ell)$ and $\Comm(G_k)$, respectively. For any $B\in M_d(\mathbb C)$, not necessarily Hermitian,
\begin{equation}
\begin{split}
    &\mathbb E\Big[\big|\Tr[B\,U_\ell\rho\,U_k\dagg]\big|^2\Big]\\
    &\qquad=\sum_{i,j}\Tr\big[\rho\,P_i\dagg\,\rho\,Q_j\big]\,\Tr\big[B\,P_i\,B\dagg\,Q_j\dagg\big].
    \label{eq:cross_commutant_gen}
\end{split}
\end{equation}
$\mathbb E[|m_{\ell k}|^2]$ is the case $B=O$. If both commutants admit HS-orthonormal bases whose elements are proportional to unitaries, then, with $\kappa_\ell=\dim_{\mathbb C}\Comm(G_\ell)$,
\begin{equation}
    \mathbb E\big[|m_{\ell k}|^2\big]\leq\frac{\kappa_\ell\,\kappa_k\,\Tr(\rho^2)\Tr(O^2)}{d^2}.
    \label{eq:cross_bound}
\end{equation}
\end{lemma}
\noindent\hyperref[proof:lem:cross_moments]{Proof} in Appendix~\ref{app:select}.

For full matchgate branches and $B=O$, this formula recovers Lemma~1 of Khatri et al.~\cite{Khatri2025PLCU}. Their Appendix~E.1 also uses a commutant projector to evaluate this moment; the statement here allows different branch groups and a non-Hermitian test matrix.

The phase-dependent moment is evaluated in the same spirit, but with the invariant forms. A group may have no nonzero forms, in which case this contribution vanishes. When forms are present, their overlaps with the input and observable decide whether a phase effect actually survives.

\begin{lemma}[Form-pairing formula for unconjugated cross moments]\label{lem:form_pairing}
For distinct branches $\ell\neq k$ with independent Haar-random unitaries (or the same required moments), let $\{J_s\}_s$ and $\{J_t\}_t$ be HS-orthonormal bases of $\Jforms(G_\ell)$ and $\Jforms(G_k)$, respectively. For any $B\in M_d(\mathbb C)$,
\begin{equation}
\begin{split}
    &\mathbb E\Big[\Tr[B\,U_\ell\rho\,U_k\dagg]^2\Big]\\
    &\qquad=\sum_{s,t}\Tr\big[J_s\dagg\,\rho\,J_t\,\rho^{T}\big]\,\Tr\big[J_t\dagg\,B\,J_s\,B^{T}\big].
    \label{eq:form_pairing}
\end{split}
\end{equation}
The moment $\mathbb E(m_{\ell k}^2)$ is the case $B=O$. With $\nu_\ell=\max_s\|J_s\|_\infty$, the modulus of \eqref{eq:form_pairing} is at most $\tau_\ell\tau_k\nu_\ell^2\nu_k^2\,\Tr(\rho^2)\Tr(B\dagg B)$; for a block-supported orthonormal form basis of the matchgate group, $\nu^2=2/d$, which is the same dependence on $1/d^2$ as \eqref{eq:cross_bound}.
\end{lemma}
\noindent\hyperref[proof:lem:form_pairing]{Proof} in Appendix~\ref{app:select}.

For irreducible branch groups, the commutants contain only scalar matrices and the bound in Lemma~\ref{lem:cross_moments} is exact. Larger commutants allow more pairings, but their contributions depend on the contractions with $\rho$ and $O$. Dimension alone does not determine the cross moment.

The invariant-form integrals in Lemma~\ref{lem:form_pairing} are familiar from integration with respect to Haar measure on orthogonal and symplectic groups \citep{CollinsSniady2006}. Their pairing here retains the coefficient-phase dependence even when the two branch groups differ. The bound depends on the operator norms of the normalized forms as well as their number.

The invariant-form contribution need not be present. It vanishes for $\U(d)$ and for $\SU(d)$ with $d\geq3$, and an independent uniform global phase removes it from any branch ensemble. Lemma~\ref{lem:vanishing} gives additional input-dependent criteria. For full matchgate branches, $\tau_\ell=2$ and a block-supported form basis has $\nu_\ell^2=2/d$. Pure fermionic Gaussian inputs, including computational-basis states, eliminate the form term, whereas suitable non-Gaussian parity-even inputs can retain it; the following example uses a four-qubit Greenberger--Horne--Zeilinger (GHZ) state. Invariant forms permit phase dependence but do not guarantee a nonzero moment for every input and observable.

The following four-qubit example evaluates the phase-dependent term explicitly. Majorana operators give a compact matrix description of the matchgate group and its invariant forms.

\begin{example}[A nonzero unconjugated matchgate moment]\label{ex:ghz}
On four qubits, define the Majorana operators
$c_{2r-1}=Z^{\otimes(r-1)}\otimes X\otimes\id^{\otimes(4-r)}$ and
$c_{2r}=Z^{\otimes(r-1)}\otimes Y\otimes\id^{\otimes(4-r)}$ for $1\leq r\leq4$, where $X,Y,Z$ are Pauli matrices. They satisfy $\{c_j,c_k\}=2\delta_{jk}\id$. The full matchgate group is generated by the quadratic directions $c_jc_k/2$.
Take independent Haar-random branches from this group and set
\begin{gather}
 \ket\psi=(\ket{0000}+\ket{1111})/\sqrt2,\qquad
 \rho=\ketbra{\psi}{\psi},\\
 O=ic_1c_2+\tfrac12ic_3c_6.
\end{gather}
Let $P=Z^{\otimes4}$, $Q_+=(\id+P)/2$, and
$J_+=Q_+c_1c_3c_5c_7/\sqrt8$. This is a Hilbert--Schmidt normalized invariant form on the even-parity subspace: every quadratic generator $X_0$ satisfies $X_0J_++J_+X_0^T=0$. The input has support only in that subspace: $Q_+\rho=\rho$, and the projector onto odd parity annihilates it. Consequently the odd-block form makes no contribution to the input contraction in Lemma~\ref{lem:form_pairing}; only $J_+$ is needed here. Direct multiplication gives
\begin{align}
 \Tr[J_+^\dagger\rho J_+\rho^T]&=1/8,\\
 \Tr[J_+^\dagger OJ_+O^T]&=-5/4.
\end{align}
The form-pairing lemma therefore yields
\begin{equation}
 \mathbb E(m_{12}^2)=-5/32,\qquad
 \mathbb E|m_{12}|^2=5/32.
\end{equation}
For $L=2$ the interference contribution to the loss variance is
\begin{equation}
 \frac{5\lambda^4w_1^2w_2^2}{16}
 \big[1-\cos 2(\gamma_1-\gamma_2)\big].
\end{equation}
It vanishes for equal phases and is maximal for a relative phase of $\pi/2$. This cancellation appears only when the unconjugated moment is retained.

Invariance of Haar measure permits multiplication only by elements of the sampled group. Here $i\id$ is excluded: it would send any nonzero invariant form $J$ to $(i\id)J(i\id)^T=-J$. For $\SU(d)$, $i\id$ belongs to the group precisely when $4$ divides $d$; its absence alone does not imply a nonzero second moment. In fact that moment vanishes for the defining representation of $\SU(d)$ for $d\geq3$, as stated above. For $\SU(2)$, the invariant form is antisymmetric, so pure inputs still give a vanishing moment by Lemma~\ref{lem:vanishing}(ii); mixed inputs need not. Thus the applicability of the phase-free formula depends on both the branch group and the input.
\end{example}

To compare interference with the inherited branch signal in a particular example, we use the fraction of total variance contributed by the cross terms. This diagnostic is computed from the actual moments, not from the dimensions of the invariant spaces alone.

\begin{definition}[Variance interference fraction]\label{def:chi}
Write $\Var[\calL]=V_{\mathrm{diag}}+V_{\mathrm{int}}$, using the first and second sums of \eqref{eq:exact_select}. Define $\chi=V_{\mathrm{int}}/\Var[\calL]$ when the total variance is positive, and set $\chi=0$ when it is zero. Since $|\mathbb E(m_{\ell k}^2)|\leq\mathbb E|m_{\ell k}|^2$, both contributions are nonnegative and $\chi\in[0,1]$. Independence gives $V_{\mathrm{diag}}=\Var(\sum_\ell|\alpha_\ell|^2m_\ell)$, the variance after deleting coherent cross terms. These squared coefficient weights are not generally the probabilities $w_\ell$ of a classical branch mixture.
\end{definition}

\subsection{Exact gradient moments}\label{sec:grad_select}

Loss fluctuations do not determine the gradient signal: differentiation changes both the overlap operator and the sign of its phase-dependent contribution. We evaluate these changes for a \emph{measurement-adjacent} gate, which acts last in its branch. Its derivative is a fixed matrix acting on the branch output. The gradient floor in Section~\ref{sec:floor} applies at any gate position.

\begin{theorem}[Exact SELECT-frame gradient variance]\label{thm:grad_select}
Fix coefficients $\alpha_\ell=\lambda w_\ell e^{i\gamma_\ell}$ and let Assumption~\ref{asm:moments}(a) hold, with each undifferentiated branch Haar-random on $G_k$. Let $\theta$ be a parameter of branch $\ell$ in measurement-adjacent form $U_\ell=e^{i\theta H}V_\ell$, with $H\in\Herm(d)$, $iH\in\mathfrak g_\ell$, and $V_\ell$ Haar-random on $G_\ell$. The variance is over the independent branch ensembles at fixed $\theta$; it also holds if $\theta$ is randomized independently of them. Write $\Gamma_{\ell k}=\Tr[(OH)\,U_\ell\rho\,U_k\dagg]$. Then
\begin{equation}
\begin{split}
    \Var\big(\partial_\theta \calL\big)
    &=\lambda^4w_\ell^4\,\Var\big(\Tr[\,i[O,H]\,U_\ell\rho U_\ell\dagg]\big)\\
    &\;+2\lambda^4\!\sum_{k\neq\ell}w_\ell^2w_k^2\Big(\mathbb E\big[|\Gamma_{\ell k}|^2\big]\\[-2pt]
    &\hspace{2em}-\mathrm{Re}\big[e^{2i(\gamma_\ell-\gamma_k)}\mathbb E(\Gamma_{\ell k}^2)\big]\Big).
    \label{eq:grad_select}
\end{split}
\end{equation}

The moments $\mathbb E|\Gamma_{\ell k}|^2$ and $\mathbb E(\Gamma_{\ell k}^2)$ are given in closed form by Lemmas~\ref{lem:cross_moments} and~\ref{lem:form_pairing} with $B=OH$. Whenever the branchwise membership hypothesis holds, the diagonal factor resolves by sector as in \eqref{eq:diag_sector} with observable $i[O,H]$:
\begin{equation}
 \Var\big(\Tr[i[O,H]U\rho U^\dagger]\big)
 =\sum_j\frac{\calP_{\mathfrak g_\ell^j}(\rho)\calP_{\mathfrak g_\ell^j}(i[O,H])}{\dim\mathfrak g_\ell^j}.
 \end{equation}
\end{theorem}
\noindent\hyperref[proof:thm:grad_select]{Proof} in Appendix~\ref{app:gradsel}.

The matrix $OH$ records the insertion of the differentiated gate into a cross-branch overlap. It need not be Hermitian, which is why the two pairing lemmas allow a general test matrix $B$.

The sign change in \eqref{eq:grad_select} comes from differentiating the overlap. The loss uses its real part; the derivative introduces a factor of $i$ and hence its imaginary part. The overlap operator changes as well, from $O$ in the loss to $OH$ in the gradient.

\subsection{Haar-random preparation in a single-pass LCU}\label{sec:random_prep}

Sampling the coefficients introduces another source of randomness without adding a branch layer. A useful baseline uses matched random preparation and unpreparation: the resulting nonnegative branch weights are uniform over the set of weights summing to one.

Specifically, let $C$ be Haar-random on $\U(L)$ independently of the branches, and use the same $C$ for preparation and unpreparation. Then
\begin{equation}
 W=(C^\dagger\otimes\id_s)S(C\otimes\id_s),\qquad
 w_\ell=|\bra\ell C\ket{\mathrm{ref}}|^2.
\end{equation}
The weights follow the uniform Dirichlet law $\mathrm{Dir}(1,\ldots,1)$ on the probability simplex, the set of nonnegative weights summing to one. A column of a Haar-random unitary can be obtained by normalizing a vector of independent standard complex Gaussian variables. Their squared magnitudes are independent exponential variables; dividing by their sum gives the stated weights. The selected block is $\sum_\ell w_\ell U_\ell$, so $\lambda=1$ and the coefficients are nonnegative. Independent left and right preparations would give a different law.

The law of total variance separates branch randomness from coefficient randomness. It adds the average variance at fixed coefficients to the variance of the conditional mean as those coefficients change. Both terms enter the baseline below; Appendix~\ref{app:select} gives the proof.

\begin{theorem}[Exact consistency with the Dirichlet-averaged formula]\label{thm:KZM_consistency}
Let the branches be independent Haar-random draws from the same compact group $G$, satisfying Assumption~\ref{asm:moments}(a), and suppose $\mathbb E(m_{\ell k}^2)=0$ for every $\ell\neq k$. Condition (b) is sufficient but not necessary for this vanishing. Set $\lambda=1$ and $\alpha_\ell=c_\ell\geq0$, and draw the coefficient vector $c=(c_1,\dots,c_L)$ from the uniform Dirichlet distribution $\mathrm{Dir}(1,\ldots,1)$, the uniform distribution on the probability simplex $\{c_\ell\geq0,\sum_\ell c_\ell=1\}$, independently of the branch parameters. With $\mu=\mathbb E[m_\ell]$,
\begin{equation}
\begin{split}
    \Var[\calL]
    &=\frac{24\,\mathbb E[m_\ell^2]+4(L-1)\big(\mu^2+\mathbb E[|m_{\ell k}|^2]\big)}{(L+1)(L+2)(L+3)}\\
    &\quad-\frac{4\mu^2}{(L+1)^2}.
\end{split}
\label{eq:KZM_exact}
\end{equation}

The expression coincides term by term with Theorem~2, Eq.~(9), of Khatri et al.~\cite{Khatri2025PLCU} specialized to identical groups.
\end{theorem}
\noindent\hyperref[proof:thm:KZM_consistency]{Proof} in Appendix~\ref{app:select}.

The denominators here come from averaging powers of the coefficient weights. They are not dimensions of a larger joint circuit group.

\begin{remark}[The additional form term after coefficient averaging]\label{rem:KZM_completion}
When the shared branch group carries invariant forms ($\tau>0$) and the inputs do not satisfy Lemma~\ref{lem:vanishing}, the Dirichlet-averaged variance \eqref{eq:KZM_exact} acquires the additional exact term
\begin{equation}
 \frac{4(L-1)\,\mathrm{Re}\,\mathbb E(m_{\ell k}^2)}{(L+1)(L+2)(L+3)},
\end{equation}
by the same fourth-moment bookkeeping with $\mathbb E[c_\ell^2c_k^2]$ weights. For pure fermionic Gaussian inputs and Haar-random full-matchgate branches, the term vanishes by Lemma~\ref{lem:vanishing}(iv), so the comparison with Eq.~(9) of Khatri et al.~\cite{Khatri2025PLCU} holds in this case even though the group itself has invariant forms. More general inputs require evaluating the form pairing: it need not vanish for mixed inputs or for entangled parity-even inputs such as the four-qubit example above.
\end{remark}

\section{Algebra generated by coefficient control}\label{sec:sandwich}

Repeatedly alternating branch operations and ancilla mixing allows products of branch operations beyond a single LCU sum. We now determine the transformations these controls generate. Dimension bounds first limit the possible algebra; a saturation theorem then identifies it under full special-unitary ancilla control. Section~\ref{sec:joint_variance} adds the sampling assumptions needed for variance predictions.

\subsection{Dimension bounds}

How much control can mixing add to the branch circuits? Before determining the complete generated algebra, we bound the number of independent directions it can contain. The branch and ancilla controls give a lower bound, with shared directions counted once. An upper bound comes from a matrix algebra containing every generated product, the \emph{associative envelope}. Comparing them shows how much additional control remains possible.

\begin{definition}[Interleaved-circuit DLA]\label{def:gBE}
The \emph{interleaved-circuit DLA}, denoted $\gint$, for branch DLAs $\{\mathfrak g_\ell\}_{\ell=1}^L$ and coefficient-layer DLA $\gcoeff$ is
\begin{equation}
    \gint=\Lie_{\mathbb R}\Big(\{K\otimes\id_s:K\in\gcoeff\}
    \cup\bigcup_{\ell\in[L]}P_\ell\otimes\mathfrak g_\ell\Big),
    \label{eq:gBE}
\end{equation}
a real Lie subalgebra of $\fu(Ld)$, where $P_\ell\otimes\mathfrak g_\ell=\{P_\ell\otimes X:X\in\mathfrak g_\ell\}$.
\end{definition}

In \eqref{eq:gBE}, $K\otimes\id_s$ acts on the ancilla alone, whereas $P_\ell\otimes X$ applies a branch generator $X\in\mathfrak g_\ell$ conditional on label $\ell$. The Lie closure includes arbitrary repetition of these controls. Fixed gates between trainable gates must be accounted for by conjugating the associated generator families before taking this closure \citep{Larocca2022DLA,DAlessandro2007}. Fixed single-pass LCU preparations are instead handled by the frame transformation in Section~\ref{sec:lifted}: setting $\gcoeff=0$ describes the SELECT branch algebra, not the group containing the full prepared unitary.

As a compact reductive algebra, $\gint$ decomposes as $\gint=\mathfrak z_{\INT}\oplus\bigoplus_{j=1}^{r}\mathfrak g_j^{\INT}$ into a center $\mathfrak z_{\INT}$ and simple ideals $\mathfrak g_j^{\INT}$, where $r$ is the number of ideals; $\calP_j^{\INT}$ denotes the projected purity \eqref{eq:projected_purity} onto $\mathfrak g_j^{\INT}$, and $\Pi_j^{\INT}$ the corresponding projection. Four subspaces recur in what follows; the list distinguishes ancilla spaces from their embeddings in the joint space.
\begin{itemize}
    \item $\Delta_a=\Span_{\mathbb R}\{iP_\ell\}_{\ell=1}^L\subseteq\fu(L)$ is the \emph{diagonal ancilla algebra}, consisting of the ancilla generators that are diagonal in the branch basis and hence do not move amplitude between branches.
    \item $\mathfrak h=\bigoplus_{\ell=1}^L P_\ell\otimes\mathfrak g_\ell$, where $\bigoplus_\ell$ denotes the internal direct sum of the pairwise-commuting subalgebras $P_\ell\otimes\mathfrak g_\ell$ (the sum is direct because different ancilla projectors have disjoint supports). The subalgebra $\mathfrak h$ is generated by the branch gates alone; it is block diagonal in the ancilla basis and contains no generator coupling distinct branches.
    \item $\mathfrak c=\gcoeff\otimes\id_s=\{K\otimes\id_s:K\in\gcoeff\}$ is the embedded coefficient-layer algebra.
    \item $\mathfrak g_{\mathrm{abs}}=\{K\in\gcoeff:K\otimes\id_s\in\mathfrak h\}$ is the subalgebra of $\gcoeff$ consisting of the coefficient directions $K$ whose action $K\otimes\id_s$ on the joint space is already contained in $\mathfrak h$; we say that such directions are \emph{absorbed} by the branch layer. The following theorem identifies these directions explicitly.
\end{itemize}

The absorption condition identifies coefficient controls already supplied by the branches. Subtracting this overlap avoids counting a direction twice. Diagonal coefficient controls can add relative phases but cannot move amplitude between labels.

\begin{theorem}[Direct-sum lower bound and exact absorption criterion]\label{thm:lower_bound}
For every coefficient layer, $\mathfrak h+\mathfrak c\subseteq\gint$, and hence
\begin{equation}
\begin{split}
    \dim\gint
    &\geq\sum\nolimits_{\ell=1}^{L}\dim\mathfrak g_\ell+\dim\gcoeff-\dim\mathfrak g_{\mathrm{abs}}\\
    &\geq\sum\nolimits_{\ell=1}^{L}\dim\mathfrak g_\ell.
\end{split}
\label{eq:lower_bound}
\end{equation}
Equality $\dim\gint=\sum_\ell\dim\mathfrak g_\ell$ holds iff $\gint=\mathfrak h$, iff $\mathfrak c\subseteq\mathfrak h$, iff every $K\in\gcoeff$ is diagonal with $i\id_s\in\mathfrak g_\ell$ wherever it acts (the exact absorption criterion). If $\gcoeff\subseteq\Delta_a$, then $\gint=\mathfrak h+\mathfrak c$ exactly, so the first inequality in \eqref{eq:lower_bound} is an equality; if $\gcoeff\not\subseteq\Delta_a$, the second is strict.
\end{theorem}
\noindent\hyperref[proof:thm:lower_bound]{Proof} in Appendix~\ref{app:sandwich}.

For a diagonal coefficient layer, Theorem~\ref{thm:lower_bound} gives $\gint=\mathfrak h+\mathfrak c$, and the combined algebra satisfies
\begin{equation}
    \dim\gint-\dim\mathfrak h\;\leq\;L,
    \label{eq:diag_defect}
\end{equation}
because $\mathfrak c\subseteq\Delta_a\otimes\id_s$ has dimension at most $L$. Each additional direction is represented by an element of $\mathfrak c$. Diagonal ancilla operators commute with both the branch generators and one another, so these added directions are central in $\gint$. Individual projectors $iP_\ell$ need not belong to a restricted diagonal coefficient algebra. These additional directions do not change the simple ideals appearing in \eqref{eq:ragone_formula} (Remark~\ref{rem:central_defect}). This is a statement about the algebra. The exact SELECT-frame loss can still depend on coefficient phases through the interference terms derived in Section~\ref{sec:exact_select}. Table~\ref{tab:notation}, placed after the saturation result, summarizes all the structural algebras and illustrates them on two small examples.

The remaining issue is what repeated mixing can create. Commutators of operations on the joint system can contain ordinary products of system generators. We therefore bound the Lie algebra by a matrix space closed under such products. The anticommutator notation $\{X,Y\}=XY+YX$ collects the symmetric product of two matrices. For $A,C\in M_L(\mathbb C)$ and $B,D\in M_d(\mathbb C)$, the following identity shows the mechanism:

\begin{equation}
    [A\otimes B,\,C\otimes D]=\tfrac12[A,C]\otimes\{B,D\}+\tfrac12\{A,C\}\otimes[B,D].
    \label{eq:tensor_comm}
\end{equation}

The first term couples an ancilla commutator to a symmetric system product; the second couples a symmetric ancilla product to a system commutator. Thus the ancilla can make products of branch directions relevant even when those products do not lie in an individual branch Lie algebra.
The containing algebra must therefore be closed under ordinary multiplication as well as commutators.

\begin{definition}[Associative envelopes]\label{def:envelopes}
For a set $S\subset M_D(\mathbb C)$, let $\Alg^*(S)$ be the smallest unital, complex, $*$-closed associative subalgebra of $M_D(\mathbb C)$ containing $S$ (associative: closed under matrix products; $*$-closed: containing $X\dagg$ whenever it contains $X$; unital: containing $\id_D$). Define
\begin{gather}
    \calA_a=\Alg^*\big(\gcoeff\cup\{P_\ell\}_{\ell\in[L]}\big)\subseteq M_L(\mathbb C),\notag\\
    \calA_s=\Alg^*\Big(\bigcup_{\ell\in[L]}\mathfrak g_\ell\Big)\subseteq M_d(\mathbb C),\\
    \calA_{\INT}=\Alg^*\Big(\mathfrak c\cup\bigcup_{\ell\in[L]}P_\ell\otimes\mathfrak g_\ell\Big)\subseteq M_{Ld}(\mathbb C),\notag
\end{gather}
the ancilla, system, and interleaved associative envelopes. For a $*$-subalgebra $\calA$, $\Skew(\calA)=\{X\in\calA:X\dagg=-X\}$ denotes its anti-Hermitian part.
\end{definition}

The anti-Hermitian part of the envelope contains every admissible generator. Passing to this part ensures that its dimension is compared with a real Lie-algebra dimension, rather than with the real dimension of the entire complex matrix space.

\begin{theorem}[Associative-envelope upper bound]\label{thm:upper_bound}
$\calA_{\INT}\subseteq\calA_a\otimes\calA_s$ and $\gint\subseteq\Skew(\calA_{\INT})\subseteq\Skew(\calA_a\otimes\calA_s)$. Consequently $\dim\gint\leq\dim_{\mathbb C}\calA_{\INT}\leq\dim_{\mathbb C}\calA_a\cdot\dim_{\mathbb C}\calA_s$, and the product bound is attained if and only if $\calA_{\INT}=\calA_a\otimes\calA_s$ and $\gint=\Skew(\calA_{\INT})$.
\end{theorem}
\noindent\hyperref[proof:thm:upper_bound]{Proof} in Appendix~\ref{app:sandwich}.

Combining Theorems~\ref{thm:lower_bound} and~\ref{thm:upper_bound} yields the dimension bounds
\begin{equation}
    \sum_{\ell=1}^L\dim\mathfrak g_\ell\;\leq\;\dim\gint\;\leq\;\dim_{\mathbb C}\calA_a\cdot\dim_{\mathbb C}\calA_s.
\end{equation}
The two bounds become explicit in different regimes. Fixed coefficients give $\gint=\mathfrak h$. A diagonal trainable layer attains the \emph{refined} lower bound in \eqref{eq:lower_bound}, which includes any unabsorbed central directions; it need not attain $\sum_\ell\dim\mathfrak g_\ell$ itself. To determine what non-diagonal control adds, we next examine the envelope blocks and then prove saturation. State and observable projections will still be needed for variance predictions.

\subsection{Envelope--commutant duality}\label{sec:duality}

Restrictions shared by all branches also restrict what ancilla mixing can achieve. To identify them, we relate two spaces: the envelope, which contains the matrix operations generated by the branches, and the commutant, which contains matrices commuting with all of them. Their relation reveals the system blocks that mixing must preserve and connects the control calculation to the invariant matrices used for interference moments. A Pauli example first shows why a small branch Lie algebra can have a large associative envelope.

\begin{example}[Complementary Pauli branches]\label{ex:pauli_envelope}
Take $L=2$, $\mathfrak g_1=\Span_{\mathbb R}\{iX_r\}_{r=1}^n$, and $\mathfrak g_2=\Span_{\mathbb R}\{iZ_r\}_{r=1}^n$, where $X_r,Z_r$ are Pauli operators on qubit $r$. Each branch algebra is abelian with dimension $n$. Their joint Lie closure adds $iY_r$ on each qubit but no multi-qubit directions, so its dimension is $3n$. In contrast, products of $X_r$ and $Z_r$ generate all Pauli strings, giving the full associative envelope $\calA_s=M_{2^n}(\mathbb C)$. The system envelope can therefore be full even when the joint branch Lie algebra is small.
\end{example}

Let $G_s$ be the \emph{joint branch group}, generated by all the branch groups. Its commutant is contained in every individual branch commutant, because commuting with all branches is a stronger requirement than commuting with one.

\begin{lemma}[Envelope--commutant duality]\label{lem:duality}
Write $\mathbb C^d\cong\bigoplus_{i=1}^{q}\mathbb C^{d_i}\otimes\mathbb C^{m_i}$ for the \emph{isotypic decomposition} of the joint branch group $G_s$: the decomposition of $\mathbb C^d$ into blocks on which $G_s$ acts as $m_i$ identical copies of one irreducible representation of dimension $d_i$, where $q$ is the number of inequivalent irreducible representations that occur, $d_i$ is the dimension of the $i$-th one, and $m_i$ is its multiplicity; then $\dim_{\mathbb C}\calA_s=\sum_id_i^2$ and $\kappa_s=\dim_{\mathbb C}\Comm(G_s)=\sum_im_i^2$. The two dimensions obey
\begin{equation}
    \dim_{\mathbb C}\calA_s\cdot\dim_{\mathbb C}\Comm(G_s)\;\geq\;d^2,
    \label{eq:envelope_commutant}
\end{equation}
with equality iff $d_i/m_i$ is independent of $i$. In particular, equality holds for an irreducible joint branch representation, for the matchgate group, and for a multiplicity-free abelian joint branch representation; more generally, an abelian joint branch representation attains equality exactly when all of its distinct characters occur with the same multiplicity.
\end{lemma}
\noindent\hyperref[proof:lem:duality]{Proof} in Appendix~\ref{app:sandwich}.

Adding the ancilla enlarges the active factor of block $i$ from dimension $d_i$ to $Ld_i$; it leaves the multiplicity $m_i$ unchanged. Since $\Comm(G_s)\subseteq\Comm(G_\ell)$, the joint commutant dimension also bounds each individual branch commutant dimension from below.

This relates the sizes of two spaces; it does not say that increasing one size must increase the variance. The commutant formula in Lemma~\ref{lem:cross_moments} contracts basis elements with $\rho$ and $O$; these overlaps can vanish even when the commutant is large. For comparison, an irreducible joint representation, a multiplicity-free abelian joint representation, and the full matchgate representation attain equality with $\kappa_s=1$, $d$, and $2$, respectively. The word \emph{joint} is essential: the complementary $X$ and $Z$ branches of Example~\ref{ex:pauli_envelope} are individually abelian but have a full joint envelope.

\subsection{Saturation and universality}\label{sec:saturation}

An upper bound on the number of controls leaves a practical question: can the available gates generate all of them? Here we determine when ancilla mixing reaches every operation permitted by the branch structure, apart from possible phase restrictions. This is the meaning of \emph{saturation}. If the branches impose no restriction on system operations, the result gives \emph{universality}: arbitrary joint unitaries up to an overall phase.

The system blocks determine which operations saturation can provide. The branches may preserve separate system subspaces, so no amount of ancilla-only mixing can connect them. Within an allowed block, traceless anti-Hermitian generators describe all unitary control after removing a common phase. Scalar generators supply that block's phase. Relative phases between different blocks can matter for operators connecting them, so we keep the central phase directions in the dimension count.

The required ancilla control is $\gcoeff\supseteq\su(L)$: every traceless ancilla generator is available through Lie closure. We call such a coefficient layer \emph{Lie-expressive}. To state the result, write the system envelope in its \emph{Wedderburn form},
\begin{equation}
 \calA_s\cong\bigoplus_{i=1}^{q}M_{d_i}(\mathbb C)\otimes\id_{m_i}.
 \label{eq:wedderburn_explained}
\end{equation}
This is the matrix-algebra version of the block decomposition just described. In block $i$, matrices can act freely on a $d_i$-dimensional factor but must act identically on its $m_i$ repeated copies. Adding the ancilla enlarges the active factor to dimension $Ld_i$. Let $e_i$ be the orthogonal projector onto system block $i$. The following small example separates the non-phase directions from the phases that remain to be counted.

\begin{example}[Two permitted blocks]\label{ex:wedderburn}
Take $L=2$, a one-qubit system, and $\mathfrak g_1=\mathfrak g_2=\mathbb R\,iZ$. Products of branch generators span $\calA_s=\operatorname{span}_{\mathbb C}\{\id,Z\}\cong\mathbb C\oplus\mathbb C$. The system blocks are the $Z$ eigenspaces: $q=2$ and $d_i=m_i=1$. With full ancilla control the joint envelope is $M_2(\mathbb C)\oplus M_2(\mathbb C)$. Its blockwise traceless part is $\su(2)\oplus\su(2)$, of dimension six; its anti-Hermitian part is $\fu(2)\oplus\fu(2)$, of dimension eight. Thus generating every traceless block direction can still leave two central directions to check.
\end{example}

\begin{theorem}[Saturation]\label{thm:saturation}
Let $L\geq2$ and $\gcoeff\supseteq\su(L)$. Then
\begin{equation}
    \gint=\bigoplus_{i=1}^{q}\su(Ld_i)\otimes\id_{m_i}\;\oplus\;\mathfrak z_{\INT},
    \qquad
    \mathfrak z_{\INT}=\pi_\zeta(\mathfrak h+\mathfrak c),
    \label{eq:saturation}
\end{equation}
where $\su(Ld_i)$ acts on $\mathbb C^L\otimes\mathbb C^{d_i}$, $\zeta=\Span_{\mathbb R}\{i(\id_L\otimes e_i)\}_{i=1}^q$ is the center of $\Skew(M_L(\mathbb C)\otimes\calA_s)$, and $\pi_\zeta$ is the HS-orthogonal projection onto $\zeta$. In particular $\sum_{i}(L^2d_i^2-1)\leq\dim\gint\leq\sum_{i}L^2d_i^2$: the algebra $\gint$ saturates its associative envelope up to an at most $q$-dimensional center.
\end{theorem}
\noindent\hyperref[proof:thm:saturation]{Proof} in Appendix~\ref{app:saturation}.

If the branch envelope is the full system matrix algebra, there is only one block and no repeated-copy restriction. The same saturation statement then gives full joint control up to global phase. The corollary identifies when the global phase direction is present as well.

\begin{corollary}[Universality]\label{cor:universality}
If, in addition, the branches \emph{jointly} generate the full matrix algebra, $\calA_s=M_d(\mathbb C)$, then $\su(Ld)\subseteq\gint\subseteq\fu(Ld)$, with $\gint=\fu(Ld)$ if and only if some generator has a nonvanishing trace component (a nonzero trace), such as $i\id_L\in\gcoeff$ or $i\id_s\in\mathfrak g_\ell$ for some $\ell$. In this regime the joint algebra contains all traceless directions. Under the joint Haar-random sampling hypothesis, its loss variance has the universal denominator $L^2d^2-1$ (Corollary~\ref{cor:universal_variance}, discussed in Section~\ref{sec:reduction}); the scaling of the full expression also depends on $\lambda$, $\rho$, and $O$.
\end{corollary}
\noindent\hyperref[proof:cor:universality]{Proof} in Appendix~\ref{app:saturation}.

There are $(Ld_i)^2-1$ traceless directions in each active block. To attain the full upper dimension bound, the circuit must also generate every block phase, so $\mathfrak z_{\INT}=\zeta$.

The center formula gives a way to check these remaining directions: project the original branch and coefficient generators onto the block identities and take their real span. Commutators cannot add a new block-trace component, so this finite calculation completes the dimension count. In Example~\ref{ex:wedderburn}, $\gcoeff=\su(2)$ yields the one-dimensional center $\mathbb R\,i(\id_a\otimes Z)$ and hence dimension seven. Adding the ancilla global phase supplies $i\id_a\otimes\id_s$ and reaches dimension eight.

Applying Corollary~\ref{cor:universality} to Example~\ref{ex:pauli_envelope}, $\gcoeff=\su(2)$ gives $\gint=\su(2^{n+1})$, because every generator is traceless. With $\gcoeff=\fu(2)$, the global phase is also present and $\gint=\fu(2^{n+1})$, attaining the product upper bound $4^{n+1}$. Appendix~\ref{sec:worked} gives additional single-pass LCU calculations.

\begin{remark}[Control through branch products]\label{rem:LARC}
Symmetry criteria already provide ways to decide whether available interactions generate a target control algebra without explicitly computing its entire Lie closure \citep{Zimboras2015Symmetry}. The saturation result here identifies the algebra block by block for the specified ancilla--branch architecture, including its exact central directions. Its constructive step is to generate branch products through ancilla commutators.

Corollary~\ref{cor:universality} is a controllability statement in the sense of Jurdjevic and Sussmann~\cite{JurdjevicSussmann1972}, but it requires less than pairwise branch controllability. In Example~\ref{ex:pauli_envelope}, the joint Lie closure has dimension only $3n$, yet the joint associative envelope is full and universality follows. Ancilla mixing therefore makes the products in the branch envelope accessible as generated control directions. Individually weak branches can consequently generate full joint-space control under repeated interleaving and, under the moment and scale assumptions of the joint variance results, an exponentially concentrated block objective.
\end{remark}

\paragraph{Restricting the mixer by a graph.}
Restricted hardware connections can be described by a graph on the branch labels. An edge joins two labels that the coefficient layer can couple; a connected component consists of labels joined by paths of such edges. For every edge $(j,k)$, assume both generators $i(E_{jk}+E_{kj})$ and $E_{jk}-E_{kj}$ are available. Their Lie closure is $\su(|C|)$ on each component $C$ with at least two vertices, while isolated vertices have no mixer direction. We also call these components \emph{clusters}.

Proposition~\ref{prop:graph} proves this decomposition and applies saturation to each component's system envelope. The two-generator assumption matters: a graph with only one specified coupling per edge need not have the same closure. If preparation starts at $\ket{\mathrm{ref}}$ and every gate preserves the components, only the component containing that state contributes to the projected loss. Parameter derivatives outside it vanish.

Thus a sparse graph can permit the same control as a dense graph with the same components. Corollary~\ref{cor:cluster} states the loss restriction and the additional assumptions for a variance formula on the prepared component.

\begin{table*}[t]
\centering
\footnotesize
\setlength{\tabcolsep}{4pt}
\renewcommand{\arraystretch}{1.08}
\begin{tabularx}{\linewidth}{@{}p{.085\linewidth}p{.095\linewidth}>{\raggedright\arraybackslash}p{.33\linewidth}>{\raggedright\arraybackslash}X@{}}
\toprule
Symbol & Lives in & Definition & Role\\
\midrule
$\mathfrak g_\ell$ & $\fu(d)$ & $\Lie_{\mathbb R}\{iH_{\ell,r}\}_r$ & DLA of branch $\ell$ (branch layer)\\
$\gcoeff$ & $\fu(L)$ & $\Lie_{\mathbb R}\{iK_\xi\}_\xi$; equal to $\{0\}$ for fixed coefficients & DLA of the coefficient layer, acting on the ancilla alone\\
$\Delta_a$ & $\fu(L)$ & $\Span_{\mathbb R}\{iP_\ell\}_\ell$ & diagonal ancilla generators: no mixing between branches\\
$\mathfrak c$ & $\fu(Ld)$ & $\gcoeff\otimes\id_s$ & coefficient layer embedded in the joint space\\
$\mathfrak h$ & $\fu(Ld)$ & $\bigoplus_\ell P_\ell\otimes\mathfrak g_\ell$ & branch layer embedded in the joint space; block diagonal in the ancilla index\\
$\mathfrak g_{\mathrm{abs}}$ & $\gcoeff$ & $\{K\in\gcoeff:K\otimes\id_s\in\mathfrak h\}$ & coefficient directions already supplied by the branch layer\\
$\gint$ & $\fu(Ld)$ & $\Lie_{\mathbb R}(\mathfrak c\cup\mathfrak h)$ & interleaved-circuit DLA (Definition~\ref{def:gBE})\\
$\calA_a$ & $M_L(\mathbb C)$ & $\Alg^*(\gcoeff\cup\{P_\ell\}_\ell)$ & ancilla associative envelope\\
$\calA_s$ & $M_d(\mathbb C)$ & $\Alg^*(\bigcup_\ell\mathfrak g_\ell)$ & system associative envelope, with Wedderburn blocks $(q;d_i,m_i)$\\
$\calA_{\INT}$ & $M_{Ld}(\mathbb C)$ & $\Alg^*(\mathfrak c\cup\bigcup_\ell P_\ell\otimes\mathfrak g_\ell)$ & interleaved associative envelope; $\gint\subseteq\Skew(\calA_{\INT})$\\
\bottomrule
\end{tabularx}

\medskip
{\setlength{\tabcolsep}{3pt}
\begin{tabular}{@{}llcccccc@{}}
\toprule
Example & $\gcoeff$ & $\dim\mathfrak c$ & $\dim\mathfrak g_{\mathrm{abs}}$ & $\dim_{\mathbb C}\calA_a$ & $\dim\gint$ & lower & upper\\
\midrule
\multirow{4}{*}{A}
 & $\{0\}$ (fixed) & 0 & 0 & 2 & 2 & 2 & 4\\
 & $\Delta_a$ (phases) & 2 & 0 & 2 & 4 & 2 & 4\\
 & $\su(2)$ (mixer) & 3 & 0 & 4 & 7 & 2 & 8\\
 & $\fu(2)$ (mixer) & 4 & 0 & 4 & 8 & 2 & 8\\
\midrule
\multirow{4}{*}{B}
 & $\{0\}$ (fixed) & 0 & 0 & 2 & 2 & 2 & 8\\
 & $\Delta_a$ (phases) & 2 & 0 & 2 & 4 & 2 & 8\\
 & $\su(2)$ (mixer) & 3 & 0 & 4 & 15 & 2 & 16\\
 & $\fu(2)$ (mixer) & 4 & 0 & 4 & 16 & 2 & 16\\
\bottomrule
\end{tabular}}
\caption{Top: the algebras of Section~\ref{sec:sandwich}. Bottom: their dimensions for two examples with $L=2$ branches on $n=1$ qubit ($d=2$), for four coefficient layers each; ``lower'' and ``upper'' are the lower and upper dimension bounds, $\sum_\ell\dim\mathfrak g_\ell$ and $\dim_{\mathbb C}\calA_a\cdot\dim_{\mathbb C}\calA_s$. Example~A has $\mathfrak g_1=\mathfrak g_2=\Span_{\mathbb R}\{iZ\}$; Example~B has $\mathfrak g_1=\Span_{\mathbb R}\{iX\}$ and $\mathfrak g_2=\Span_{\mathbb R}\{iZ\}$, with $X,Z$ the single-qubit Pauli matrices. Both have $\dim\mathfrak h=2$. In Example~A, $\calA_s$ has $q=2$ one-dimensional blocks; in Example~B, $\calA_s=M_2(\mathbb C)$ is a single block with $d_1=2$. In Example~A the branches share an abelian algebra, so a mixer can only reach $\su(2)\oplus\su(2)$ plus a center; in Example~B the branches are complementary and the mixer saturates $\su(4)$, respectively $\fu(4)$, even though the joint Lie closure of the two branches is only $\su(2)$ (Example~\ref{ex:pauli_envelope}). A diagonal layer adds exactly the central directions counted by \eqref{eq:diag_defect}, and $\mathfrak g_{\mathrm{abs}}=\{0\}$ throughout because no branch algebra contains $i\id_s$.}
\label{tab:notation}
\end{table*}

\section{Variance in the joint Haar-random regime}\label{sec:joint_variance}

Knowing what a circuit can realize does not tell us how often an initialization procedure realizes it. Even a universal gate family can be initialized close to the identity. We now add a sampling assumption: the first two moments agree with uniform sampling over the generated joint group. Under this assumption and the stated operator condition, the control structure determines an explicit loss variance.

\subsection{Joint-space sector reduction}\label{sec:reduction}

We now turn the control structure into a prediction for measured loss fluctuations. A direction matters only if both the input and the observable are sensitive to it. Under the joint moment assumption, the sector formula combines these overlaps with the number of directions in each sector. Applying it on the ancilla--system space retains the measurement dependence that a total dimension count would miss.

We use the block-frame operators from Section~\ref{sec:lifted}, so the input and measurement are fixed while $W$ is sampled. The membership condition says that at least one operator consists of Hermitian control directions plus an operator unchanged by every circuit in the group. The latter part contributes a constant. This is the same principle as Section~\ref{sec:dla_variance}, now applied on the ancilla--system space.

\begin{theorem}[Sector reduction]\label{thm:reduction}
Let $\lambda>0$ be fixed and let $W$ be an exact second-moment design over the compact connected group $G_{\INT}=e^{\gint}$ (the \emph{joint-design hypothesis}). Use the deterministic block-frame operators $\widetilde\rho=\widetilde\rho_{\mathrm{ref}}$ and $\widetilde O=\widetilde O_{\mathrm{ref}}$ from \eqref{eq:lift_BE}. Suppose
\begin{equation}
\begin{split}
    &\widetilde\rho\in i\gint+\Herm(\Comm(G_{\INT}))\\
    &\qquad\text{or}\quad
    \widetilde O\in i\gint+\Herm(\Comm(G_{\INT})).
\end{split}
    \label{eq:membership}
\end{equation}
Write $\gint=\mathfrak z_{\INT}\oplus\bigoplus_{j=1}^{r}\mathfrak g_j^{\INT}$ and $\calP_j^{\INT}=\calP_{\mathfrak g_j^{\INT}}$. Then
\begin{equation}
    \Var[\LBE]=\lambda^4\sum_{j=1}^{r}\frac{\calP_j^{\INT}(\widetilde\rho)\,\calP_j^{\INT}(\widetilde O)}{\dim\mathfrak g_j^{\INT}}.
    \label{eq:main_reduction_formula}
\end{equation}
For an approximate design, this equality is replaced by a comparison with the value under Haar measure; Proposition~\ref{prop:lifted_design_error} gives the error in a specified norm.
\end{theorem}
\noindent\hyperref[proof:thm:reduction]{Proof} in Appendix~\ref{app:reduction}.

Appendix~\ref{app:reduction} gives the covariance form and bounds using the smallest and largest simple-ideal dimensions. The quantity $Q=\sum_j\calP_j^{\INT}(\widetilde\rho)\calP_j^{\INT}(\widetilde O)$ records how strongly the two lifted operators overlap the same sectors; a large algebra alone does not fix $Q$.

The covariance extension is useful when a measured objective combines several observables: their variances cannot simply be added unless their covariances vanish.

Membership is automatic for Hermitian operators when $\gint\supseteq\su(Ld)$. With fixed coefficients, however, $\gint$ is ancilla-block-diagonal, while the SELECT-frame lifted operators generally have off-diagonal blocks whenever at least two weights are nonzero. The single-pass LCU calculation in Section~\ref{sec:exact_select} retained these blocks explicitly. Gradient formulas require additional attention to the position of the differentiated gate; the fixed-coefficient result was established in Section~\ref{sec:grad_select}, and the joint-space edge and independent-factor results in Appendix~\ref{app:gradient}.

When $\gint\supseteq\su(Ld)$ and the joint ensemble is Haar-random over $G_{\INT}$, the sum \eqref{eq:main_reduction_formula} collapses to a single sector, and an exact computation under Haar measure (Theorem~\ref{thm:full_haar_cov}) gives the closed form
\begin{equation}
    \Var(\LBE)=\lambda^4\,
    \frac{\big(\Tr(\rho^2)-\tfrac1{Ld}\big)\big(\Tr(O^2)-\tfrac{\Tr(O)^2}{Ld}\big)}{L^2d^2-1}
    \label{eq:universal_LCQNN_variance}
\end{equation}
(Corollary~\ref{cor:universal_variance}). The numerator in \eqref{eq:universal_LCQNN_variance} must be included when extracting a scaling law. For a pure input and a traceless Pauli observable, $\Tr(O^2)=d$, so the variance is $\Theta(\lambda^4/(L^2d))$ as $d$ grows, rather than $\Theta(\lambda^4/(L^2d^2))$. With $\lambda$ at most polynomial in $n$, this is exponentially small in $n$.

\section{Matchgate branches: fixed coefficients and mixing}\label{sec:matchgate}

Matchgate circuits let us compare the two architectures explicitly. Their branch algebras have polynomial dimension, and the specified single-pass LCU losses and gradients retain inverse-polynomial variance. Repeated ancilla mixing greatly enlarges control within each system-parity subspace, while preserving parity itself.

Majorana operators represent fermionic degrees of freedom as matrices on qubits. Let $c_1,\dots,c_{2n}$ be Hermitian matrices satisfying $\{c_j,c_k\}=2\delta_{jk}\id$ \citep{Diaz2023Matchgate}. A \emph{matchgate}, or free-fermionic, circuit has gate Hamiltonians quadratic in these operators. Its branch DLA lies in $\so(2n)\cong\Span_{\mathbb R}\{\tfrac12c_jc_k\}_{j<k}$, whose dimension is $n(2n-1)$. Such branches are classically simulable and trainable in previously studied settings \citep{TerhalDiVincenzo2002,JozsaMiyake2008,Diaz2023Matchgate,Khatri2025PLCU}.

In the computational basis, fermion parity distinguishes bit strings with an even number of ones from those with an odd number. The parity operator $P=(-i)^nc_1\cdots c_{2n}=Z^{\otimes n}$ has eigenvalues $+1$ and $-1$ on these subspaces. An operator is \emph{parity-even} if it commutes with $P$, and thus preserves both subspaces. Products of an even number of Majorana operators have this property. Their linear combinations form the even Clifford algebra, $\mathrm{Cl}^0_{2n}=\Alg^*\{c_jc_k\}_{j<k}\cong M_{2^{n-1}}(\mathbb C)\oplus M_{2^{n-1}}(\mathbb C)$. Its two inequivalent irreducible blocks, called \emph{half-spinor} blocks, are the eigenspaces of $P$, with projectors $(\id\pm P)/2$.

A fermionic Gaussian state is characterized by its two-point Majorana correlations: higher even correlations follow by summing signed products of two-point correlations, the fermionic Wick rule. This property makes the stated moment calculations compact. A bounded-degree Majorana observable uses products containing at most a fixed number of Majorana factors, independent of $n$.

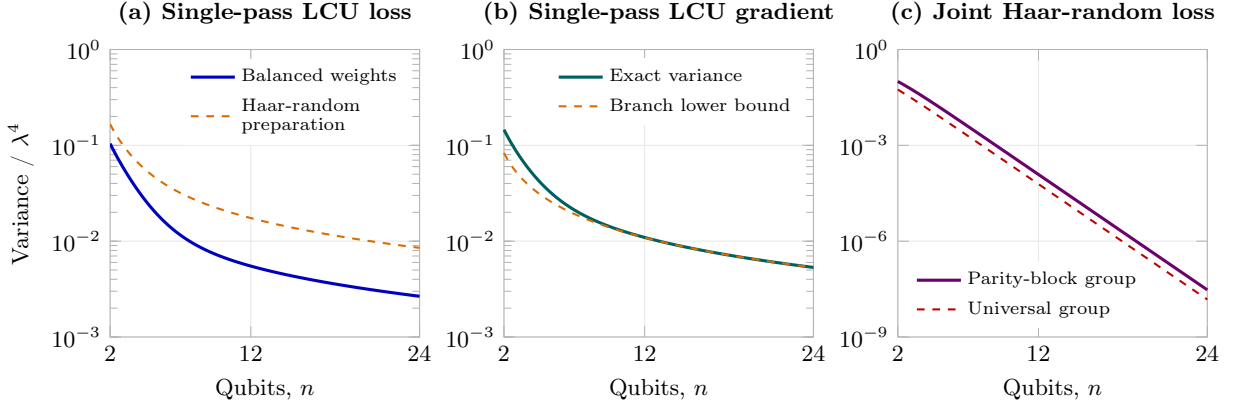
\begin{figure*}[t]
\centering
\resizebox{.94\textwidth}{!}{\begin{tikzpicture}
\begin{groupplot}[
  group style={group size=3 by 1,horizontal sep=1.15cm},
  width=5.8cm,height=5.5cm,
  xmin=2,xmax=24,xtick={2,12,24},
  ymode=log,grid=major,grid style={gray!18},
  axis line style={gray!60},tick style={gray!60},
  xlabel={Qubits, $n$},
  label style={font=\small},tick label style={font=\small},
  title style={font=\small\bfseries},
  legend style={font=\scriptsize,draw=none,fill=white,
    cells={anchor=west},row sep=1pt},
  samples=180,domain=2:24,
]
\nextgroupplot[
  title={(a) Single-pass LCU loss},
  ylabel={Variance / $\lambda^4$},
  ymin=0.001,ymax=1,ytick={0.001,0.01,0.1,1},
  legend pos=north east,
]
\addplot[blue!75!black,very thick] {((2*x-1)^(-1)+2^(1-x))/8};
\addlegendentry{Balanced weights}
\addplot[orange!85!black,thick,dashed] {2/(5*(2*x-1))+2^(1-x)/15};
\addlegendentry{\shortstack[l]{Haar-random\\preparation}}
\nextgroupplot[
  title={(b) Single-pass LCU gradient},
  ymin=0.001,ymax=1,ytick={0.001,0.01,0.1,1},
  legend pos=north east,
]
\addplot[teal!75!black,very thick] {1/(4*(2*x-1))+1/(4*2^x)};
\addlegendentry{Exact variance}
\addplot[orange!85!black,thick,dashed] {1/(4*(2*x-1))};
\addlegendentry{Branch lower bound}
\nextgroupplot[
  title={(c) Joint Haar-random loss},
  ymin=0.000000001,ymax=1,ytick={0.000000001,0.000001,0.001,1},
  legend pos=south west,
]
\addplot[violet!85!black,very thick] {1/(2*(2^x+1))};
\addlegendentry{Parity-block group}
\addplot[red!75!black,thick,dashed] {1/(2*(2^(x+1)+1))};
\addlegendentry{Universal group}
\end{groupplot}
\end{tikzpicture}
}
\caption{Loss and gradient benchmarks for $L=2$, $n\geq2$, $\rho=\ketbra00^{\otimes n}$, and $O=Z_1$. (a) Single-pass LCU loss with independent full-matchgate Haar-random branches: balanced coefficients give $\Var(\calL)/\lambda^4=\tfrac18[(2n-1)^{-1}+2^{1-n}]$; matched Haar-random preparation gives uniform simplex weights and $\Var(\calL)=\tfrac25(2n-1)^{-1}+\tfrac1{15}2^{1-n}$ at $\lambda=1$, recovering the $L=2$ specialization of Eq.~(17) in Khatri et al.~\cite{Khatri2025PLCU}. (b) Balanced single-pass LCU gradient for $U_1=e^{i\theta X_1X_2}V_1$: the exact variance and its branch floor are evaluated in \eqref{eq:matchgate_gradient_benchmark}. (c) Joint Haar-random loss of the interleaved model: parity-block and universal values are $[2(2^n+1)]^{-1}$ and $[2(2^{n+1}+1)]^{-1}$ after dividing by $\lambda^4$. The universal curve is a separate benchmark for branches with a full joint system envelope; matchgate branches retain parity and cannot reach that algebra. The vertical ranges differ; the vertical axes are logarithmic. The horizontal axis is qubit count, not depth.}
\label{fig:scaling}
\end{figure*}

\begin{theorem}[Matchgate variance and control regimes]\label{thm:matchgate}
Let $n\geq2$ and $\mathfrak g_\ell\subseteq\so(2n)$ for all $\ell\in[L]$.
\begin{enumerate}
    \item[(i)] \textbf{Diagonal coefficient layer.} If $\gcoeff\subseteq\Delta_a$, then $\gint=\mathfrak c+\mathfrak h$ with $\dim\gint\leq L+Ln(2n-1)=O(Ln^2)$.

    \emph{Simulation:} if $\rho$ is a fermionic Gaussian state, as defined above (every computational-basis state is one) and $O$ is a polynomial of bounded degree in the Majorana operators, then $\calL$ and its first parameter derivatives can be evaluated classically in $\mathrm{poly}(n,L)$ time.

    \emph{Variance:} if Assumption~\ref{asm:moments}(a) holds for the branch groups $G_\ell=e^{\mathfrak g_\ell}$, then for arbitrary Hermitian $O$, Theorem~\ref{thm:exact_select} gives $\Var[\calL]$ exactly without an observable-membership restriction. When the branches are Haar-random, their cross moments are evaluated using the invariant spaces of $G_\ell$.

    For \emph{Haar-random full-matchgate} branches, $\mathfrak g_\ell=\so(2n)$ for all $\ell$, the additional condition $O\in i\,\so(2n)+\mathbb R\id$ ensures branchwise membership, so under that condition the diagonal terms resolve by \eqref{eq:diag_sector}, and $\kappa_\ell=\tau_\ell=2$, so every cross term is $O(\Tr(\rho^2)\Tr(O^2)/d^2)$ and the form terms vanish for pure fermionic Gaussian inputs (Lemma~\ref{lem:vanishing}(iv)); coefficient averaging connects this case to the loss-variance analysis of Khatri et al.~\cite{Khatri2025PLCU} as specified in Appendix~\ref{app:select}.

    For a proper subalgebra $\mathfrak g_\ell\subsetneq\so(2n)$ the invariants of $G_\ell$ can be much larger than $2$ (Remark~\ref{rem:matchgate_sub}), and the cross-term bounds of Lemmas~\ref{lem:cross_moments} and~\ref{lem:form_pairing} scale with them.
    \item[(ii)] \textbf{Mixing coefficient layer.} If $L\geq2$, $\gcoeff\supseteq\su(L)$ and the branches jointly generate $\mathrm{Cl}^0_{2n}$, then $\su(L2^{n-1})^{\oplus2}\subseteq\gint\subseteq\fu(L2^{n-1})^{\oplus2}$, so $\dim\gint\geq2(L^24^{n-1}-1)$ grows exponentially in $n$. This dimension removes a polynomial-in-$n$ guarantee for direct full-algebra simulation. Under the exact joint-design hypothesis, for every input $\rho$ and parity-even $O$ (or every parity-even $\rho$ and arbitrary $O$), the loss \eqref{eq:lifted_loss_general} obeys:
    \begin{equation}
        \Var[\LBE]\leq\frac{\lambda^4\,\Tr(\rho^2)\,\Tr(O^2)}{L^24^{n-1}-1}.
        \label{eq:matchgate_BP}
    \end{equation}
\end{enumerate}
In both regimes every generator commutes with $\id_a\otimes P$, so the system fermion parity survives as a superselection grading: every reachable unitary is block diagonal with respect to the two eigenspaces of $\id_a\otimes P$, and $\fu(L2^n)$ is never reached.
\end{theorem}
\noindent\hyperref[proof:thm:matchgate]{Proof} in Appendix~\ref{app:matchgate}.

The coefficient layer acts only on the ancilla, so it preserves system parity in both regimes. What changes is the algebra within each parity block. With fixed coefficients, Theorem~\ref{thm:exact_select} retains the branch and interference terms. With a Lie-expressive mixer, saturation produces two large simple ideals, and the joint-design calculation gives \eqref{eq:matchgate_BP}. Thus parity preservation and variance suppression are compatible.

The variance bound is exponentially small in $n$ when $\lambda$ and $\|O\|_\infty$ grow at most polynomially, because $\Tr(O^2)\leq d\|O\|_\infty^2$ and $d=2^n$. It need not be attained by every input and observable. For gradients, the position of the gate and its surrounding random factors also matter. Under the edge or independent-factor hypotheses of Appendix~\ref{app:gradient}, a parity-even observable and fixed joint generator $\widehat H$ give the same denominator $L^24^{n-1}-1$. Polynomial bounds on $\lambda$, $\|O\|_\infty$, and $\|\widehat H\|_\infty$ then imply exponential gradient suppression.

The enlarged algebra also affects direct Lie-algebra simulation, which tracks one coordinate per basis direction: its dimension no longer gives a polynomial cost guarantee. This does not rule out other simulation methods. Likewise, random coefficients in a single-pass LCU still require the single-pass LCU moment calculation; a trainable mixer does not by itself supply joint-design moments.

\paragraph{An evaluated single-pass LCU gradient.}
Use the input and observable in Figure~\ref{fig:scaling}, balanced coefficients, and $U_1=e^{i\theta H}V_1$ with $H=X_1X_2=-ic_2c_3$. Both $V_1$ and $U_2$ are independent Haar-random full-matchgate draws. The branch-gradient observable is $i[O,H]=-2Y_1X_2$, with variance $4/(2n-1)$. The commutant and form pairings give $\mathbb E|\Gamma_{12}|^2=2/2^n$ and $\mathbb E(\Gamma_{12}^2)=0$. Hence
\begin{equation}
 \frac{\Var(\partial_\theta\calL)}{\lambda^4}
 =\underbrace{\frac{1}{4(2n-1)}}_{\text{branch floor}}
  +\underbrace{\frac{1}{4\cdot2^n}}_{\text{interference}}.
 \label{eq:matchgate_gradient_benchmark}
\end{equation}
Panel (b) plots this exact variance and the floor separately. The calculation is detailed in \hyperref[proof:benchmark]{Appendix~\ref*{app:matchgate}}; it makes the polynomial gradient signal explicit without identifying the single-pass LCU ensemble with Haar measure on the joint group.

\paragraph{Fixed depth and the joint Haar-random variance bound.}
For its stated input and traceless quadratic observable, the free-fermion stacked-LCU analysis of Khatri et al.~\cite{Khatri2026Stacked} gives an unnormalized loss-variance lower bound $\Omega(1/(nL^{3T}))$, where $T$ is the number of stacked layers and $L$ the number of terms per layer. At fixed $T$, this is inverse-polynomial in $n$ and $L$. When $T$ grows with $n$, the bound can become exponentially small, so it does not protect arbitrary depths.

This fixed-depth result is compatible with concentration under joint Haar-random sampling. The architectures also differ: a product of postselected LCU maps need not realize the reused-ancilla ensemble in \eqref{eq:interleaved}. If an interleaved family converges to the moments of Haar measure on the joint group, a polynomial finite-depth signal must cross over to the variance predicted under the joint sampling law. Neither analysis locates that crossover or proves convergence for the other architecture. The finite-depth crossover for the interleaved family remains a conditional prediction here, without numerical verification. Figure~\ref{fig:scaling} therefore labels the ensembles separately and distinguishes coefficient randomization from circuit depth.

\begin{remark}[Invariants of matchgate subalgebras]\label{rem:matchgate_sub}
The counts $\kappa_\ell=\tau_\ell=2$ are properties of the full matchgate group, whose representation on the qubits splits into two inequivalent irreducible half-spinor blocks. On restriction to a proper subgroup, those blocks may split or become equivalent, and the invariant spaces can grow. For example, on $n=2$ qubits the one-dimensional algebra $\mathfrak g=\Span_{\mathbb R}\{iZ_1\}$, where $Z_1=-ic_1c_2$ is quadratic in the Majorana operators, generates the one-parameter phase-rotation group $\{e^{i\theta Z_1}\}$, whose two two-dimensional eigenspaces acquire opposite phases; a direct computation gives $\kappa=\tau=8$, while Assumption~\ref{asm:moments}(a) still holds. Theorem~\ref{thm:matchgate}(i) therefore applies to such branches with their own invariants, and the audit computes $(\kappa_\ell,\tau_\ell)$ from the actual branch generators (step A2 in Section~\ref{app:audit_protocol}) rather than assuming the full-matchgate values.
\end{remark}

\section{Coefficient design and practical predictions}\label{sec:audit}\label{supp:guide}

A circuit-design prediction must specify the signal to be measured, the initialization law, and the applicable variance formula. We organize these choices into a calculation, then examine coefficient tuning, measurement cost, postselection, and approximate moments.

\begin{table*}[t]
\centering\small
\setlength{\tabcolsep}{5pt}\renewcommand{\arraystretch}{1.12}
\begin{tabularx}{\textwidth}{@{}p{.26\textwidth}Xp{.17\textwidth}@{}}
\toprule
Argument & Hypotheses to check & Proof location\\
\midrule
Gradient floor, Proposition~\ref{cor:floor} & Fixed coefficients; independent branch data; zero means of the other branches; finite second moments & Section~\ref{sec:floor}\\
Single-pass LCU loss and pairings & Independence and zero means for the decomposition; required Haar-measure moments for invariant pairings & Appendix~\ref{app:select}\\
Single-pass LCU and joint gradients & Stated gate placement and ensemble factorization for closed forms; membership for sector formulas & Appendix~\ref{app:gradients}\\
Bounds, saturation, graph control & Specified generator families; full special-unitary mixer for saturation; central directions for the exact upper bound & Appendix~\ref{app:control}\\
Joint-space variance & Joint first and second moments; state or observable membership for the sector formula & Appendix~\ref{app:joint}\\
Matchgate and Pauli examples & Actual branch groups and envelopes; specified input, observable, and ensemble & Appendix~\ref{app:examples}\\
Postselection and sampling & Positive success factor for normalization; stated gate spectrum and independent shots for the estimator & Appendix~\ref{app:postselect}\\
Approximate moments & Controlled branch errors or ordinary joint-space errors in the stated norm & Appendix~\ref{app:design}\\
Coefficient optimization & Applicable moments; fixed scale; specified loss or gradient objective & Appendix~\ref{app:audit}\\
QFIM rank & Pure output state; circuit generators in the indicated Lie algebra & Appendix~\ref{app:qfim}\\
\bottomrule
\end{tabularx}
\caption{Assumptions and proof locations. Structural results require no ensemble law. A large algebra alone is not a concentration hypothesis. Statements in the main text also link directly to their proofs.}
\label{tab:map}
\end{table*}

\subsection{A five-step audit}\label{app:audit_protocol}

A useful prediction must be traceable to the circuit, the initialization procedure, and the measured objective. The audit below records these inputs before recommending coefficients. Supply the branch generators $\{H_{\ell,r}\}$, coefficient generators $\{K_\xi\}$, input $\rho$, observable $O$, and coefficients $(w,\gamma)$. Also specify the parameter to be differentiated and whether the objective is $\calL$, $\calL/\lambda^2$, or a normalized postselected expectation.

For a single-pass LCU, check the independence and moment assumptions needed for the chosen result. For an interleaved-circuit prediction, check the joint-design and membership assumptions. Closed-form joint gradients additionally require the gate placement or independent-factor conditions in Appendix~\ref{app:gradient}. Then proceed as follows:
\begin{enumerate}
    \item[\textbf{A1.}] \emph{Closure.} Compute $\dim\mathfrak g_\ell$ by commutator closure, bound $\dim\gint$ using Theorems~\ref{thm:lower_bound} and~\ref{thm:upper_bound} (exactly, via Proposition~\ref{prop:graph}, for graph mixers).
    \item[\textbf{A2.}] \emph{Invariants.} Find HS-orthonormal bases of the commutant and invariant-form space by solving, for every branch generator $H_{\ell,r}$,
    \begin{equation}
     [H_{\ell,r},Y]=0,\qquad
     H_{\ell,r}J+JH_{\ell,r}^{T}=0.
    \end{equation}
    Their solution-space dimensions are $\kappa_\ell$ and $\tau_\ell$.

These equations are the infinitesimal versions of $UYU^\dagger=Y$ and $UJU^T=J$, obtained by differentiating at the identity. They let the calculation start from gate generators rather than from an explicit parameterization of Haar measure. Retain the basis matrices for the moment contractions; dimensions alone are insufficient.
    \item[\textbf{A3.}] \emph{Moment table.} Evaluate $v_\ell=\Var[m_\ell]$ by \eqref{eq:diag_sector} only when branchwise membership holds; otherwise compute or bound the required branch second moment separately. Evaluate the commutant pairings \eqref{eq:cross_commutant_gen} and form pairings \eqref{eq:form_pairing} for $B=O$ (loss) and $B=OH_{\ell,r}$ (measurement-adjacent gradients), together with the commutator purities $\calP_{\mathfrak g_\ell^j}(i[O,H_{\ell,r}])$.
    \item[\textbf{A4.}] \emph{Predictions.} Apply Proposition~\ref{cor:floor} at any gate position when its hypotheses suffice. For closed-form interference, assemble the applicable loss and measurement-adjacent gradient formulas. Use Proposition~\ref{prop:shots} for its Pauli-type generator setting, and report the error scale and confidence level. For normalized training, also evaluate success statistics and identify any missing normalized-gradient information. Record the bound on the rank of the quantum Fisher information matrix (QFIM), which counts locally distinguishable state directions, and the computational cost of the audit.
    \item[\textbf{A5.}] \emph{Tuning.} Specify a single-parameter or aggregate gradient objective. Use the explicit two-branch formulas where applicable; for more branches check phase compatibility and optimize the resulting coupled objective. Report loss-variance choices separately. For a trainable graph mixer, compare candidate prepared components using their actual system envelopes and the applicable variance regime.
\end{enumerate}

For a pure joint output state, the quantum Fisher information matrix (QFIM) measures local distinguishability along parameter directions. The Lie-algebra bound $\operatorname{rank}F\leq\dim\gint$ follows from Theorem~1 of Larocca et al.~\cite{Larocca2023Overparam}. Their orbit argument also gives a state-dependent refinement. Corollary~\ref{cor:qfim} writes this refinement explicitly using the stabilizer of the joint input, with global phase removed. This bounds the number of independent directions; it does not assert that a given circuit attains the bound or that this many parameters suffice for successful optimization.

Solving for invariant matrices as dense linear systems involves $d^2$ unknown entries, where $d=2^n$. Efficient audits therefore require compact operator representations and accessible moment contractions, as in the stated Gaussian matchgate setting. Report these representation costs along with the predictions \citep{Cerezo2024Sim,Goh2023LieSim}.

\subsection{Choosing phases and weights}

Coefficient tuning requires a choice of objective. Loss variance, the gradient variance of one parameter, and the sum of gradient variances across several branches can favor different coefficients. We keep these objectives separate below.

\paragraph{Choose the gradient objective first.}
A single-parameter objective favors a designated branch; an aggregate objective includes a chosen set of parameters. Their optima need not coincide, and neither guarantees a favorable optimization trajectory \citep{Grant2019,Arrasmith2022}.

\paragraph{Phase choices.}
The phase-dependent terms in \eqref{eq:exact_select} and \eqref{eq:grad_select} have opposite signs. The loss contribution of a nonzero pair moment is maximized when $2(\gamma_\ell-\gamma_k)\equiv-\arg\mathbb E(m_{\ell k}^2)$ modulo $2\pi$. For a measurement-adjacent gradient, the maximizing condition is instead $2(\gamma_\ell-\gamma_k)\equiv\pi-\arg\mathbb E(\Gamma_{\ell k}^2)$ modulo $2\pi$. If the moment arguments agree, the loss-maximizing phases minimize the added gradient contribution. A zero moment makes the corresponding pair phase independent.

For two branches these are explicit relative-phase choices. For more branches, pairwise choices must be compatible: phase differences must sum to zero around any closed cycle of branch labels; Theorem~\ref{thm:tuning} gives sufficient conditions for simultaneous alignment. An aggregate gradient objective can introduce further competing phase preferences. The audit must therefore optimize the chosen aggregate and check these cycle constraints, rather than assuming every pair can be optimized independently. In all cases covered by the floor, phase tuning cannot reduce the total gradient variance below its single-branch contribution.

The size of the phase effect also matters. For irreducible Haar-random branches or Haar-random full-matchgate branches, Lemmas~\ref{lem:cross_moments} and~\ref{lem:form_pairing} bound each loss interference term by $O(\|O\|_\infty^2/d)$, and each measurement-adjacent gradient term by $O(\|O\|_\infty^2\|H\|_\infty^2/d)$. With fixed $\lambda$, polynomially bounded operator norms, and an inverse-polynomial branch variance, phase tuning changes only an exponentially small part of the signal as $d=2^n$ grows. In that regime it controls the fine structure of the variance; it does not change its leading polynomial scale. The balanced-weight thresholds below consequently fail for sufficiently large $n$ under these assumptions.

\paragraph{Weight choices.}
Hold $\lambda$ fixed during coefficient tuning. For $L=2$, let $V$ be the common single-branch loss variance and let $C_{12}=\mathbb E|m_{12}|^2+\mathrm{Re}[e^{2i(\gamma_1-\gamma_2)}\mathbb E(m_{12}^2)]$. Balanced weights maximize the loss variance globally when $C_{12}\geq7V$; below this threshold assigning all weight to one branch is optimal.

Assigning all weight to one branch sets the other coefficient to zero. It optimizes the stated loss-variance criterion but removes that branch from the model.

For a parameter of branch $1$, let $V_\nabla=\Var(\Tr[i[O,H]U_1\rho U_1\dagg])$ and $C_\nabla=\mathbb E|\Gamma_{12}|^2-\mathrm{Re}[e^{2i(\gamma_1-\gamma_2)}\mathbb E(\Gamma_{12}^2)]$. Writing $w=w_1=1-w_2$ gives the gradient-variance objective $\lambda^4[V_\nabla w^4+2C_\nabla w^2(1-w)^2]$. Its derivative at balance is $\lambda^4V_\nabla/2$. Thus, if $V_\nabla>0$, increasing the differentiated branch's weight slightly improves this objective. Theorem~\ref{thm:tuning}(iv) gives the interior candidate and the threshold at which it outperforms assigning all weight to branch $1$.

When the two designated branch gradients have the same $V_\nabla$ and $C_\nabla$, their summed variance has a balanced optimum when $2C_\nabla\geq7V_\nabla$.

A large summed variance can conceal a branch with no gradient signal. At a boundary optimum, one coefficient is zero and that branch contributes nothing. Report the chosen objective with the coefficients: optimizing the sum does not guarantee that every selected parameter remains trainable.

\subsection{Measurement and normalized postselection}\label{sec:postselection}

A variance describes variation across circuit initializations; an experiment must also estimate the gradient from finitely many measurements. We first translate the unnormalized gradient scale into a sufficient number of circuit repetitions, called \emph{shots}. We then consider what changes when only successful ancilla outcomes are retained and the loss is normalized by their success factor.

For a gate that occurs once and has $H^2=\id$, Proposition~\ref{prop:shots} uses an exact four-point parameter-shift rule: the derivative is a fixed linear combination of losses measured at four shifted parameter values \citep{Wierichs2022}. It gives a sufficient shot budget to control estimation error relative to the ensemble standard deviation. The estimator for $\calL$ counts failed ancilla outcomes as zero. Its precision guarantee does not require discarding them and does not assert that every initialization has a gradient as large as that standard deviation.

For the normalized (postselected) loss we write $\pacc=\Tr[A\rho A\dagg]$ for the \emph{postselection factor}, a random variable through its dependence on the branch parameters, related to the success probability of Section~\ref{sec:lcu} by $\psucc=\pacc/\lambda^2$; the normalized loss is $\calL/\pacc$. Theorem~\ref{thm:postselect_train} assumes only $\pacc>0$ almost surely and gives, with $\bpacc=\mathbb E[\pacc]$,
\begin{equation}
    \big|\sd(\calL/\pacc)-\sd(\calL)/\bpacc\big|\leq\|O\|_\infty\,\sd(\pacc)/\bpacc;
\end{equation}
if $O\neq0$ and $\Var(\pacc)\leq\Var(\calL)/(4\|O\|_\infty^2)$, then $\Var(\calL/\pacc)\geq\Var(\calL)/(4\lambda^4)$. This transfers a lower bound on \emph{loss variance} when the fluctuations of $\pacc$ are sufficiently small relative to those of $\calL$. The required moments of $\pacc$ can be evaluated by setting $O=\id_s$. The theorem does not by itself transfer a gradient floor: differentiating $\calL/\pacc$ also introduces $\partial_\theta\pacc$, whose joint behavior with the loss gradient must be controlled.

Normalization also changes the measurement cost. At a fixed parameter point, an accepted shot requires on average $\lambda^2/\pacc$ circuit repetitions. This overhead can be large even when the normalized loss has appreciable variance. The mean $\bpacc$ is useful to report, but does not by itself bound the typical or average reciprocal cost $\lambda^2/\pacc$. This accepted-shot cost is distinct from the unnormalized estimator in Proposition~\ref{prop:shots}, which uses all shots.

In particular,
\begin{equation}
 \partial_\theta(\calL/\pacc)
 =\frac{\partial_\theta\calL}{\pacc}
  -\frac{\calL\,\partial_\theta\pacc}{\pacc^2}.
\end{equation}
The first term rescales the loss gradient, while the second accounts for the changing normalization. A normalized-gradient prediction must control both terms, including their covariance.

\subsection{Allowing moment errors}

At finite depth, the circuit averages may differ from the ideal moments used above. A variance-error bound tells us whether this discrepancy is small relative to the predicted signal. Such a bound takes the moment error as input; determining the depth needed to reach that error requires a separate convergence argument.

For a single-pass LCU, the comparison must retain the phase information in the branch overlaps. Definition~\ref{def:moment_error} does this by adjoining a two-level flag that can switch a branch unitary on or off. The resulting controlled-unitary moment channels retain terms with unequal numbers of unitary and conjugate-unitary factors. For each branch, $\varepsilon_\ell$ bounds the distance of its first two such channels from the Haar-measure reference. The distance is measured in the diamond norm, which bounds output trace-norm errors even in the presence of an auxiliary reference. Proposition~\ref{prop:hybrid} then gives a variance-error bound weighted by $\sum_\ell w_\ell\varepsilon_\ell$. Since the weights sum to one, a uniform branch-error bound does not acquire an additional factor of the number of branches.

For an interleaved circuit, matching ordinary first and second conjugation moments is sufficient for the block-loss variance. If their diamond-norm errors are at most $\varepsilon$, Proposition~\ref{prop:lifted_design_error} bounds the variance error by $3\lambda^4\varepsilon\|O\|_\infty^2$. Both bounds compare ensembles at a stated moment error; neither supplies a depth at which that error is reached. Appendix~\ref{app:design} defines the metrics and proves the bounds.

\section{Conclusion}\label{sec:outlook}

A single coherent combination can retain a useful branch gradient signal: under the stated independence and zero-mean assumptions, its gradient variance is bounded below by the differentiated branch's variance multiplied by the fourth power of that branch's coefficient magnitude. This bound holds at any gate position. Consequently, an inverse-polynomial branch variance and an inverse-polynomial coefficient magnitude are sufficient to retain an inverse-polynomial unnormalized gradient variance.

Training the coefficients of one sum changes its interference, while repeatedly interleaving branch operations and ancilla mixing can also enlarge the family of reachable transformations. Our single-pass LCU formulas quantify the dependence of loss and gradient variances on complex coefficient phases. Our control analysis characterizes the Lie algebra generated by repeated interleaving with full special-unitary ancilla control. Predicting concentration additionally requires information about how the circuit is sampled; a large control algebra alone does not determine the initialization variance.

These results have two implications for circuit design. First, coefficient choices should reflect the quantity being optimized: phases that increase loss variance need not increase the gradient variance of a chosen parameter. Second, the unnormalized signal and the normalized postselected objective require separate treatment. The latter depends on a parameter-dependent success factor, which affects both its gradient and its measurement cost.

The matchgate applications illustrate polynomial single-pass LCU variances and exponential concentration under the stated joint Haar-random sampling assumptions. These results describe distinct circuit ensembles. Determining whether a particular interleaved initialization approaches the Haar-measure moments, and how its variance changes with depth, remains a separate problem. Extending the moment analysis to correlated branches and normalized-loss gradients would further clarify its application to architectures with shared parameters and conditional outputs.

\section*{Acknowledgements}

This research has been funded in part through the research project ``Quantum Software Consortium: Exploring Distributed Quantum Solutions for Canada'' (QSC), under grant \#ALLRP587590-23 from the Natural Sciences and Engineering Research Council of Canada (NSERC) Alliance Consortia Quantum Grants program.

\bibliographystyle{unsrtnat}
\begingroup\small\raggedright
\setlength{\bibsep}{1pt}
{\setlength{\emergencystretch}{3em}\bibliography{references}}
\endgroup

\appendix
\section{Technical preliminaries}\label{app:prelim}

Complexification and group averaging simplify several proofs below. Complexification allows complex coefficients in a real generator space, giving $\mathfrak g^{\mathbb C}=\mathfrak g+i\mathfrak g$; retaining the anti-Hermitian matrices recovers the original algebra. The Haar-measure twirl projects operators onto the commutant. We record these facts, then use the twirl to remove invariant contributions from the variance.

\begin{lemma}[Real forms and twirls]\label{lem:real_form}
(i) For any real Lie algebra $\mathfrak g\subseteq\fu(D)$ of anti-Hermitian matrices, $\mathfrak g\cap i\mathfrak g=\{0\}$, so $\mathfrak g$ is a real form of its complexification $\mathfrak g^{\mathbb C}=\mathfrak g+i\mathfrak g\subseteq\fgl(D,\mathbb C)$, where $\fgl(D,\mathbb C)$ is the space of all complex matrices with the commutator bracket, and
\begin{equation}
    \mathfrak g=\mathfrak g^{\mathbb C}\cap\fu(D).
    \label{eq:real_form}
\end{equation}
(ii) For $U$ Haar-random on a compact group $G\subseteq\U(D)$, the twirl $X\mapsto\mathbb E_U[UXU\dagg]$ is the HS-orthogonal projection onto $\Comm(G)$ \citep{Mele2024}.
\end{lemma}

\begin{proof}[Proof of Lemma~\ref{lem:real_form}]\phantomsection\label{proof:lem:real_form}
(i) If $X\in\mathfrak g\cap i\mathfrak g$, then $X\dagg=-X$ and $X=iY$ with $Y\dagg=-Y$, forcing $Y\dagg=Y$, hence $Y=0$. For \eqref{eq:real_form}, decompose any anti-Hermitian $P+iQ$ with $P,Q\in\mathfrak g$ into its anti-Hermitian part $P$ and Hermitian part $iQ$; anti-Hermiticity of the sum forces $iQ$ anti-Hermitian, hence $Q=0$ by the first claim applied to $iQ$. (ii) is standard; see Mele~\cite{Mele2024}.
\end{proof}

\begin{lemma}[DLA variance formula \citep{Ragone2024,Fontana2023Adjoint}]\label{lem:ragone}
Let $\mathfrak g\subseteq\fu(D)$ be compact and reductive with center--simple decomposition $\mathfrak g=\mathfrak z\oplus\bigoplus_j\mathfrak g_j$, let $G=e^{\mathfrak g}$, and let $W$ be Haar-random on $G$ or drawn from any exact second-moment design over $G$. Let $\sigma,X\in\Herm(D)$ and suppose the membership hypothesis \eqref{eq:ragone_hypothesis} holds. Then $\Var_W[\Tr(X\,W\sigma W\dagg)]=\sum_j\calP_{\mathfrak g_j}(\sigma)\,\calP_{\mathfrak g_j}(X)/\dim\mathfrak g_j$; that is, \eqref{eq:ragone_formula} holds. Without \eqref{eq:ragone_hypothesis}, other invariant subspaces of the $G$-action and their covariances can contribute; the displayed sector formula is then not justified \citep{Diaz2023Matchgate}.
\end{lemma}

The identity is not the only operator unchanged by a restricted group. The next lemma allows any Hermitian commutant component to be removed from the fluctuating part of the loss. It is needed, for example, when parity is conserved but its generator does not belong to the DLA.

\begin{lemma}[Generalized membership]\label{lem:gen_membership}
Lemma~\ref{lem:ragone} holds if \eqref{eq:ragone_hypothesis} is replaced by $\sigma\in i\mathfrak g+\Herm(\Comm(G))$ or $X\in i\mathfrak g+\Herm(\Comm(G))$. The sums need not be direct. The projected purities are evaluated on the full operators; commutant components have zero projection onto every simple ideal.
\end{lemma}

\begin{proof}[Proof of Lemma~\ref{lem:gen_membership}]\phantomsection\label{proof:lem:gen_membership}
Write $\sigma=\sigma_{\mathfrak g}+\sigma_c$ with $\sigma_{\mathfrak g}\in i\mathfrak g$ and $\sigma_c\in\Herm(\Comm(G))$. The term $\Tr[XW\sigma_cW\dagg]=\Tr[X\sigma_c]$ is deterministic, so Lemma~\ref{lem:ragone} applies to $\sigma_{\mathfrak g}$. To identify its sector projections, let $C\in\Herm(\Comm(G))$ and $A,B\in\mathfrak g$. Cyclicity of trace, $\Tr(AB)=\Tr(BA)$, gives $\Tr(C[A,B])=\Tr([C,A]B)=0$. Each simple ideal is spanned by its commutators, hence $C$ is HS-orthogonal to every $i\mathfrak g_j$ and $\calP_{\mathfrak g_j}(\sigma)=\calP_{\mathfrak g_j}(\sigma_{\mathfrak g})$. The observable-side case follows by cyclicity.
\end{proof}

\section{Single-pass LCU loss moments}\label{app:select}

We prove the single-pass LCU loss formula by expanding its second moment. Independence removes terms containing a branch only once, leaving the two pair moments. We evaluate those moments, give conditions for the phase-dependent term to vanish, and then average over random coefficients.

\begin{proof}[Proof of Theorem~\ref{thm:exact_select}]\phantomsection\label{proof:thm:exact_select}
This expansion extends the second-moment calculation in Appendix~A.1 of Khatri et al.~\cite{Khatri2025PLCU} to fixed complex coefficients while retaining the unconjugated pairing. Fix the coefficients and use only independent branches with zero first moments; no Haar-random sampling or operator-membership assumption is needed for this expansion.

Since $\rho,O$ are Hermitian, $m_{k\ell}=\overline{m_{\ell k}}$. Expand
\begin{equation}
    \mathbb E[\calL^2]=\sum_{\ell,k,\ell',k'}\alpha_\ell\bar\alpha_k\alpha_{\ell'}\bar\alpha_{k'}\,\mathbb E[m_{\ell k}m_{\ell'k'}].
\end{equation}
By independence, any branch index appearing exactly once contributes a factor $\mathbb E[U]$ or $\mathbb E[U\dagg]$, which vanishes by (a). The surviving index patterns are:
\begin{itemize}
    \item $\ell=k$, $\ell'=k'$ (diagonal--diagonal): contributes, by independence,
    \begin{equation}
      \sum_\ell|\alpha_\ell|^4\mathbb E[m_\ell^2]
      +\sum_{\ell\neq\ell'}|\alpha_\ell|^2|\alpha_{\ell'}|^2\mathbb E[m_\ell]\mathbb E[m_{\ell'}].
    \end{equation}
    \item $\ell=k'\neq k=\ell'$ (conjugate pairing): contributes $\sum_{\ell\neq k}|\alpha_\ell|^2|\alpha_k|^2\,\mathbb E[|m_{\ell k}|^2]$, using $m_{\ell k}m_{k\ell}=|m_{\ell k}|^2$.
    \item $\ell=\ell'\neq k=k'$ (unconjugated pairing): contributes $\sum_{\ell\neq k}\alpha_\ell^2\bar\alpha_k^2\,\mathbb E[m_{\ell k}^2]$. Since $m_{k\ell}=\overline{m_{\ell k}}$, the $(k,\ell)$ term is the complex conjugate of the $(\ell,k)$ term, so this family is real and equals $\sum_{\ell\neq k}\mathrm{Re}\big[\alpha_\ell^2\bar\alpha_k^2\,\mathbb E(m_{\ell k}^2)\big]$. Under Assumption~\ref{asm:moments}(b) it vanishes, since $m_{\ell k}^2$ is bilinear in $U_\ell\otimes U_\ell$.
\end{itemize}
Meanwhile $\mathbb E[\calL]=\sum_\ell|\alpha_\ell|^2\mathbb E[m_\ell]$ by (a), so subtracting $\mathbb E[\calL]^2$ cancels the $\ell\neq\ell'$ products and converts $\mathbb E[m_\ell^2]$ into $\Var[m_\ell]$. Substituting $\alpha_\ell=\lambda w_\ell e^{i\gamma_\ell}$ gives \eqref{eq:exact_select}: the first two families involve only the moduli, and the phases enter only through $\alpha_\ell^2\bar\alpha_k^2=\lambda^4w_\ell^2w_k^2e^{2i(\gamma_\ell-\gamma_k)}$ in the unconjugated pairing.
\end{proof}

\begin{proof}[Proof of Lemma~\ref{lem:cross_moments}]\phantomsection\label{proof:lem:cross_moments}

Independence lets us average the branches in succession. Averaging each branch's conjugation with respect to Haar measure projects onto its commutant, so the result is a sum over pairs of invariant matrices. The argument allows a non-Hermitian test operator $B$.

Put $\xi=\Tr[BU_\ell\rho U_k\dagg]$. We prove \eqref{eq:cross_commutant_gen} for arbitrary $B$; setting $B=O$ gives the loss moment. The adjoints of an HS-orthonormal commutant basis form another HS-orthonormal basis, whether or not the original list is closed under adjoints.

Write $|\xi|^2=\Tr[U_\ell(\rho V\dagg B)]\cdot\Tr[U_\ell\dagg(B\dagg V\rho)]$ with $V=U_k$ fixed, using cyclicity. For $U$ Haar-random on a compact $G$ with commutant basis $\{P_i\}$, the twirl identity in index form reads $\mathbb E[U_{ab}\bar U_{dc}]=\sum_i(P_i)_{ad}(\bar P_i)_{bc}$, where subscripts denote matrix entries in a fixed orthonormal basis. Summing the repeated matrix-entry indices converts the entrywise twirl back into a trace expression:
\begin{equation}
    \mathbb E_{U_\ell}\big[|\xi|^2\big]
    =\sum_i\Tr\big[(BP_iB\dagg)V(\rho P_i\dagg\rho)V\dagg\big].
\end{equation}
Applying the twirl over $V=U_k$ using the basis $\{Q_j\dagg\}$, so that $\mathbb E[VXV\dagg]=\sum_j\Tr[Q_jX]Q_j\dagg$, and cycling the traces gives \eqref{eq:cross_commutant_gen}.

For \eqref{eq:cross_bound}, write $P_i=V_i/\sqrt d$, $Q_j=W_j/\sqrt d$ with $V_i,W_j$ unitary; the Cauchy--Schwarz inequality for the HS inner product bounds each factor, $|\Tr[\rho V_i\dagg\rho W_j]|\leq\Tr(\rho^2)$ and $|\Tr[OV_iOW_j\dagg]|\leq\Tr(O^2)$, and there are $\kappa_\ell\kappa_k$ terms, each carrying the prefactor $1/d^2$.
\end{proof}

\begin{proof}[Proof of Lemma~\ref{lem:form_pairing}]\phantomsection\label{proof:lem:form_pairing}
\emph{Step 1: averaging two unconjugated copies.}
The map $U\mapsto U\otimes U$ is a unitary representation of $G$, so $\mathbb E[U\otimes U]$ is the orthogonal projector onto its fixed subspace. Identifying $v\in\mathbb C^d\otimes\mathbb C^d$ with the matrix $J$ via $v=\sum_{ac}J_{ac}\ket a\otimes\ket c$, the fixed-point condition $(U\otimes U)v=v$ reads $UJU^{T}=J$, and $\ell_2$-orthonormality of the $v$'s is HS-orthonormality of the $J$'s. Hence, in index form,
\begin{equation}
    \mathbb E\big[U_{ab}U_{cd}\big]=\sum_s(J_s)_{ac}\,\overline{(J_s)_{bd}} .
    \label{eq:UU_index}
\end{equation}

\emph{Step 2: first branch.}

Hold the second branch fixed. The squared overlap is a quadratic expression in entries of the first branch, so the projector identity just obtained supplies its conditional average.
Let $\xi=\Tr[B\,U_\ell\rho\,U_k\dagg]=\sum_{ab}(U_\ell)_{ab}T_{ba}$ with $T=\rho\,U_k\dagg B$. Using \eqref{eq:UU_index} for $G_\ell$,
\begin{equation}
\begin{split}
    \mathbb E_{U_\ell}\big[\xi^2\,\big|\,U_k\big]
    &=\!\!\sum_{ab,cd}\!\!T_{ba}T_{dc}\,\mathbb E[U_{ab}U_{cd}]\\
    &=\sum_s\Tr\big[J_s\dagg TJ_sT^{T}\big],
\end{split}
\end{equation}
where we contracted $\sum_{ac}T_{ba}(J_s)_{ac}T_{dc}=(TJ_sT^{T})_{bd}$ and $\sum_{bd}\overline{(J_s)_{bd}}X_{bd}=\Tr[J_s\dagg X]$.

\emph{Step 3: second branch.}

The remaining random matrices now occur as an adjoint and an entrywise conjugate. Reordering the trace exposes exactly the conjugate version of the same invariant-form average.
Insert $T=\rho\,U_k\dagg B$ and cycle:
\begin{equation}
 \Tr[J_s\dagg TJ_sT^{T}]
 =\Tr\big[(\rho^{T}J_s\dagg\rho)\,U_k\dagg(BJ_sB^{T})\bar U_k\big].
 \end{equation} The mixed twirl over $V=U_k$ follows from the conjugate of \eqref{eq:UU_index} for $G_k$:
\begin{equation}
    \mathbb E\big[V\dagg Y\bar V\big]=\sum_t\Tr\big[J_t\dagg Y\big]\,J_t ,
\end{equation}
which contracted against $\rho^{T}J_s\dagg\rho$ and reordered cyclically gives \eqref{eq:form_pairing}.

\emph{Step 4: the bound.}

To bound the pairing, replace each basis matrix by its operator norm and sum over the $\tau_\ell\tau_k$ pairs. The Hilbert--Schmidt Cauchy--Schwarz inequality gives
\begin{align}
 |\Tr[J_s\dagg\rho J_t\rho^{T}]|
 &\leq\|J_s\|_\infty\|J_t\|_\infty\Tr(\rho^2),\\
 |\Tr[J_t\dagg BJ_sB^{T}]|
 &\leq\|J_s\|_\infty\|J_t\|_\infty\Tr(B\dagg B).
\end{align}
Here we use $\|XY\|_{\HS}\leq\|X\|_\infty\|Y\|_{\HS}$ and invariance of the HS norm under transpose. Summing the $\tau_\ell\tau_k$ products proves the bound. For the matchgate group, each element of a block-supported orthonormal form basis is $(d/2)^{-1/2}$ times a unitary embedding between half-spinor spaces (Lemma~\ref{lem:vanishing}(iv)), so $\nu^2=2/d$.
\end{proof}

The phase term vanishes if the group has no invariant forms or if the input contracts to zero with every relevant form. The next lemma gives criteria for both cases.

\begin{lemma}[When the form terms vanish]\label{lem:vanishing}
Let $\tau_\ell=\tau(G_\ell)$.
\begin{itemize}
    \item[(i)] For $G_\ell=\U(d)$ or $\SU(d)$ with $d\geq3$, $\tau_\ell=0$ and hence $\mathbb E(m_{\ell k}^2)=0$ for every other branch $k$. Independently, augmenting any ensemble by a uniform global phase gives $\mathbb E[U_\ell\otimes U_\ell]=0$, even without a Haar-random sampling assumption. The phase-extended group $\U(1)G_\ell$ has zero form dimension; this does not change the form dimension assigned to the original group $G_\ell$.
    \item[(ii)] If $\rho$ is pure and every $J\in\Jforms(G_\ell)$ is antisymmetric, then $\mathbb E[\Tr(BU_\ell\rho U_k\dagg)^2]=0$ for every other branch $k$ and every $B$.
    \item[(iii)] If $\rho=\mathsf P\rho \mathsf P$ for an orthogonal projection $\mathsf P$ satisfying $\mathsf P J\mathsf P^T=0$ for every $J\in\Jforms(G_\ell)$, then $\mathbb E(m_{\ell k}^2)=0$.
    \item[(iv)] \emph{Matchgates.} For $G_\ell=e^{\so(2n)}$, $\tau_\ell=2$: the forms live on the half-spinor blocks and are antisymmetric for $n\equiv2$, symmetric for $n\equiv0\ (\mathrm{mod}\ 4)$, and block-off-diagonal for odd $n$. Moreover $\bra{x}J\ket{x'}=0$ unless $x'=\bar x$ (the bitwise complement of the bit string $x$), so $\mathbb E(m_{\ell k}^2)=0$ for every computational-basis state and every $n$. The same holds for every pure fermionic Gaussian input. These vanishing claims do not extend to general mixed inputs, or to all pure inputs with support on both parities when $n$ is odd; for $n\equiv0\ (\mathrm{mod}\ 4)$, entangled parity-even pure states can give $\mathbb E(m_{\ell k}^2)\neq0$ for suitable observables; Example~\ref{ex:ghz} gives the case $n=4$.
\end{itemize}
\end{lemma}

\begin{proof}[Proof of Lemma~\ref{lem:vanishing}]\phantomsection\label{proof:lem:vanishing}
(i) By \eqref{eq:UU_index}, $\tau(G)=\dim\mathrm{Fix}(U\otimes U)$, which, by orthogonality of irreducible characters, counts pairs of mutually dual irreducible subrepresentations. For $\U(d)$, $\int(\Tr U)^2dU=0$; for $\SU(d)$ the defining representation is dual to its conjugate, and $\bar d\cong d$ only at $d=2$. Augmenting any ensemble with an independent uniform phase multiplies $\mathbb E[U\otimes U]$ by $\mathbb E[e^{2i\varphi}]=0$.

(ii) For $\rho=vv\dagg$, $\Tr[J_s\dagg\rho J_t\rho^{T}]=(v\dagg J_t\bar v)(v^{T}J_s\dagg v)$, and $v^{T}J_s\dagg v$ is the quadratic form of the antisymmetric matrix $J_s\dagg$ at $v$, which vanishes.

(iii) Since $\rho=\mathsf P\rho \mathsf P$ and $\rho^T=\mathsf P^T\rho^T\mathsf P^T$, cycling gives $\Tr[J_s\dagg\rho J_t\rho^T]=\Tr[(\mathsf P^TJ_s\dagg \mathsf P)\rho J_t\rho^T]$. The first factor is $(\mathsf P J_s\mathsf P^T)\dagg=0$, since $\mathsf P$ and $\mathsf P^T$ are Hermitian. A non-self-dual isotypic block satisfies this condition when invariant forms couple it only to a distinct dual block. The criterion does not require a real choice of basis.

(iv) The representation of $e^{\so(2n)}$ on the qubits is $S_+\oplus S_-$, where $S_+$ and $S_-$ are the inequivalent irreducible half-spinor representations, each of multiplicity one, and the classification of invariant bilinear forms on spinor representations gives: $S_\pm$ self-dual with a symmetric form when $2n\equiv0\ (\mathrm{mod}\ 8)$, self-dual with an antisymmetric form when $2n\equiv4\ (\mathrm{mod}\ 8)$, and $S_\pm^*\cong S_\mp$ when $n$ is odd; in each case exactly two invariant forms, so $\tau=2$.

A basis can be chosen with each form supported on one block (or one off-diagonal block) and proportional to a unitary between the corresponding half-spinor spaces, giving the $\nu^2=2/d$ claim.

For the basis-state claim, consider the commuting subgroup of independent mode-phase rotations, or Cartan torus, $\{\exp(\sum_r\theta_r\,\tfrac12c_{2r-1}c_{2r})\}$. It acts diagonally on the computational basis with weights $(\pm\tfrac12)^{\times n}$, the coefficients of the rotation angles in each basis vector's phase. Invariance forces $J_{xy}=0$ unless the weights of $x$ and $y$ cancel, that is, $y=\bar x$ (bitwise complement); since $x\neq\bar x$, all diagonal entries vanish, and (ii)'s factorization $\Tr[J_s\dagg\rho J_t\rho^{T}]=(J_t)_{xx}\overline{(J_s)_{xx}}$ for $\rho=\ketbra{x}{x}$ gives zero.

For a pure fermionic Gaussian input $\rho=V\ketbra{x}{x}V^\dagger$ with $V$ a matchgate unitary, invariance gives $V^\dagger J\bar V=J$. Hence the pure-state contraction reduces to the same zero diagonal entry as for $\ket{x}$. Equivalently, when both branches are Haar-random full-matchgate unitaries, right invariance absorbs $V$ into the two independent branches. For mixed inputs the contraction need not factor into such diagonal entries; for odd $n$, coherence between the two parity blocks can couple to the off-diagonal invariant forms.

For $n\equiv0\ (\mathrm{mod}\ 4)$ the forms are symmetric with $J_{x\bar x}\neq0$, so states with support on complementary pairs, including GHZ states, generically give nonzero values; at $n=4$, $\rho_{\mathrm{GHZ}}$ and $O=ic_1c_2+\tfrac12ic_3c_6$ give the exact value $\mathbb E(m^2)=-5/32$.
\end{proof}

\begin{remark}[Why the cross terms are invisible to the joint-DLA formula]\label{rem:why_cross}
With fixed coefficients, $\gint=\mathfrak h$ is ancilla-block-diagonal, while $\widetilde\rho_\alpha,\widetilde O_\alpha$ carry off-diagonal blocks of weight $\sqrt{w_\ell w_k}$; those blocks violate \eqref{eq:membership} and contribute exactly the pairings \eqref{eq:cross_commutant_gen} and \eqref{eq:form_pairing}. The cross terms are thus not a correction \emph{to} the DLA framework but the precise content of its membership failure: $\kappa$ counts the conjugation-invariant directions of the branch group, $\tau$ its invariant directions under the congruence action $J\mapsto UJU^T$, and the off-diagonal blocks excite exactly these two channels.
\end{remark}

\begin{proof}[Proof of Theorem~\ref{thm:KZM_consistency}]\phantomsection\label{proof:thm:KZM_consistency}

Condition on the independently sampled Dirichlet coefficients. Theorem~\ref{thm:exact_select} supplies the conditional variance, since the identical Haar-random branches have zero first moments and vanishing unconjugated pair moments. Variation of the conditional mean supplies the terms proportional to $\mu^2$.

By the law of total variance, $\Var[\calL]=\mathbb E_c[\Var_\theta[\calL|c]]+\Var_c[\mathbb E_\theta[\calL|c]]$. Theorem~\ref{thm:exact_select} gives the conditional variance, and $\mathbb E_\theta[\calL|c]=\mu\sum_\ell c_\ell^2$ by (a). The uniform Dirichlet moments are
\begin{gather}
    \mathbb E[c_\ell^4]=\frac{24}{L(L{+}1)(L{+}2)(L{+}3)},\qquad
    \mathbb E[c_\ell^2]=\frac{2}{L(L{+}1)},\notag\\
    \mathbb E[c_\ell^2c_k^2]=\frac{4}{L(L{+}1)(L{+}2)(L{+}3)}\quad(\ell\neq k).
\end{gather}
Hence
\begin{equation}
    \mathbb E_c\big[\Var_\theta[\calL|c]\big]
    =\frac{24\Var[m_\ell]+4(L-1)\,\mathbb E[|m_{\ell k}|^2]}{(L+1)(L+2)(L+3)},
\end{equation}
and, using $\mathbb E[\sum_\ell c_\ell^2]=\tfrac{2}{L+1}$ and $\mathbb E[(\sum_\ell c_\ell^2)^2]=\tfrac{4L+20}{(L+1)(L+2)(L+3)}$,
\begin{equation}
    \Var_c\big[\mathbb E_\theta[\calL|c]\big]
    =\mu^2\Big(\frac{4L+20}{(L{+}1)(L{+}2)(L{+}3)}-\frac{4}{(L{+}1)^2}\Big).
\end{equation}
Adding the two, writing $\Var[m_\ell]=\mathbb E[m_\ell^2]-\mu^2$, and using $-24+4L+20=4(L-1)$ gives \eqref{eq:KZM_exact}, which matches Theorem~2, Eq.~(9), of Khatri et al.~\cite{Khatri2025PLCU} in its identical-groups form.
\end{proof}

\section{Gradient moments}\label{app:gradients}

Gradient moments depend on where the parameter enters the circuit. We first differentiate single-pass LCU overlaps, then treat joint circuits by identifying the surrounding random factor that supplies the required averaging.

\subsection{Exact single-pass LCU gradients}\label{app:gradsel}

A lower bound tells us which branch signal survives, but not how much interference adds to the gradient. Here we calculate that additional contribution and its dependence on coefficient phases. For a measurement-adjacent gate with independent Haar-random branches, differentiation produces the overlap operator $OH$ and the minus sign in the phase-dependent term; the floor in Section~\ref{sec:floor} needs neither restriction.

\begin{proof}[Proof of Theorem~\ref{thm:grad_select}]\phantomsection\label{proof:thm:grad_select}
Haar measure is invariant under both left and right multiplication by fixed group elements. Consequently, for any fixed $\theta$ the pair formed by $e^{i\theta H}V_\ell$ and $\{U_k\}_{k\neq\ell}$ has the same joint law as the pair formed by $U_\ell$ and $\{U_k\}_{k\neq\ell}$, so it suffices to compute the variance of the derivative at $\theta=0$. Since only $U_\ell$ depends on $\theta$ and $\partial_\theta U_\ell=iHU_\ell$,
\begin{equation}
\begin{gathered}
    \partial_\theta m_{\ell k}=i\Gamma_{\ell k},\qquad
    \partial_\theta m_{k\ell}=-i\overline{\Gamma_{\ell k}}\ \ (k\neq\ell),\\
    \partial_\theta m_\ell=g_\ell^{\mathrm{branch}}:=\Tr\big[i[O,H]\,U_\ell\rho U_\ell\dagg\big],
\end{gathered}
\end{equation}
using $\partial_\theta m_\ell=i\Tr[OHU\rho U\dagg]-i\Tr[OU\rho U\dagg H]=\Tr[i[O,H]U\rho U\dagg]$ by cyclicity, and $m_{k\ell}=\overline{m_{\ell k}}$. Hence
\begin{equation}
    \partial_\theta\calL=|\alpha_\ell|^2g_\ell^{\mathrm{branch}}
    +\sum_{k\neq\ell}2\,\mathrm{Re}\big[i\,\alpha_\ell\bar\alpha_k\,\Gamma_{\ell k}\big].
    \label{eq:dN_decomp}
\end{equation}

This has the same branch-plus-interference structure as the loss, but the interference overlaps carry a factor of $i$. Squaring that factor is the source of the minus sign in the phase-sensitive gradient term.
Now expand $\Var(\partial_\theta\calL)=\mathbb E[(\partial_\theta\calL)^2]-\mathbb E[\partial_\theta\calL]^2$ and use independence with Assumption~\ref{asm:moments}(a), exactly as in the proof of Theorem~\ref{thm:exact_select}: (i) any term in which some $U_k$ ($k\neq\ell$) appears only once, including all diagonal--cross products and all cross--cross products with $k\neq k'$, has zero expectation; (ii) the diagonal--diagonal family contributes $|\alpha_\ell|^4\Var(g_\ell^{\mathrm{branch}})$ after subtracting the mean, since $\mathbb E[\partial_\theta\calL]=|\alpha_\ell|^2\mathbb E[g_\ell^{\mathrm{branch}}]$; (iii) for each $k$, writing $\omega=i\alpha_\ell\bar\alpha_k\Gamma_{\ell k}$, the square $(2\,\mathrm{Re}\,\omega)^2=\omega^2+\bar \omega^2+2|\omega|^2$ contributes
\begin{equation}
    2|\alpha_\ell|^2|\alpha_k|^2\,\mathbb E|\Gamma_{\ell k}|^2
    +2\,\mathrm{Re}\big[-\alpha_\ell^2\bar\alpha_k^2\,\mathbb E(\Gamma_{\ell k}^2)\big].
\end{equation}
Substituting $\alpha_\ell=\lambda w_\ell e^{i\gamma_\ell}$ gives \eqref{eq:grad_select}. The closed forms for $\mathbb E|\Gamma_{\ell k}|^2$ and $\mathbb E(\Gamma_{\ell k}^2)$ are Lemmas~\ref{lem:cross_moments} and~\ref{lem:form_pairing} with $B=OH$.

For the sector form of the diagonal factor: if $\rho\in i\mathfrak g_\ell+\mathbb R\id$, \eqref{eq:diag_sector} applies directly; if instead $O\in i\mathfrak g_\ell+\mathbb R\id$, then $i[O,H]\in i\mathfrak g_\ell$ because $iH\in\mathfrak g_\ell$ and $\mathfrak g_\ell$ is a Lie algebra, so the membership hypothesis holds for the pair $(\rho,i[O,H])$ and \eqref{eq:diag_sector} applies with observable $i[O,H]$.
\end{proof}

\subsection{Joint-space edge and interior gradients}\label{app:gradient}

A circuit's output distribution does not determine its parameter derivatives. At an edge, differentiating inserts a fixed generator and permits a direct variance calculation. For an interior gate, we condition on one surrounding factor and average the other, provided they are independent.

\begin{theorem}[Joint-space edge-gradient variance]\label{thm:grad_variance}
Fix $\lambda>0$, $\theta\in\mathbb R$, a Hermitian $\widehat H$ with $i\widehat H\in\gint$, and the block-frame operators $\widetilde\rho=\widetilde\rho_{\mathrm{ref}}$, $\widetilde O=\widetilde O_{\mathrm{ref}}$. Let
$W(\theta)=e^{i\theta\widehat H}V$, where $V$ is an exact second-moment design over $G_{\INT}$ whose distribution is independent of $\theta$. If $\widetilde O\in i\gint+\Herm(\Comm(G_{\INT}))$, set $X=i[\widetilde O,\widehat H]$. Then $\mathbb E\,\partial_\theta\LBE=0$ and
\begin{equation}
 \Var(\partial_\theta\LBE)=\lambda^4\sum_j
       \frac{\calP_j^{\INT}(\widetilde\rho)\calP_j^{\INT}(X)}{\dim\mathfrak g_j^{\INT}}.
 \label{eq:grad_variance}
\end{equation}
For $Ld\geq2$ and $\gint\supseteq\su(Ld)$, this implies
\begin{equation}
 \Var(\partial_\theta\LBE)\leq
 \frac{\lambda^4\|[\widetilde O,\widehat H]\|_{\HS}^2\Tr(\rho^2)}{(Ld)^2-1}.
 \label{eq:local_BP}
\end{equation}
For a state-adjacent gate $W=Ve^{i\theta\widehat H}$, assume instead $\widetilde\rho\in i\gint+\Herm(\Comm(G_{\INT}))$. The exact formula becomes $\lambda^4\sum_j\calP_j^{\INT}(\widetilde O)\calP_j^{\INT}(i[\widehat H,\widetilde\rho])/\dim\mathfrak g_j^{\INT}$.
\end{theorem}

\begin{proof}[Proof of Theorem~\ref{thm:grad_variance}]\phantomsection\label{proof:thm:grad_variance}
At a measurement-adjacent edge, differentiating gives $\partial_\theta\LBE=\lambda^2\Tr[XW\widetilde\rho W\dagg]$. The commutant component of $\widetilde O$ commutes with $\widehat H$, so $X\in i[\gint,\gint]$. Its mean under Haar measure is zero: the twirl of $\widetilde\rho$ commutes with $\widehat H$, and cyclicity kills the commutator. Left multiplication by the fixed group element $e^{i\theta\widehat H}$ preserves design moments. Applying Lemmas~\ref{lem:ragone}--\ref{lem:gen_membership} to the Hermitian pair $(\widetilde\rho,X)$ proves \eqref{eq:grad_variance}. In the universal case there is one simple ideal, and its projected norms are bounded by the full HS norms, giving \eqref{eq:local_BP}.

At the state edge, differentiate $e^{i\theta\widehat H}\widetilde\rho e^{-i\theta\widehat H}$; its derivative is the conjugate of $i[\widehat H,\widetilde\rho]$. Right invariance of the design moments gives the stated symmetric formula.
\end{proof}

For an interior gate, the operations after the gate rotate the observable seen by its generator. The next result treats this rotated observable as fixed while averaging the independent circuit factor before the gate. This converts the interior calculation into the edge calculation just proved.

\begin{corollary}[An independent interior factor]\label{cor:interior_gradient}
Let $W=V_L e^{i\theta\widehat H}V_R$, with the same fixed operators, scale, and generator as above. Suppose $V_R$ is an exact second-moment design over $G_{\INT}$, independent of $V_L\in G_{\INT}$, and the distributions of both factors are independent of $\theta$. If $\widetilde O$ satisfies the observable membership condition, define $O_L=V_L\dagg\widetilde O V_L$ and $X_L=i[O_L,\widehat H]$. Then
\begin{equation}
 \Var(\partial_\theta\LBE)=\lambda^4\sum_j
 \frac{\calP_j^{\INT}(\widetilde\rho)\,\mathbb E_{V_L}\calP_j^{\INT}(X_L)}{\dim\mathfrak g_j^{\INT}}.
 \label{eq:interior_gradient}
\end{equation}
If the smallest simple-ideal dimension is $D_{\min}$, then
\begin{equation}
 \Var(\partial_\theta\LBE)\leq
 \frac{4\lambda^4\|\widehat H\|_\infty^2\Tr(\rho^2)\Tr(O^2)}{D_{\min}}.
\end{equation}
When there are no simple ideals, this derivative vanishes under membership.
\end{corollary}

\begin{proof}[Proof of Corollary~\ref{cor:interior_gradient}]\phantomsection\label{proof:cor:interior_gradient}
Condition on $V_L$. Conjugation within $G_{\INT}$ preserves the Lie algebra and fixes its commutant, so it preserves membership, and the remaining ensemble is the edge ensemble of Theorem~\ref{thm:grad_variance}. Its conditional mean derivative is zero. The law of total variance therefore gives \eqref{eq:interior_gradient}. Bound the sector sum by $\|\widetilde\rho\|_{\HS}^2\|X_L\|_{\HS}^2/D_{\min}$ and use $\|X_L\|_{\HS}\leq2\|\widehat H\|_\infty\|\widetilde O\|_{\HS}$.
\end{proof}

For a single-pass LCU branch parameter $U_\ell=e^{i\theta H}V_\ell$ with fixed preparations, the block-frame derivative is $\partial_\theta W=i\widehat H W$, where
\begin{equation}
 \widehat H=(\PREP_L\dagg\otimes\id_s)(P_\ell\otimes H)(\PREP_L\otimes\id_s).
\end{equation}
This conjugated generator, rather than $P_\ell\otimes H$ in the SELECT frame, is the generator needed for a block-frame edge calculation. The theorem still requires its algebra membership and the stated design law for the remaining factor. With random preparations, $\widehat H$ can be correlated with $W$; a marginal design law for $W$ alone does not justify \eqref{eq:grad_variance}. Corollary~\ref{cor:interior_gradient} states one sufficient independent-factor condition explicitly.

\section{Interleaved control}\label{app:control}

We first bound the generated algebra and relate its envelope to the commutant. We then construct every allowed traceless block direction under full ancilla control. Restricted graph connections divide this construction into separate components. The arguments concern available controls and require no initialization law.

\subsection{Dimension bounds and envelope duality}\label{app:sandwich}

Before calculating the full control algebra, we need to know how large it could be and which restrictions it must respect. We count the directions already supplied by the branch and ancilla layers, subtracting their overlap, then place all generated commutators in an associative algebra. Relating this envelope to the commutant identifies the blocks behind the upper bound.

\begin{remark}[Central defect]\label{rem:central_defect}
Diagonality alone does \emph{not} force $\dim\gint=\sum_\ell\dim\mathfrak g_\ell$: a diagonal generator contributes the central directions $iP_\ell\otimes\id_s$, which lie outside $\mathfrak h$ unless $i\id_s\in\mathfrak g_\ell$. The discrepancy is at most $L$-dimensional and entirely central, so it never affects \eqref{eq:ragone_formula}.
\end{remark}

\begin{proof}[Proof of Theorem~\ref{thm:lower_bound}]\phantomsection\label{proof:thm:lower_bound}
The absorption criterion of Theorem~\ref{thm:lower_bound} asserts the equivalence of: (1) $\dim\gint=\sum_\ell\dim\mathfrak g_\ell$; (2) $\gint=\mathfrak h$; (3) $\mathfrak c\subseteq\mathfrak h$; (4) every $K\in\gcoeff$ is diagonal, $K=i\sum_\ell c_\ell P_\ell$ with $c_\ell\in\mathbb R$ and $i\id_s\in\mathfrak g_\ell$ whenever $c_\ell\neq0$. The proof below establishes the bounds and this equivalence.

\emph{Step 1: $\mathfrak h$ is a Lie algebra and its defining sum is direct.}
From $(P_\ell\otimes X)(P_m\otimes Y)=\delta_{\ell m}P_\ell\otimes XY$ we get
\begin{equation}
    [P_\ell\otimes X,\,P_m\otimes Y]=\delta_{\ell m}\,P_\ell\otimes[X,Y].
    \label{eq:branch_bracket}
\end{equation}
If $X,Y\in\mathfrak g_\ell$ then $[X,Y]\in\mathfrak g_\ell$, so each $P_\ell\otimes\mathfrak g_\ell$ is a Lie subalgebra, and distinct summands commute; thus $\mathfrak h$ is a Lie subalgebra. If $\sum_\ell P_\ell\otimes X_\ell=0$, multiplying by $P_m\otimes\id_s$ on both sides gives $X_m=0$, so the sum is direct and $\dim\mathfrak h=\sum_\ell\dim\mathfrak g_\ell$.

\emph{Step 2: the lower bounds.}
Every element of $\mathfrak h$ and of $\mathfrak c$ lies in the real span of the generators in \eqref{eq:gBE}, so $\mathfrak h+\mathfrak c\subseteq\gint$. Since $K\mapsto K\otimes\id_s$ is injective, $\dim\mathfrak c=\dim\gcoeff$, and $\mathfrak h\cap\mathfrak c=\mathfrak g_{\mathrm{abs}}\otimes\id_s$ by definition. The dimension formula for sums of subspaces gives
\begin{equation}
    \dim\gint\geq\dim\mathfrak h+\dim\mathfrak c-\dim(\mathfrak h\cap\mathfrak c),
\end{equation}
which is the first inequality in \eqref{eq:lower_bound}; the second follows since $\mathfrak g_{\mathrm{abs}}\subseteq\gcoeff$.

\emph{Step 3: characterization of $\mathfrak h\cap\mathfrak c$.}

An ancilla-only generator belongs to the branch algebra only if its blocks have the controlled-branch form. Write $K=(K_{\ell m})_{\ell,m\in[L]}$ in the ancilla basis. The $(\ell,m)$ block of $K\otimes\id_s$ is $K_{\ell m}\id_s$, whereas elements of $\mathfrak h$ have zero off-diagonal blocks and an $\ell$-th diagonal block in $\mathfrak g_\ell$. Thus membership requires $K_{\ell m}=0$ for $\ell\ne m$ and $K_{\ell\ell}\id_s\in\mathfrak g_\ell$ for every $\ell$. Anti-Hermiticity gives $K_{\ell\ell}=ic_\ell$ with $c_\ell\in\mathbb R$, proving the equivalence of (3) and (4).

\emph{Step 4: equivalence of (1)--(3).}
(1)$\Rightarrow$(2): $\mathfrak h\subseteq\gint$ with equal dimensions forces equality. (2)$\Rightarrow$(3): all coefficient generators lie in $\gint=\mathfrak h$. (3)$\Rightarrow$(1): if $\mathfrak c\subseteq\mathfrak h$, all generators lie in the Lie algebra $\mathfrak h$, so minimality of the Lie closure gives $\gint\subseteq\mathfrak h$, and the reverse inclusion is Step~2.

\emph{Step 5: the diagonal case.}
If $\gcoeff\subseteq\Delta_a$, every $K\in\gcoeff$ commutes with every $P_\ell$, so $[K\otimes\id_s,P_\ell\otimes X]=[K,P_\ell]\otimes X=0$ and therefore $[\mathfrak c,\mathfrak h]=0$. Since $[\mathfrak h,\mathfrak h]\subseteq\mathfrak h$ and $[\mathfrak c,\mathfrak c]\subseteq\mathfrak c$, the subspace $\mathfrak h+\mathfrak c$ is a Lie algebra containing the generators, whence $\gint=\mathfrak h+\mathfrak c$. Finally, if $\gcoeff\not\subseteq\Delta_a$, choose $K\in\gcoeff$ with a nonzero off-diagonal entry; Step~3 shows $K\otimes\id_s\notin\mathfrak h$, so $\dim\gint\geq\dim\mathfrak h+1$.
\end{proof}

\begin{proof}[Proof of Theorem~\ref{thm:upper_bound}]\phantomsection\label{proof:thm:upper_bound}
\emph{Step 1: envelope inclusion.}
$\calA_a\otimes\calA_s$ is a unital $*$-closed subalgebra of $M_{Ld}(\mathbb C)$ containing every generator of $\calA_{\INT}$ ($K\otimes\id_s$ and $P_\ell\otimes X$ are elementary tensors of members), so minimality gives $\calA_{\INT}\subseteq\calA_a\otimes\calA_s$.

\emph{Step 2: $\Skew(\calA)$ is a Lie algebra, of real dimension $\dim_{\mathbb C}\calA$.}

This step reconciles the two dimension conventions in the bound. Each complex matrix decomposes into Hermitian and anti-Hermitian parts, and multiplication by $i$ pairs these two real spaces.
For $X,Y\in\Skew(\calA)$, $[X,Y]\in\calA$ by associativity and $[X,Y]\dagg=YX-XY=-[X,Y]$, so $\Skew(\calA)$ is a real Lie algebra. The decomposition $Z=\tfrac12(Z+Z\dagg)+\tfrac12(Z-Z\dagg)$ gives $\calA=\Herm(\calA)\oplus_{\mathbb R}\Skew(\calA)$, and $H\mapsto iH$ is a real-linear bijection $\Herm(\calA)\to\Skew(\calA)$; hence $\dim_{\mathbb R}\Skew(\calA)=\tfrac12\dim_{\mathbb R}\calA=\dim_{\mathbb C}\calA$.

\emph{Step 3: conclusion.}
Every generator of $\gint$ is anti-Hermitian and lies in $\calA_{\INT}$, so minimality of the Lie closure gives $\gint\subseteq\Skew(\calA_{\INT})$, and Step~1 gives $\Skew(\calA_{\INT})\subseteq\Skew(\calA_a\otimes\calA_s)$. Applying Step~2 and $\dim_{\mathbb C}(\calA_a\otimes\calA_s)=\dim_{\mathbb C}\calA_a\dim_{\mathbb C}\calA_s$ yields the dimension chain in the statement. If $\calA_{\INT}=\calA_a\otimes\calA_s$ and $\gint=\Skew(\calA_{\INT})$, the product bound is attained; conversely, equality of the outer terms forces equality throughout the nested chain.
\end{proof}

\begin{lemma*}[Envelope--commutant duality; restatement of Lemma~\ref{lem:duality}]
Let $\mathbb C^d\cong\bigoplus_{i=1}^{q}\mathbb C^{d_i}\otimes\mathbb C^{m_i}$ be the isotypic decomposition of the joint branch group, so that $\calA_s\cong\bigoplus_i M_{d_i}(\mathbb C)\otimes\id_{m_i}$ and $\Comm(G_s)\cong\bigoplus_i\id_{d_i}\otimes M_{m_i}(\mathbb C)$, with $\dim_{\mathbb C}\calA_s=\sum_id_i^2$, $\dim_{\mathbb C}\Comm(G_s)=\sum_im_i^2$, and $\sum_id_im_i=d$. The two dimensions obey
\begin{equation}
    \dim_{\mathbb C}\calA_s\cdot\dim_{\mathbb C}\Comm(G_s)\;\geq\;d^2,
\end{equation}
with equality if and only if $d_i/m_i$ is independent of $i$. In particular, equality holds for an irreducible joint branch representation, for the matchgate group, and for a multiplicity-free abelian joint branch representation; more generally, an abelian joint branch representation attains equality exactly when all of its distinct characters occur with the same multiplicity.
\end{lemma*}

\begin{proof}[Proof of Lemma~\ref{lem:duality}]\phantomsection\label{proof:lem:duality}
The double-commutant theorem states that a finite-dimensional unital $*$-closed algebra is the commutant of its commutant. Applying it to the $*$-closed algebra $\calA_s$ (commuting with all elements of $G_s$ is equivalent to commuting with the unital $*$-algebra they generate) gives the stated block decomposition. Counting matrix entries in each active or multiplicity factor gives the two dimension identities.

For \eqref{eq:envelope_commutant}, the Cauchy--Schwarz inequality gives $d^2=(\sum_id_im_i)^2\leq(\sum_id_i^2)(\sum_im_i^2)$, with equality iff $(d_i)_i\propto(m_i)_i$. For an irreducible joint representation, $q=1$ and equality is immediate; for the matchgate group, $q=2$ with $d_1=d_2=2^{n-1}$ and $m_1=m_2=1$.

For an abelian joint group every irreducible representation is a character, so $d_i=1$ for all $i$, $\dim_{\mathbb C}\calA_s=q$, $\kappa_s=\sum_im_i^2$, and $d=\sum_im_i$; the equality condition becomes $q\sum_im_i^2=(\sum_im_i)^2$, that is, $m_1=\dots=m_q$. The multiplicity-free case $m_i=1$ is the special case in which $\kappa_s=d$. Being abelian alone does not suffice: for $G_s=\{\operatorname{diag}(1,1,e^{it}):t\in\mathbb R\}$ on $\mathbb C^3$ the two characters occur with multiplicities $(m_1,m_2)=(2,1)$, so $\dim_{\mathbb C}\calA_s=2$, $\kappa_s=5$, and $\dim_{\mathbb C}\calA_s\cdot\kappa_s=10>9=d^2$.
\end{proof}

\subsection{Saturation and universality}\label{app:saturation}

The dimension bound does not yet establish that the gates can realize every allowed operation. This proof closes that gap under full special-unitary ancilla control. We use ancilla transitions to generate products of branch directions, filling every traceless block, then count the remaining phase directions. When the system envelope is full, this construction gives universality.

\begin{proof}[Proof of Theorem~\ref{thm:saturation}]\phantomsection\label{proof:thm:saturation}
Assume $L\geq2$ and full $\su(L)$ ancilla control. The branch directions are independently available under repeated interleaving.

Under the hypothesis, $\calA_a=M_L(\mathbb C)$, since $\Alg^*\{\su(L)\}=M_L(\mathbb C)$ for $L\geq2$. By Theorem~\ref{thm:upper_bound} and the Wedderburn form of $\calA_s$,
\begin{equation}
    \gint\subseteq\Skew(M_L(\mathbb C)\otimes\calA_s)\cong\bigoplus_i\fu(Ld_i)\otimes\id_{m_i},
\end{equation}
whose derived part is $\bigoplus_i\su(Ld_i)\otimes\id_{m_i}$ and whose center is $\zeta$. We prove the reverse inclusion of the derived part, working with the complexification $\gint^{\mathbb C}$ and using Lemma~\ref{lem:real_form} to descend.

Complexification lets us use individual matrix units, which need not be anti-Hermitian, during the construction. Intersecting with the anti-Hermitian matrices at the end recovers the unitary generators.

\emph{Step 0: ancilla matrix units.}

The notation $\fsl(D,\mathbb C)$ denotes all traceless complex $D\times D$ matrices. In this space, off-diagonal matrix units move between labels and diagonal differences supply relative phases.
$\gint\supseteq\gcoeff\otimes\id_s\supseteq\su(L)\otimes\id_s$, so $\gint^{\mathbb C}\supseteq\fsl(L,\mathbb C)\otimes\id_s$: every off-diagonal matrix unit $E_{ab}\otimes\id_s$ ($a\neq b$) and every difference $(E_{aa}-E_{bb})\otimes\id_s$ is available.

\emph{Step 1: words of length one.}

We first move a single branch direction from its original ancilla label to every matrix position. This provides the ingredients needed to multiply system directions in the next step.
Fix $\ell$ and $X\in\mathfrak g_\ell^{\mathbb C}$. The generator gives $E_{\ell\ell}\otimes X\in\gint^{\mathbb C}$, and bracketing with ancilla units yields the following identities. The first and last use $a\ne\ell$; the second uses any $b\ne\ell$; the third uses $a,b\ne\ell$ with $a\ne b$:
\begin{align}
    [E_{a\ell}\otimes\id_s,\,E_{\ell\ell}\otimes X]&=E_{a\ell}\otimes X,\notag\\
    [E_{\ell\ell}\otimes X,\,E_{\ell b}\otimes\id_s]&=E_{\ell b}\otimes X,\notag\\
    [E_{a\ell}\otimes\id_s,\,E_{\ell b}\otimes X]&=E_{ab}\otimes X,\\
    [E_{a\ell}\otimes\id_s,\,E_{\ell a}\otimes X]&=(E_{aa}-E_{\ell\ell})\otimes X;\notag
\end{align}
adding the generator to the last line gives $E_{aa}\otimes X$ for every $a$. Hence
\begin{equation}
    E_{ab}\otimes X\in\gint^{\mathbb C}
    \quad\text{for all }a,b\in[L],\ X\in\textstyle\bigcup_\ell\mathfrak g_\ell^{\mathbb C}.
    \label{eq:length_one}
\end{equation}

\emph{Step 2: off-diagonal words and diagonal differences.}

A word is a finite ordered product of branch generators; its length is the number of factors. The induction below appends one factor at a time using a commutator on the joint space. This is how associative products become accessible through Lie brackets.
Let $\calW$ consist of finite products of elements of $\bigcup_\ell\mathfrak g_\ell^{\mathbb C}$, including the empty word $\id_s$. These words span $\calA_s$. We claim only that $E_{ab}\otimes w\in\gint^{\mathbb C}$ for $a\neq b$ and every $w\in\calW$. Step~0 gives the empty word. If the claim holds for $w$ and $X$ is a branch generator, Step~1 supplies $E_{aa}\otimes X$, and
\begin{equation}
    [E_{aa}\otimes X,E_{ab}\otimes w]=E_{ab}\otimes Xw
    \qquad(a\neq b).
\end{equation}
This proves the off-diagonal induction. Bracketing with $E_{ba}\otimes\id_s$ then gives $(E_{aa}-E_{bb})\otimes w$. Individual diagonal blocks are not yet available for arbitrary words: their traces may lie outside the generated algebra.
For $a\neq b$, the available off-diagonal words also give
\begin{equation}
\begin{split}
 &[E_{ab}\otimes w,E_{ba}\otimes w']+[E_{ba}\otimes w,E_{ab}\otimes w']\\
 &\qquad=(E_{aa}+E_{bb})\otimes[w,w'].
\end{split}
\end{equation}
Summing over $a<b$ yields $(L-1)\id_L\otimes[w,w']$, and $L\geq2$ therefore implies
\begin{equation}
    \id_L\otimes[\calA_s,\calA_s]\subseteq\gint^{\mathbb C},
    \label{eq:identity_slot}
\end{equation}
where $[\calA_s,\calA_s]$ denotes the linear span of commutators.

\emph{Step 3: assembling the semisimple part.}

The off-diagonal units and diagonal differences give $\fsl(L,\mathbb C)\otimes\calA_s\subseteq\gint^{\mathbb C}$. Together with \eqref{eq:identity_slot}, these directions contain
\begin{equation}
 \bigoplus_i\left(\fsl(L)\otimes M_{d_i}
       +\id_L\otimes\fsl(d_i)\right)\otimes\id_{m_i}.
\end{equation}
Here each summand is independently available: the proven inclusions hold for every element of $\calA_s$ and $[\calA_s,\calA_s]$, including elements supported on a single Wedderburn block. No closure under multiplication of arbitrary Lie elements is assumed. Within block $i$, the displayed sum is traceless and has dimension $(L^2-1)d_i^2+(d_i^2-1)=(Ld_i)^2-1$, so it equals $\fsl(Ld_i,\mathbb C)$. Intersecting with $\fu(Ld)$ by Lemma~\ref{lem:real_form} yields $\bigoplus_i\su(Ld_i)\otimes\id_{m_i}\subseteq\gint$.
Denote this semisimple algebra by $\mathfrak s$; the center remains to be determined.

\emph{Step 4: exact center.}

Block traces detect the remaining phase directions and vanish on commutators. To use this, write $\Skew(M_L(\mathbb C)\otimes\calA_s)=\mathfrak s\oplus\zeta$. For each block, the linear functional $\varphi_i(Y)=\Tr[Y(\id_L\otimes e_i)]$ satisfies $\varphi_i([Y,Y'])=\Tr[Y[Y',\id_L\otimes e_i]]=0$, because $\id_L\otimes e_i$ is central in the envelope. Every bracket therefore lies in $\mathfrak s$. The central component of any generated direction is a real combination of central components of the original generators. Since Step~3 gives $\mathfrak s\subseteq\gint$, we obtain $\gint=\mathfrak s\oplus\pi_\zeta(\mathfrak h+\mathfrak c)$, proving \eqref{eq:saturation}. The dimension bounds follow from $\dim\zeta=q$.
\end{proof}

\begin{proof}[Proof of Corollary~\ref{cor:universality}]\phantomsection\label{proof:cor:universality}
Apply Theorem~\ref{thm:saturation} with $q=1$, $d_1=d$, $m_1=1$: the center $\zeta=\mathbb R\,i\id_{Ld}$ is one-dimensional, and $\pi_\zeta(\mathfrak h+\mathfrak c)\neq\{0\}$ exactly when some generator has a nonvanishing trace component.
\end{proof}

The following count is used in the graph-control calculation: it gives the contribution of each connected cluster directly from its system-envelope blocks and its available central directions.

\begin{corollary}[Dimension at saturation]\label{cor:iota_saturated}
Under Theorem~\ref{thm:saturation},
\begin{equation}
 \dim\gint=L^2\dim_{\mathbb C}\calA_s-q+\dim\mathfrak z_{\INT}.
 \label{eq:dim_saturated}
\end{equation}
The center defect satisfies $0\leq q-\dim\mathfrak z_{\INT}\leq q$.
\end{corollary}
\begin{proof}[Proof of Corollary~\ref{cor:iota_saturated}]\phantomsection\label{proof:cor:iota_saturated}
Sum the simple-ideal dimensions $(Ld_i)^2-1$ and add the center dimension. Since $\sum_i d_i^2=\dim_{\mathbb C}\calA_s$, the displayed formula follows.
\end{proof}

\subsection{Graph-generated coefficient control}\label{app:graph}

Limited connections between branch labels can restrict which circuits participate in the output. We identify those restrictions from the graph of allowed ancilla couplings. Its connected components can be treated separately: saturation determines control within each nontrivial component, while the prepared reference state determines which component contributes to the loss.

\begin{proposition}[Connectivity governs the interleaved DLA]\label{prop:graph}
Let $\Gamma$ be a graph on the vertex set $[L]$ with edge set $E(\Gamma)$, and let
\begin{equation}
\begin{split}
    \gcoeff(\Gamma)=\Lie_{\mathbb R}\big\{&i(E_{jk}+E_{kj}),\ E_{jk}-E_{kj}\\[-2pt]
    &:(j,k)\in E(\Gamma)\big\}.
\end{split}
\end{equation}
Let $\calC(\Gamma)$ be the set of connected components of $\Gamma$, and split it into the components with at least two vertices, $\calC_{\geq2}(\Gamma)$, and the isolated vertices, $\calC_1(\Gamma)$. Then
\begin{equation}
    \gcoeff(\Gamma)=\bigoplus_{C\in\calC_{\geq2}(\Gamma)}\su(|C|),
    \label{eq:graph_closure}
\end{equation}
where $\su(|C|)$ acts on $\Span\{\ket\ell:\ell\in C\}$; an isolated vertex carries no coefficient generator. Consequently: (i) the interleaved-circuit DLA splits over the components. Each component with $|C|\geq2$ saturates Theorem~\ref{thm:saturation} on its own ancilla block, whereas an isolated branch retains exactly its own algebra:
\begin{equation}
\begin{split}
    \gint=&\bigoplus_{C\in\calC_{\geq2}(\Gamma)}\Big(\bigoplus_i\su\big(|C|\,d_i^C\big)\otimes\id_{m_i^C}\oplus\mathfrak z_C\Big)\\
    &\oplus\bigoplus_{\{\ell\}\in\calC_1(\Gamma)}P_\ell\otimes\mathfrak g_\ell,
\end{split}
    \label{eq:graph_gBE}
\end{equation}
where $\calA_s^C\cong\bigoplus_{i=1}^{q_C}M_{d_i^C}(\mathbb C)\otimes\id_{m_i^C}$ is the joint envelope of the branches in $C$ and $\mathfrak z_C$, of dimension at most $q_C$, is the center that Theorem~\ref{thm:saturation} assigns to the $|C|$-branch sub-architecture on $C$. Hence
\begin{equation}
\begin{split}
    \dim\gint=&\sum_{C\in\calC_{\geq2}(\Gamma)}\big(|C|^2\dim_{\mathbb C}\calA_s^C-q_C+\dim\mathfrak z_C\big)\\
    &+\sum_{\{\ell\}\in\calC_1(\Gamma)}\dim\mathfrak g_\ell.
\end{split}
    \label{eq:graph_dim}
\end{equation}
(ii) $\dim\gint$ is unchanged by edge changes that preserve the connected-component partition.
\end{proposition}

\begin{proof}[Proof of Proposition~\ref{prop:graph}]\phantomsection\label{proof:prop:graph}
For a single edge $(j,k)$, the two generators and their bracket
\begin{equation}
    [\,i(E_{jk}+E_{kj}),\,E_{jk}-E_{kj}\,]=-2i(E_{jj}-E_{kk})
\end{equation}
span $\su(2)_{jk}$ on $\Span\{\ket j,\ket k\}$. For two edges sharing a vertex, the bracket computation $[E_{jk}-E_{kj},\,E_{kl}-E_{lk}]=E_{jl}-E_{lj}$ produces the missing edge, and iterating yields $\su(3)$ on $\{j,k,l\}$. Induction along a spanning tree, a connected subset of edges without cycles that reaches every vertex of the component, with at least two vertices gives $\su(|C|)$; an isolated vertex belongs to no edge and therefore carries no generator; generators on distinct components have disjoint supports and commute, proving \eqref{eq:graph_closure}.

For claim (i), every generator of $\gint$ is supported on a single ancilla block $\Span\{\ket\ell:\ell\in C\}\otimes\calH_s$, and operators supported on distinct blocks have zero product, so the Lie closure is the direct sum of the closures computed block by block. On a block with $|C|\geq2$ the coefficient generators close to $\su(|C|)$, so Theorem~\ref{thm:saturation} applies to the $|C|$-branch sub-architecture on that block; its hypothesis $L\geq2$ is exactly $|C|\geq2$. On a block $C=\{\ell\}$ the only generators are $P_\ell\otimes\mathfrak g_\ell$, which is already a Lie algebra, so the closure is $P_\ell\otimes\mathfrak g_\ell$ itself.

The dimension count \eqref{eq:graph_dim} follows from Corollary~\ref{cor:iota_saturated} applied to each saturated block. Claim (ii) is immediate since \eqref{eq:graph_closure} and \eqref{eq:graph_gBE} depend only on the partition.
\end{proof}

Consider $L=4$ branches on one qubit with $\mathfrak g_\ell=\fu(2)$. For the partitions $\{1\}\{2\}\{3\}\{4\}$, $\{12\}\{3\}\{4\}$, $\{12\}\{34\}$, and $\{1234\}$, equation~\eqref{eq:graph_dim} gives dimensions $16$, $24$, $32$, and $64$. Every connected graph, from a path to the complete graph, generates $\fu(8)$. Disconnected components mix independently.

Here an isolated branch contributes dimension four, which happens to equal the dimension of the full one-qubit algebra. Saturation cannot be assumed for an isolated vertex in general. For example, take two branches on two qubits with $\mathfrak g_1=\mathfrak g_2=\su(2)\oplus\su(2)$, the six-dimensional algebra of independent single-qubit rotations. Their joint envelope is $M_4(\mathbb C)$. Without an edge, the generated dimension is $6+6=12$; incorrectly applying saturation to each vertex would give $2\dim\su(4)=30$. With the edge, the pair generates $\su(8)$, of dimension $63$.

\begin{corollary}[Restriction to the prepared component]\label{cor:cluster}
Use the graph-generated coefficient algebra of Proposition~\ref{prop:graph}, with both specified generators present on each edge. Let $C_0$ be the component containing the reference basis state $\ket{\mathrm{ref}}$, and let every circuit gate be generated by the indicated coefficient and branch algebras. Then $\LBE$ and all its parameter derivatives equal those of the circuit restricted to $C_0$; derivatives with respect to parameters supported outside $C_0$ vanish.

If $|C_0|\geq2$, $\calA_s^{C_0}=M_d(\mathbb C)$, and the restricted joint ensemble is an exact second-moment design over its generated group, then
\begin{equation}
 \Var[\LBE]=\lambda^4
 \frac{\big(\Tr(\rho^2)-\tfrac1{|C_0|d}\big)
       \big(\Tr(O^2)-\tfrac{\Tr(O)^2}{|C_0|d}\big)}{|C_0|^2d^2-1}.
 \label{eq:cluster_variance}
\end{equation}
For a general system envelope, use the component's simple ideals in Theorem~\ref{thm:reduction}, subject to membership and the restricted design hypothesis. Component saturation alone does not imply either a universal variance formula or design moments.

For a single-pass LCU whose preparations use only these graph gates, all coefficients outside $C_0$ vanish, $\lambda=\sum_{\ell\in C_0}|\alpha_\ell|$, and $\calL=\LBE$. If $C_0=\{\ell\}$, this gives $\calL=|\alpha_\ell|^2m_\ell$ and $\Var[\calL]=|\alpha_\ell|^4\Var[m_\ell]$. Resolve the remaining branch variance by \eqref{eq:diag_sector} only under its branchwise hypotheses.
\end{corollary}

\begin{proof}[Proof of Corollary~\ref{cor:cluster}]\phantomsection\label{proof:cor:cluster}
Every allowed generator, and hence every gate, is block diagonal in the ancilla components. Since both $P_{\mathrm{ref}}\otimes\rho$ and $P_{\mathrm{ref}}\otimes O$ are supported on $C_0$, the trace in \eqref{eq:lifted_loss_general} involves only that block. This identity persists under parameter differentiation.

For a nontrivial component, Proposition~\ref{prop:graph} and the full-envelope assumption give a restricted algebra containing $\su(|C_0|d)$. Its exact design moments imply \eqref{eq:cluster_variance} by Corollary~\ref{cor:universal_variance}; no traceful generator is needed for conjugation variance. With a smaller envelope, saturation instead leaves the specified system blocks, so the sector theorem and its membership condition apply. Finally, graph-generated preparations cannot move $\ket{\mathrm{ref}}$ outside $C_0$, proving the single-pass LCU coefficient support and isolated-branch statements.
\end{proof}

\section{Joint-space variance}\label{app:joint}

With the sampling and membership assumptions in place, we apply the unitary variance formula to the joint loss. We also derive its covariance form and evaluate the full-control case.

\subsection{Sector reduction and covariance}\label{app:reduction}

When an objective combines several measured quantities, their fluctuations can reinforce or offset one another. We therefore need covariance as well as individual variances. Starting from the joint-space sector formula, we derive this covariance by considering a sum of observables, then obtain simpler bounds that retain the input and measurement overlaps.

\begin{proof}[Proof of Theorem~\ref{thm:reduction}]\phantomsection\label{proof:thm:reduction}
By \eqref{eq:lifted_loss_general}, $\LBE=\lambda^2\tcalL$, where $\tcalL=\Tr[\widetilde O_{\mathrm{ref}}W\widetilde\rho_{\mathrm{ref}}W\dagg]$. The deterministic scale gives $\Var[\LBE]=\lambda^4\Var[\tcalL]$. Lemmas~\ref{lem:ragone} and~\ref{lem:gen_membership} apply to the fixed lifted operators under \eqref{eq:membership}, yielding \eqref{eq:main_reduction_formula}. Exact matching of the first two conjugation moments suffices because both the mean and the second moment of $\tcalL$ depend only on these channels.
\end{proof}

\begin{theorem}[Joint covariance]\label{thm:joint_cov}
Under the hypotheses of Theorem~\ref{thm:reduction}, for Hermitian $X,Y\in i\gint+\mathbb R\,\id$,
\begin{equation}
\begin{split}
    &\Cov\big(\Tr[X\,W\widetilde\rho W\dagg],\,\Tr[Y\,W\widetilde\rho W\dagg]\big)\\
    &\qquad=\sum_{j=1}^{r}\frac{\calP_j^{\INT}(\widetilde\rho)\,\big\langle\Pi_j^{\INT}(X),\Pi_j^{\INT}(Y)\big\rangle_{\HS}}{\dim\mathfrak g_j^{\INT}}.
\end{split}
\label{eq:joint_cov}
\end{equation}
\end{theorem}

\begin{proof}[Proof of Theorem~\ref{thm:joint_cov}]\phantomsection\label{proof:thm:joint_cov}

Apply the variance formula to the sum of the observables. Its cross term is twice their covariance, an identity called polarization. The stated membership condition holds for both observables and their sum.

Let $f_X(W)=\Tr[X\,W\widetilde\rho W\dagg]$; $f$ is linear in the observable and $X+Y\in i\gint+\mathbb R\,\id$, so Theorem~\ref{thm:reduction} applies to $f_X$, $f_Y$, $f_{X+Y}$. Polarization, $\Cov(f_X,f_Y)=\tfrac12(\Var f_{X+Y}-\Var f_X-\Var f_Y)$, together with $\mathbb R$-linearity of $\Pi_j^{\INT}$ on Hermitian operators and HS-orthogonality of distinct simple ideals, yields \eqref{eq:joint_cov}.
\end{proof}

When there are many ideals, bounding their denominators by the smallest and largest dimensions simplifies the sector sum. The common overlap factor $Q$ retains the dependence on the input and observable.

\begin{proposition}[Variance bounds from sector dimensions]\label{prop:variance_structural}
Under the hypotheses of Theorem~\ref{thm:reduction}, let $Q=\sum_j\calP_j^{\INT}(\widetilde\rho)\,\calP_j^{\INT}(\widetilde O)\in[0,\Tr(\rho^2)\Tr(O^2)]$ and suppose $\gint$ has at least one simple ideal. Let $\dim_{\min}\leq\dim_{\max}$ be its extreme simple-ideal dimensions. If there are no simple ideals, membership makes the loss deterministic instead. Then:
\begin{enumerate}
    \item[(i)] (sector sandwich)
    $\lambda^4Q/\dim_{\max}\leq\Var[\LBE]\leq\lambda^4Q/\dim_{\min}$;
    \item[(ii)] (dimension bound)
    $\Var[\LBE]\geq\lambda^4Q/\dim\gint$. For $Q>0$, equality requires that every contributing ideal have dimension $\dim\gint$; one active ideal in a larger direct sum is not sufficient;
    \item[(iii)] (envelope--commutant window) the guaranteed floor $\Var_\flat=\lambda^4Q/(\dim_{\mathbb C}\calA_a\dim_{\mathbb C}\calA_s)$ satisfies, with $\kappa_s=\dim_{\mathbb C}\Comm(G_s)$,
    \begin{equation}
        \frac{\lambda^4Q}{\dim_{\mathbb C}\calA_a\,d^2}
        \;\leq\;\Var_\flat\;\leq\;
        \frac{\kappa_s\,\lambda^4Q}{\dim_{\mathbb C}\calA_a\,d^2}.
        \label{eq:var_commutant_window}
    \end{equation}
\end{enumerate}
\end{proposition}

\begin{proof}[Proof of Proposition~\ref{prop:variance_structural}]\phantomsection\label{proof:prop:variance_structural}
Each summand of \eqref{eq:main_reduction_formula} is nonnegative with denominator in $[\dim_{\min},\dim_{\max}]$, giving (i). For the bound on $Q$: with $a_j=\calP_j^{\INT}(\widetilde\rho)$, $b_j=\calP_j^{\INT}(\widetilde O)$, $Q\leq(\sum_ja_j)(\sum_jb_j)\leq\|\widetilde\rho\|_{\HS}^2\|\widetilde O\|_{\HS}^2=\Tr(\rho^2)\Tr(O^2)$, since $\sum_j\calP_j^{\INT}(X)\leq\|X\|_{\HS}^2$. (ii): every ideal has dimension at most $\dim\gint$; equality in the summed bound requires equality for every summand with nonzero numerator. (iii): $\Var[\LBE]\geq\lambda^4Q/\dim\gint\geq\Var_\flat$ by Theorem~\ref{thm:upper_bound}, and $d^2/\kappa_s\leq\dim_{\mathbb C}\calA_s\leq d^2$ by Lemma~\ref{lem:duality}.
\end{proof}

Part (iii) bounds the envelope-based floor between two quantities involving $\kappa_s$. A larger commutant widens the allowed window; it does not by itself prove a larger realized variance. If the duality inequality is saturated, then $\dim_{\mathbb C}\calA_s=d^2/\kappa_s$ and $\Var_\flat$ equals the upper endpoint of the displayed window.

For $\kappa_s=1$ the endpoints coincide. In either case the bound retains $Q$, which may be small or zero.

\subsection{Universal Haar-measure covariance}\label{app:haar}

Full unitary control provides a reference case in which the loss fluctuations can be calculated without resolving separate sectors. Sampling from Haar measure treats all traceless operator directions symmetrically, giving a covariance formula with one denominator. We separate the identity components to keep the result valid for operators with nonzero trace.

\begin{theorem}[Universal joint Haar-random covariance]\label{thm:full_haar_cov}
Let $D=Ld\geq2$ and let $W$ be Haar-random on $\U(D)$ or $\SU(D)$. For Hermitian $\sigma,X,Y$ and $f_X(W)=\Tr[X\,W\sigma W\dagg]$,
\begin{equation}
\begin{split}
    \Cov(f_X,f_Y)
    ={}&\frac{\Tr(\sigma^2)-\Tr(\sigma)^2/D}{D^2-1}\\
    &\times\Big(\Tr(XY)-\tfrac{\Tr(X)\Tr(Y)}{D}\Big).
\end{split}
\label{eq:full_haar_cov}
\end{equation}
\end{theorem}

\begin{proof}[Proof of Theorem~\ref{thm:full_haar_cov}]\phantomsection\label{proof:thm:full_haar_cov}
For the conjugation-type moments used here, $\SU(D)$ and $\U(D)$ agree for every $D\geq2$: by Schur--Weyl duality, which identifies the operators commuting with every tensor-power unitary, the fixed-operator space of the two-copy conjugation action is $\Span\{\id^{\otimes2},\Pi\}$ for both, with $\Pi$ the swap operator, defined by $\Pi(x\otimes y)=y\otimes x$. Decompose $\sigma=\sigma_0+\tfrac{\Tr\sigma}{D}\id$, and similarly $X,Y$, with traceless $\sigma_0,X_0,Y_0$; the identity parts are conjugation-invariant and cancel in the covariance, leaving
\begin{equation}
    \Cov(f_X,f_Y)=\mathbb E_W\big[\Tr(X_0W\sigma_0W\dagg)\Tr(Y_0W\sigma_0W\dagg)\big].
\end{equation}
\emph{Step 1: combine the two traces.} The notation $\Tr_{12}$ means the trace over both tensor factors. Using $\Tr(A)\Tr(B)=\Tr_{12}[A\otimes B]$, the right side equals $\Tr_{12}[(X_0\otimes Y_0)\,\mathbb E_W W^{\otimes2}(\sigma_0\otimes\sigma_0)(W\dagg)^{\otimes2}]$.
\emph{Step 2: average the two-copy operator.} Only the identity and swap directions survive this group average. The second-moment integral on $\U(D)$ \citep{Collins2003,Roberts2017} gives, for $M=\sigma_0\otimes\sigma_0$ (so $\Tr M=0$, $\Tr[M\Pi]=\|\sigma_0\|_{\HS}^2$),
\begin{equation}
    \mathbb E_W\big[W^{\otimes2}M(W\dagg)^{\otimes2}\big]
    =\frac{\|\sigma_0\|_{\HS}^2}{D^2-1}\Big(\Pi-\tfrac1D\id^{\otimes2}\Big).
\end{equation}
\emph{Step 3: contract with the observables.} The swap converts a tensor-product trace back into an ordinary matrix product. With $\Tr_{12}[(A\otimes B)\Pi]=\Tr(AB)$ and tracelessness,
\begin{equation}
    \Cov(f_X,f_Y)=\frac{\|\sigma_0\|_{\HS}^2}{D^2-1}\,\Tr(X_0Y_0).
\end{equation}
\emph{Step 4: restore the original operators.} Substitute
\begin{align}
 \|\sigma_0\|_{\HS}^2&=\Tr(\sigma^2)-\Tr(\sigma)^2/D,\\
 \Tr(X_0Y_0)&=\Tr(XY)-\Tr(X)\Tr(Y)/D
\end{align}
to obtain \eqref{eq:full_haar_cov}.
\end{proof}

\begin{corollary}[Universal joint-space variance]\label{cor:universal_variance}
Let $Ld\geq2$, $\gint\supseteq\su(Ld)$, and let the joint ensemble be an exact second-moment design over $G_{\INT}$. Then \eqref{eq:universal_LCQNN_variance} holds for $\LBE$.
\end{corollary}
\begin{proof}[Proof of Corollary~\ref{cor:universal_variance}]\phantomsection\label{proof:cor:universal_variance}
Set $D=Ld$, $\sigma=P_{\mathrm{ref}}\otimes\rho$, and $X=Y=P_{\mathrm{ref}}\otimes O$ in Theorem~\ref{thm:full_haar_cov}. Their traces and squared Hilbert--Schmidt norms are those of $\rho$ and $O$. Finally $\Var(\LBE)=\lambda^4\Var(\tcalL)$.
\end{proof}

\section{Matchgate and Pauli examples}\label{app:examples}

We verify the matchgate theorem and the plotted gradient benchmark, then calculate Pauli examples with small branch algebras and a full joint envelope.

\subsection{Matchgate regimes}\label{app:matchgate}

Matchgate circuits let us test how added mixing changes control and training signals while a shared symmetry remains intact. We verify the claims about control dimension, classical evaluation cost, and initialization variance under their respective assumptions. The proof also shows why parity preservation does not prevent extensive control within each parity subspace.

\begin{proof}[Proof of Theorem~\ref{thm:matchgate}]\phantomsection\label{proof:thm:matchgate}
The structural parts use the specified matchgate subalgebras and mixer. The simulation claim additionally uses a Gaussian input and bounded-degree observable; the variance claims use their stated moment and membership conditions.

\emph{(i).} The structure of $\gint$ is Theorem~\ref{thm:lower_bound} (diagonal case); the dimension bound follows from $\dim\mathfrak g_\ell\leq n(2n-1)$.

\emph{Simulation.}

The diagonal contributions are ordinary expectation values; the off-diagonal contributions are complex overlaps between different branch outputs. Reconstructing the coherent loss requires both their magnitudes and their phases.

With a diagonal coefficient layer the coefficient gates only change the values of $\alpha$, so it suffices to evaluate $\calL=\sum_{\ell,k}\alpha_\ell\bar\alpha_k\Tr[O\,U_\ell\rho\,U_k\dagg]$ and its derivatives for given $\alpha$. Each of the $L^2$ terms equals $\Tr[(U_k\dagg OU_k)(U_k\dagg U_\ell)\rho]$, where $U_k\dagg OU_k$ is again a Majorana polynomial of the same degree, because conjugation by a matchgate unitary maps Majorana operators to linear combinations of Majorana operators, and $U_k\dagg U_\ell$ is a matchgate unitary.

A parameter derivative inserts a quadratic generator $H$ inside the product and therefore raises the degree by at most two per derivative. For a Gaussian input, each term is thus a matrix element of a bounded-degree Majorana polynomial between fermionic Gaussian states, and such quantities are computable in time polynomial in $n$ by free-fermion methods \citep{TerhalDiVincenzo2002,JozsaMiyake2008,DiasKoenig2024}; there are $L^2$ terms.

For the off-diagonal contributions, the relative phase must be tracked as well as the covariance matrix; the overlap and evolution algorithms of Dias and K\"onig~\cite{DiasKoenig2024} provide this information for Gaussian state vectors.

\emph{Variance.} Under Assumption~\ref{asm:moments}(a) the exact formula of Theorem~\ref{thm:exact_select} applies to any branch groups, and Lemmas~\ref{lem:cross_moments} and~\ref{lem:form_pairing} express the cross moments through the commutant and the form space of each $G_\ell$. For full matchgate branches, $O\in i\,\so(2n)+\mathbb R\,\id=i\mathfrak g_\ell+\mathbb R\,\id$, so the branchwise membership hypothesis holds and \eqref{eq:diag_sector} applies; the matchgate commutant is $\Span\{\id,P\}$, giving $\kappa_\ell=\kappa_k=2$ in \eqref{eq:cross_bound}, and $\tau_\ell=2$ by Lemma~\ref{lem:vanishing}(iv), with form terms vanishing for pure fermionic Gaussian inputs and $O(\Tr(\rho^2)\Tr(O^2)/d^2)$ in general by the bound of Lemma~\ref{lem:form_pairing} with $\nu^2=2/d$. For a proper subalgebra $\mathfrak g_\ell\subsetneq\so(2n)$ the same lemmas hold with the commutant and form space of $G_\ell=e^{\mathfrak g_\ell}$, whose dimensions may exceed $2$ (Remark~\ref{rem:matchgate_sub}); the diagonal term resolves by \eqref{eq:diag_sector} only if $O\in i\mathfrak g_\ell+\Herm(\Comm(G_\ell))$, and otherwise enters Theorem~\ref{thm:exact_select} as the unresolved single-branch variance $\Var[m_\ell]$.

\emph{(ii).} Every $\mathfrak g_\ell$ consists of even Clifford elements, so $\calA_s\subseteq\mathrm{Cl}^0_{2n}$, and the joint-generation hypothesis gives equality; when some $\mathfrak g_\ell=\so(2n)$, products of quadratic Majorana monomials generate all even monomials for $n\geq2$, and the two half-spinor blocks are irreducible and inequivalent. The Wedderburn data are $q=2$, $d_1=d_2=2^{n-1}$, $m_i=1$, so Theorem~\ref{thm:saturation} gives the two-sided inclusion, with the upper end from Theorem~\ref{thm:upper_bound}.

For \eqref{eq:matchgate_BP}: a parity-even $O$ lifts to an operator $\widetilde O$ that is block diagonal with respect to $\id_a\otimes P$ (the same argument applies to $\widetilde\rho$ when $\rho$ is parity-even). Such an operator decomposes as (element of $i\gint$) $+$ (real combination of $\id$ and $\id_a\otimes P$), because $\gint$ contains $\su(L2^{n-1})^{\oplus2}$ on the two parity blocks; the latter two operators commute with every element of $G_{\INT}$ and thus lie in $\Herm(\Comm(G_{\INT}))$, so the generalized membership of Lemma~\ref{lem:gen_membership} holds for $\widetilde O$ (the commutant extension in \eqref{eq:membership} is essential when the parity direction is absent from the Lie algebra). The other lifted operator is arbitrary, since only one of the two needs to satisfy the hypothesis. Theorem~\ref{thm:reduction} then gives
\begin{equation}
\begin{split}
    \Var[\LBE]
    &\leq\frac{\lambda^4}{\min_j\dim\mathfrak g_j^{\INT}}
    \Big(\max_j\calP_j^{\INT}(\widetilde\rho)\Big)\sum_j\calP_j^{\INT}(\widetilde O)\\
    &\leq\frac{\lambda^4\,\Tr(\rho^2)\Tr(O^2)}{L^24^{n-1}-1},
\end{split}
\end{equation}
using $\sum_j\calP_j^{\INT}(X)\leq\|X\|_{\HS}^2$ and $\Tr(\widetilde\rho^2)=\Tr(\rho^2)$, $\Tr(\widetilde O^2)=\Tr(O^2)$.

The parity preservation follows directly: branch generators are even Clifford elements, so $[P_\ell\otimes X,\id_a\otimes P]=P_\ell\otimes[X,P]=0$; coefficient generators act trivially on the system factor; and commutation with $\id_a\otimes P$ passes to brackets and spans.
\end{proof}

\subsection{Proof of the gradient benchmark}\label{proof:benchmark}

To see how much of the plotted gradient signal comes from an individual branch and how much comes from interference, we evaluate both contributions explicitly. The branch calculation uses the input's two-point Majorana correlations, collected in a covariance matrix. This matrix describes state correlations; it is distinct from the statistical covariance of two losses.

Matchgate rotations determine the branch term in Figure~\ref{fig:scaling}. The even-parity subspace and the vanishing input form contraction determine the interference term.

For \eqref{eq:matchgate_gradient_benchmark}, define the real antisymmetric covariance matrix by $\Gamma_{ab}=\Tr(\rho\,ic_ac_b)$ for $a\ne b$ and $\Gamma_{aa}=0$. A pure fermionic Gaussian state satisfies $\Gamma^T\Gamma=\id_{2n}$, so
\begin{equation}
 \sum_{a<b}\Gamma_{ab}^{\,2}=n.
\end{equation}
Full-matchgate Haar-random conjugation rotates $\Gamma$ by a Haar-random orthogonal matrix of determinant one. All $n(2n-1)$ off-diagonal coordinates then have the same second moment and zero mean. Each quadratic Pauli expectation therefore has variance $1/(2n-1)$, including at $n=2$. Since $i[Z_1,X_1X_2]=-2Y_1X_2$, the diagonal gradient variance is $4/(2n-1)$.

The Haar-random output of either branch is supported on the even-parity subspace of dimension $D_+=2^{n-1}$. Its mean density matrix is $Q_+/D_+$, where $Q_+=(\id+P)/2$. With $B=OH=Z_1X_1X_2$, $B$ preserves parity and $B^\dagger B=\id$. Independent first conjugation moments therefore give
\begin{equation}
 \mathbb E|\Gamma_{12}|^2
 =\frac{\Tr(Q_+BQ_+B^\dagger)}{D_+^2}
 =\frac1{D_+}=\frac2{2^n}.
\end{equation}
Lemma~\ref{lem:vanishing}(iv) removes the unconjugated moment for this computational-basis input, for $B=OH$ as well as for $B=O$. At $w_1=w_2=1/2$, Theorem~\ref{thm:grad_select} gives the claimed expression. No joint-space design is used.

\subsection{Worked examples}\label{sec:worked}

The same branch circuits can behave differently when combined in one pass or through repeated mixing. These Pauli examples make the comparison concrete and show where the input, observable, and sampling law enter. We first identify the available control, then calculate variance only when the required experimental and ensemble data are specified.

Both examples use $L=2$ branches on $n$ qubits, with $d=2^n$, and compare fixed coefficients with a fully expressive ancilla layer. Figure~\ref{fig:scaling} supplies a separate matchgate benchmark. Its joint parity-block curve follows directly from the sector formula: $\widetilde\rho_{\mathrm{ref}}$ occupies only the even block of dimension $d$, with projected purity $1-1/d$, while the even-block restriction of $\widetilde O_{\mathrm{ref}}$ is traceless with squared HS norm $d/2$. Thus its variance divided by $\lambda^4$ is $(1-1/d)(d/2)/(d^2-1)=1/[2(d+1)]$. The universal curve is \eqref{eq:universal_LCQNN_variance}, which reduces to $1/[2(2d+1)]$ for the same operators.

\paragraph{Abelian branches with a full joint envelope.}
Let $\mathfrak g_1=\Span_{\mathbb R}\{iX_r\}_{r=1}^n$ and $\mathfrak g_2=\Span_{\mathbb R}\{iZ_r\}_{r=1}^n$, as in Example~\ref{ex:pauli_envelope}, both abelian of dimension $n$. With fixed or diagonal coefficients, the joint algebra remains ancilla-block-diagonal. The implemented LCU remains a coherent sum, however; a block-diagonal DLA does not turn it into an incoherent mixture. Each branch group has $2^n$ distinct characters, giving $\kappa_\ell=\tau_\ell=2^n$. These counts permit many cross pairings, but their values still depend on $\rho$, $O$, and the phases.

Nor does an abelian branch DLA imply zero single-branch variance for arbitrary operators. For example, a uniformly random $Z$ rotation with input $\ketbra{+}{+}$ and observable $X$ gives a fluctuating expectation; the branchwise membership condition is what determines whether \eqref{eq:diag_sector} applies. This example therefore illustrates algebraic growth, without assigning an interference share before the input and observable are fixed.

Now switch on $\gcoeff=\fu(2)$: the joint \emph{Lie} closure of the branches has dimension only $3n$, so pairwise-controllability criteria are silent, but the joint \emph{associative} envelope is full (Example~\ref{ex:pauli_envelope}), and Corollary~\ref{cor:universality} gives
\begin{equation}
    \gint=\fu(2^{n+1}),\qquad \dim\gint=4^{n+1}.
\end{equation}
Under the exact joint-design assumption, the variance of $\LBE$ equals \eqref{eq:universal_LCQNN_variance}. This is exponentially small when $\lambda$ and $\|O\|_\infty$ grow at most polynomially in $n$; the structural conclusion alone does not specify the loss or gradient scale. The branch products become accessible through ancilla commutators, as described in Remark~\ref{rem:LARC}.

\paragraph{Single-qubit-universal branches: exact numbers.}
Let instead $\mathfrak g_1=\mathfrak g_2=\fu(1)\oplus\bigoplus_{r=1}^n\su(2)_r$, where $\su(2)_r$ acts on qubit $r$, so that each branch group is $\U(1)\cdot\SU(2)^{\times n}$ (the global phase enforces Assumption~\ref{asm:moments}(b)), and take $\rho=\ketbra{+}{+}^{\otimes n}$ with $\ket+=(\ket0+\ket1)/\sqrt2$, and $O=X_1$, the Pauli $X$ operator on the first qubit. Equation \eqref{eq:diag_sector} localizes the diagonal variance in the sector $\su(2)_1$, giving $\Var[m_\ell]=\tfrac13$; irreducibility gives $\mathbb E[|m_{12}|^2]=1/d$. Theorem~\ref{thm:exact_select} is then fully explicit:
\begin{equation}
    \Var[\calL]=\lambda^4\Big(\frac{w_1^4+w_2^4}{3}+\frac{2w_1^2w_2^2}{d}\Big).
    \label{eq:worked_variance}
\end{equation}

The branch term remains finite at fixed weights while the interference term decreases with system size. Thus the loss variance is $\Theta(\lambda^4)$. At $n=1$ with balanced weights, interference accounts for $\chi=0.60$ of the total; its share decreases with $d$, as \eqref{eq:cross_bound} requires. With $\gcoeff=\fu(2)$, the full envelope of even one branch gives $\gint=\fu(2d)$. Under joint Haar-random sampling, the variance is then \eqref{eq:universal_LCQNN_variance}.

The examples distinguish an actual interference variance from the dimension of reachable control. The closures can be checked directly at small size; efficient large-$n$ evaluation depends on the representation and the structure available to the audit.

\section{Postselection and measurement}\label{app:postselect}

Postselection changes the loss and the cost of estimating it. We first compare normalized and unnormalized loss variances using the bounded observable. We then bound the measurement error of an unnormalized gradient estimator.

\begin{theorem}[Loss-variance transfer under postselection]\label{thm:postselect_train}
Let $\lambda>0$ be fixed, $\|O\|_\infty>0$, $\pacc=\Tr[A\rho A\dagg]>0$ almost surely, and $\bpacc=\mathbb E[\pacc]$. (For $O=0$, both losses vanish identically.) Then
$\big|\sd(\calL/\pacc)-\sd(\calL)/\bpacc\big|\leq\|O\|_\infty\,\sd(\pacc)/\bpacc$. In particular, if $O\neq0$ and $\Var(\pacc)\leq\Var(\calL)/(4\|O\|_\infty^2)$, then $\Var(\calL/\pacc)\geq\Var(\calL)/(4\bpacc^{\,2})\geq\Var(\calL)/(4\lambda^4)$.
\end{theorem}

\begin{proof}[Proof of Theorem~\ref{thm:postselect_train}]\phantomsection\label{proof:thm:postselect_train}

Compare the normalized loss with the loss divided by the mean success factor. Their difference is the fluctuation of that factor times a bounded normalized expectation. This bounds the error without estimating an inverse random denominator or assuming independence.

The pointwise identity
\begin{equation}
    \frac{\calL}{\pacc}=\frac{\calL}{\bpacc}+R,
    \qquad
    R=\frac{\calL}{\pacc}\cdot\frac{\bpacc-\pacc}{\bpacc},
\end{equation}
and the triangle inequality for standard deviation on the space of square-integrable random variables give $|\sd(\calL/\pacc)-\sd(\calL/\bpacc)|\leq\sd(R)$. Since $|\calL/\pacc|\leq\|O\|_\infty$ pointwise ($|\Tr[OA\rho A\dagg]|\leq\|O\|_\infty\Tr[A\rho A\dagg]$),
\begin{equation}
    \sd(R)\leq\sqrt{\mathbb E[R^2]}
    \leq\frac{\|O\|_\infty}{\bpacc}\,\sd(\pacc),
\end{equation}
and, since $\bpacc$ is a constant, $\sd(\calL/\bpacc)=\sd(\calL)/\bpacc$, which gives the first claim. Under the concentration hypothesis, $\|O\|_\infty\sd(\pacc)\leq\tfrac12\sd(\calL)$, so $\sd(\calL/\pacc)\geq\sd(\calL)/(2\bpacc)$. Squaring and using $\pacc\leq\|A\|_\infty^2\leq\lambda^2$ so that $\bpacc\leq\lambda^2$ gives the final claim.
\end{proof}

The hypothesis can hold: for independent zero-mean irreducible Haar-random branches and fixed coefficients, Lemma~\ref{lem:cross_moments} with $O=\id_s$, together with $|\mathbb E(m_{\ell k}^2)|\leq\mathbb E|m_{\ell k}|^2$, gives $\Var(\pacc)=O(\lambda^4/d)$. Thus the transfer condition holds eventually if $\Var(\calL)/\lambda^4\gg\|O\|_\infty^2/d$, for example when a polynomially weighted branch supplies an inverse-polynomial loss variance and $\|O\|_\infty$ grows at most polynomially.

The pointwise bound $|\calL/\pacc|\leq\|O\|_\infty$ makes a positive uniform lower bound on $\pacc$, higher-moment control, and a Taylor expansion unnecessary. The result complements the unnormalized analysis of Khatri et al.~\cite{Khatri2025PLCU}, where postselection appears as sub-normalization of the state.

\subsection{Sampling the unnormalized gradient}

To resolve a predicted gradient scale, measurement noise must be small relative to that scale. The proposition gives an estimator and an error-probability bound at a fixed parameter point, using the ensemble standard deviation as the target precision.

This precision bound does not give the probability that a random initialization has a gradient of that size.

\begin{proposition}[Shot-budget predictor]\label{prop:shots}
In the single-occurrence parameter setting of Theorem~\ref{thm:grad_select}, let $H^2=\id$, fix $0<\delta<1$, and let $\sigma^2=\Var(\partial_\theta\calL)>0$ denote the initialization-ensemble variance. The loss has trigonometric degree at most two, meaning that only constant terms and first and second sine and cosine harmonics occur. Estimate its derivative by the exact four-point shift rule of Wierichs et al.~\cite{Wierichs2022}, with independent shots at the four shifts. At every fixed parameter point, the probability over measurement outcomes that the estimation error exceeds $\sigma/2$ is at most $\delta$ whenever the shots per shifted evaluation satisfy
\begin{equation}
    S\;\geq\;\frac{6\lambda^4\|O\|_\infty^2}{\delta\,\Var(\partial_\theta \calL)},
    \label{eq:shots}
\end{equation}
with $\Var(\partial_\theta \calL)$ given in closed form by Theorem~\ref{thm:grad_select}. This is a sufficient precision budget relative to $\sigma$, not an anti-concentration statement about random gradients. It is polynomial when $\lambda^4\|O\|_\infty^2/\sigma^2$ is polynomial.
\end{proposition}

\begin{proof}[Proof of Proposition~\ref{prop:shots}]\phantomsection\label{proof:prop:shots}

Because the parameter occurs once and $H^2=\id$, the loss has only two positive frequencies. We use their exact shift rule, estimate each shifted loss from independent bounded outcomes, and apply Chebyshev's inequality to the resulting estimator.

With $H^2=\id$, $e^{i\theta H}=\cos\theta\,\id+i\sin\theta\,H$, so each $m_{\ell k}(\theta)$ with $k\neq\ell$ is a degree-one and $m_\ell(\theta)$ a degree-two trigonometric polynomial; hence $\calL(\theta)$ has harmonics $\{0,1,2\}$ and the equidistant parameter-shift rule of Wierichs et al.~\cite{Wierichs2022} for $R=2$ distinct positive frequencies, $\partial_\theta\calL=\sum_{\mu=1}^{4}c_\mu\,\calL(\theta+s_\mu)$ with $s_\mu=\tfrac{(2\mu-1)\pi}{4}$ and $c_\mu=\tfrac{(-1)^{\mu-1}}{8\sin^2(s_\mu/2)}$, is exact, with $\sum_\mu|c_\mu|=2$ and $\sum_\mu c_\mu^2=\tfrac32$.

Each shifted loss $\calL(\theta+s_\mu)$ is estimated by the empirical mean $\widehat{\calL}_\mu$ of $S$ independent, identically distributed single-shot outcomes $X=\lambda^2\,o\,\mathbf 1[\text{ancilla}=\mathrm{ref}]$, where $o$ is the measured eigenvalue of $O$ and $\mathbf 1[\cdot]$ the indicator of the ancilla outcome; $\mathbb E[X]=\lambda^2\Tr[(P_{\mathrm{ref}}\otimes O)W\widetilde\rho_{\mathrm{ref}}W\dagg]=\calL(\theta+s_\mu)$ and $|X|\leq\lambda^2\|O\|_\infty$, so $\Var(X)\leq\lambda^4\|O\|_\infty^2$.

The estimator $\widehat{\partial_\theta\calL}=\sum_\mu c_\mu\widehat{\calL}_\mu$ is unbiased with variance at most $\sum_\mu c_\mu^2\,\lambda^4\|O\|_\infty^2/S=\tfrac32\lambda^4\|O\|_\infty^2/S$. With $S\geq6\lambda^4\|O\|_\infty^2/(\delta\,\Var(\partial_\theta\calL))$, Chebyshev's inequality gives
\begin{equation}
    \Pr\Big[\big|\widehat{\partial_\theta\calL}-\partial_\theta\calL\big|\geq\tfrac{\sd(\partial_\theta\calL)}2\Big]
    \leq\frac{\tfrac32\lambda^4\|O\|_\infty^2/S}{\Var(\partial_\theta\calL)/4}\leq\delta.
\end{equation}
The probability here is conditional on the parameter point and concerns shot noise only. No lower-tail bound for the gradient over initializations has been used.
\end{proof}

\section{Approximate moment errors}\label{app:design}

Imperfect moment matching changes the predicted variance. For a single-pass LCU, controlled branch moments retain the relative-phase information needed to bound this change. For a joint circuit, ordinary conjugation moments suffice. Both comparisons take the moment discrepancy as a given input.

For a linear map $T$ on matrices, the diamond norm is
$\|T\|_\diamond=\sup_{r\geq1}\sup_{X\neq0}\|(T\otimes\mathrm{id}_r)(X)\|_1/\|X\|_1$.
It bounds the trace-norm error even when an auxiliary reference is retained.

Here $\mathrm{id}_r$ is the identity map on matrices of an $r$-dimensional reference system. The supremum includes all inputs and reference sizes. A \emph{moment channel} averages a conjugated operator over a circuit ensemble; bounding it in diamond norm gives the dimension-independent error estimates below.

\begin{definition}[Controlled moment error]\label{def:moment_error}
For $U\in\U(d)$, introduce an auxiliary two-level flag and the controlled unitary $\operatorname{Ctrl}(U)=\ketbra00\otimes\id_s+\ketbra11\otimes U$. This flag is only a device for defining a norm; it is not the LCU ancilla. For an ensemble $\calE$, set
\begin{equation}
    \Psi_{\calE}^{(t)}(X)=\mathbb E_{U\sim\calE}[\operatorname{Ctrl}(U)^{\otimes t}X(\operatorname{Ctrl}(U)\dagg)^{\otimes t}],
\end{equation}
and define
\begin{equation}
 \varepsilon(\calE,G)=\max_{t=1,2}
    \|\Psi_{\calE}^{(t)}-\Psi_{\mathrm{Haar}(G)}^{(t)}\|_\diamond.
\end{equation}
Off-diagonal flag blocks retain products with unequal numbers of $U$ and $\bar U$ factors, including the invariant-form moments. Ordinary conjugation moments contain equal numbers of unitary and conjugate-unitary factors; these are called balanced moments. Matching them alone does not control all of the terms retained by the flag.
\end{definition}

A \emph{hybrid} argument replaces actual branch distributions by their Haar-measure references one at a time. Summing the changes over these intermediate ensembles bounds the total error, even if the actual branches do not have exactly zero means.

\begin{proposition}[Hybrid bound]\label{prop:hybrid}
Let the actual branch ensembles $\calE_1,\ldots,\calE_L$ be independent, and compare them with independent Haar-random branches on $G_1,\ldots,G_L$. For fixed coefficients and $\varepsilon_\ell=\varepsilon(\calE_\ell,G_\ell)$ from Definition~\ref{def:moment_error},
\begin{equation}
 \big|\Var_{\calE}[\calL]-\Var_{\mathrm{Haar}}[\calL]\big|
 \leq8\lambda^4\|O\|_\infty^2\sum_{\ell=1}^Lw_\ell\varepsilon_\ell.
 \label{eq:hybrid_bound}
\end{equation}
No zero-first-moment assumption on the actual ensembles is needed for this comparison. To evaluate the reference variance by Theorem~\ref{thm:exact_select}, the reference groups must satisfy its hypotheses.
\end{proposition}

\begin{proof}[Proof of Proposition~\ref{prop:hybrid}]\phantomsection\label{proof:prop:hybrid}
Replace the branches by their Haar-random counterparts one at a time, conditioning on all other branches. The expansion of the mean contains individual overlaps $m_{ab}$, and the second moment contains products $m_{ab}m_{cd}$. We bound each such term before restoring its coefficient weight. For a fixed branch $\ell$, each factor $\Tr[OU_a\rho U_b\dagg]$ has zero or one occurrence of $U_\ell$ on each side of $\rho$. Encode their presence by flag values $u,v\in\{0,1\}$. The corresponding block of $\Psi^{(1)}(\ketbra uv\otimes\rho)$ is $\mathbb E[U_\ell^u\rho(U_\ell\dagg)^v]$. The other branch unitaries can be moved into the test operator by cyclicity, without changing its norm bound $\|O\|_\infty$.

Two loss factors use the analogous block of $\Psi^{(2)}$ with input a flag matrix unit tensored with $\rho\otimes\rho$ and a test operator of norm at most $\|O\|_\infty^2$. Both inputs have trace norm one. The controlled diamond error therefore bounds a changed mean monomial by $\varepsilon_\ell\|O\|_\infty$ and a changed second-moment monomial by $\varepsilon_\ell\|O\|_\infty^2$.

The sum of absolute coefficient weights of mean terms containing branch $\ell$ is at most $2\lambda^2w_\ell$; the corresponding second-moment sum is at most $4\lambda^4w_\ell$, by counting the terms in which that branch occupies any of the four index positions; counting a term more than once only increases this upper bound. Thus the second moment changes by at most $4\lambda^4w_\ell\varepsilon_\ell\|O\|_\infty^2$. Every intermediate ensemble has $|\mathbb E\calL|\leq\lambda^2\|O\|_\infty$, so the squared mean changes by at most another $4\lambda^4w_\ell\varepsilon_\ell\|O\|_\infty^2$. Summing these successive differences cancels the intermediate ensemble variances and leaves the difference between the actual and reference variances, proving the bound.
\end{proof}

Adding an independent uniform phase to both the actual and reference branches removes moments with unequal numbers of unitary and conjugate-unitary factors. It does not in general leave the original actual variance unchanged. The controlled norm explicitly avoids making that replacement without an ensemble assumption.

\begin{proposition}[Design error on the joint space]\label{prop:lifted_design_error}
Use the fixed operators and scale in \eqref{eq:lifted_loss_general}. If
$\|\Phi^{(t)}_{\calE}-\Phi^{(t)}_{\mathrm{Haar}(G_{\INT})}\|_\diamond\leq\varepsilon$ for $t=1,2$, where $\Phi^{(t)}_{\calE}(X)=\mathbb E_{\calE}[W^{\otimes t}X(W\dagg)^{\otimes t}]$, then
\begin{equation}
 \big|\Var_{\calE}[\LBE]-\Var_{G_{\INT}}[\LBE]\big|
 \leq3\lambda^4\varepsilon\|O\|_\infty^2.
\end{equation}
Membership is needed only to express the reference variance as the sector sum in Theorem~\ref{thm:reduction}.
\end{proposition}

\begin{proof}[Proof of Proposition~\ref{prop:lifted_design_error}]\phantomsection\label{proof:prop:lifted_design_error}
For $\tcalL=\LBE/\lambda^2$, write its variance as
$\Tr[(\widetilde O_{\mathrm{ref}}\otimes\widetilde O_{\mathrm{ref}})\Phi^{(2)}(\widetilde\rho_{\mathrm{ref}}\otimes\widetilde\rho_{\mathrm{ref}})]-(\Tr[\widetilde O_{\mathrm{ref}}\Phi^{(1)}(\widetilde\rho_{\mathrm{ref}})])^2$.
The first term changes by at most $\varepsilon\|O\|_\infty^2$. The mean changes by at most $\varepsilon\|O\|_\infty$, and each mean has magnitude at most $\|O\|_\infty$, so its square changes by at most $2\varepsilon\|O\|_\infty^2$. Multiply by $\lambda^4$.
\end{proof}

\section{Coefficient optimization}\label{app:audit}

We optimize loss variance and gradient variance separately, first over phases and then over two-branch weights. The scale $\lambda$ stays fixed so that rescaling all coefficients cannot improve the objective trivially.

The squared weights $x_\ell$ turn the variance into a quadratic form, so the matrix below is a compact table of branch variances and pair contributions. The constraint still comes from the original weights summing to one; it is not the usual simplex constraint on $x$. A phase potential means one real angle assigned to each branch, whose differences reproduce all requested pair angles.

\begin{theorem}[Loss- and gradient-signal tuning]\label{thm:tuning}
Hold $\lambda>0$ fixed while tuning. All phase statements below concern nonzero moments; a zero moment makes that pair phase independent. Let $x_\ell=w_\ell^2$, $v_\ell=\Var[m_\ell]$, and
\begin{equation}
    C_{\ell k}(\gamma)=\mathbb E|m_{\ell k}|^2
    +\mathrm{Re}\big[e^{2i(\gamma_\ell-\gamma_k)}\mathbb E(m_{\ell k}^2)\big]\;\geq\;0 .
    \label{eq:Ceff}
\end{equation}
Under Assumption~\ref{asm:moments}(a), $\Var[\calL]=\lambda^4\,x^{T}\mathsf K(\gamma)\,x$ with $\mathsf K_{\ell\ell}=v_\ell$ and $\mathsf K_{\ell k}=C_{\ell k}(\gamma)$, over the domain $\{x_\ell\geq0,\ \sum_\ell\sqrt{x_\ell}=1\}$.
\begin{itemize}
    \item[(i)] \emph{Phases, loss objective.} Each $C_{\ell k}$ is maximized, over the relative phase alone, at $2(\gamma_\ell-\gamma_k)\equiv-\arg\mathbb E(m_{\ell k}^2)\ (\mathrm{mod}\ 2\pi)$, with maximum $\mathbb E|m_{\ell k}|^2+|\mathbb E(m_{\ell k}^2)|$. For $L=2$ this optimizes $\Var[\calL]$ exactly; for $L>2$ it is a classical phase-alignment problem, solved exactly whenever $\arg\mathbb E(m_{\ell k}^2)=\phi_k-\phi_\ell$ for a potential $\phi$.
    \item[(ii)] \emph{Weights, loss objective ($L=2$, $v_1=v_2=V$, $C=C_{12}(\gamma)$).} The balanced point $w=(\tfrac12,\tfrac12)$ is a strict local maximum of $\Var[\calL]$ iff $C>3V$, and the global maximum iff $C\geq7V$; otherwise the maximum concentrates on one branch.
    \item[(iii)] \emph{Phases, gradient objective.} For a parameter of branch $\ell$ as in Theorem~\ref{thm:grad_select}, the interference contribution of the pair $(\ell,k)$ to $\Var(\partial_\theta\calL)$ is
    $C^{\nabla}_{\ell k}(\gamma)=\mathbb E|\Gamma_{\ell k}|^2-\mathrm{Re}[e^{2i(\gamma_\ell-\gamma_k)}\mathbb E(\Gamma_{\ell k}^2)]\geq0$,
    a sinusoid of the \emph{opposite} sign: it is maximized at $2(\gamma_\ell-\gamma_k)\equiv\pi-\arg\mathbb E(\Gamma_{\ell k}^2)\ (\mathrm{mod}\ 2\pi)$, with maximum $\mathbb E|\Gamma_{\ell k}|^2+|\mathbb E(\Gamma_{\ell k}^2)|$, and minimized at $2(\gamma_\ell-\gamma_k)\equiv-\arg\mathbb E(\Gamma_{\ell k}^2)$, with minimum $\mathbb E|\Gamma_{\ell k}|^2-|\mathbb E(\Gamma_{\ell k}^2)|$. If $\arg\mathbb E(\Gamma_{\ell k}^2)=\arg\mathbb E(m_{\ell k}^2)$, the loss-optimal phases of (i) minimize $C^{\nabla}_{\ell k}$: maximizing $\Var[\calL]$ can therefore minimize the added gradient contribution. For $L>2$, all pairwise gradient maxima are attained simultaneously whenever $\pi+\arg\mathbb E(\Gamma_{\ell k}^2)\equiv\psi_k-\psi_\ell$ for a potential $\psi$, by $\gamma=\psi/2$.
    \item[(iv)] \emph{Weights, gradient objective ($L=2$).} Fix the phases and a measurement-adjacent parameter of branch $1$, write $w_1=w$, $w_2=1-w$, $V_\nabla=\Var(\Tr[\,i[O,H]\,U_1\rho U_1\dagg])$, and $C_\nabla=C^{\nabla}_{12}(\gamma)$. Then \eqref{eq:grad_select} reads
    \begin{equation}
    \begin{split}
        \Var(\partial_\theta\calL)&=\lambda^4\mathcal G(w),\\
        \mathcal G(w)&=V_\nabla w^4+2C_\nabla w^2(1-w)^2 .
    \end{split}
        \label{eq:Fgrad}
    \end{equation}
    If $V_\nabla=0<C_\nabla$, the unique maximizer is $w=1/2$; if both vanish, every weight is optimal. In the remainder of (iv), assume $V_\nabla>0$. Since $\mathcal G'(\tfrac12)=V_\nabla/2$, balanced weights are not a stationary point of the single-parameter gradient objective whenever $V_\nabla>0$. If $C_\nabla\leq4V_\nabla$, $\mathcal G$ is nondecreasing on $[0,1]$ and is maximized at $w=1$. If $C_\nabla>4V_\nabla$, $\mathcal G$ has an interior local maximum at
    \begin{equation}
    \begin{gathered}
        w_-=\frac{3C_\nabla-\sqrt{C_\nabla^2-4C_\nabla V_\nabla}}{2(V_\nabla+2C_\nabla)}\in\big(\tfrac12,\tfrac23\big),\\
        \mathcal G(w_-)=C_\nabla w_-^2(1-w_-),
    \end{gathered}
        \label{eq:wminus}
    \end{equation}
    which is the global maximum iff $C_\nabla\geq r_*V_\nabla$, where $r_*=w_*^2/\big((1-w_*)(2w_*-1)\big)\approx7.41$ and $w_*\approx0.544$ is the real root of $w^3+w^2+w=1$. For $V_\nabla>0$, an optimal weight therefore favors the differentiated branch and approaches $\tfrac12$ only as $C_\nabla/V_\nabla\to\infty$. By contrast, the symmetric aggregate $\mathcal G(w)+\mathcal G(1-w)$, which adds the gradient variances of one measurement-adjacent parameter from each branch, equals the loss objective of (ii) with $(V,C)$ replaced by $(V_\nabla,2C_\nabla)$, so balanced weights maximize it globally iff $2C_\nabla\geq7V_\nabla$.
\end{itemize}
\end{theorem}

\begin{proof}[Proof of Theorem~\ref{thm:tuning}]\phantomsection\label{proof:thm:tuning}
Set $x_\ell=w_\ell^2$ in \eqref{eq:exact_select}. Then
\begin{equation}
 \sum_\ell w_\ell^4 v_\ell=\sum_\ell v_\ell x_\ell^2,
 \qquad
 \sum_{\ell\neq k}w_\ell^2w_k^2C_{\ell k}=\sum_{\ell\neq k}C_{\ell k}x_\ell x_k.
\end{equation}
The symmetry $C_{\ell k}=C_{k\ell}$ holds because $\mathbb E|m_{k\ell}|^2=\mathbb E|m_{\ell k}|^2$ and $\mathbb E(m_{k\ell}^2)=\overline{\mathbb E(m_{\ell k}^2)}$.

(i) For fixed moduli, $C_{\ell k}(\gamma)=\mathbb E|m_{\ell k}|^2+\mathrm{Re}[e^{2i\Delta}\mathbb E(m_{\ell k}^2)]$ with $\Delta=\gamma_\ell-\gamma_k$ is a sinusoid in $2\Delta$ with mean $\mathbb E|m_{\ell k}|^2$ and amplitude $|\mathbb E(m_{\ell k}^2)|$; the extrema and extremizers follow, and $C_{\ell k}\geq0$ because $|\mathbb E(m_{\ell k}^2)|\leq\mathbb E|m_{\ell k}|^2$.

For $L=2$ only one relative phase exists, so optimizing $C_{12}$ optimizes $\Var[\calL]$ for any weights. If $\arg\mathbb E(m_{\ell k}^2)=\phi_k-\phi_\ell$, setting $\gamma=\phi/2$ gives $2(\gamma_\ell-\gamma_k)+\arg\mathbb E(m_{\ell k}^2)=(\phi_\ell-\phi_k)+(\phi_k-\phi_\ell)=0$ for every pair, attaining every pairwise maximum simultaneously.

(ii) If $V=C=0$, the objective vanishes identically and every weight is optimal. Otherwise, substituting $w=\tfrac12+x$, $1-w=\tfrac12-x$ into $f(w)=(w^4+(1-w)^4)V+2w^2(1-w)^2C$:
\begin{equation}
    f=\tfrac{V+C}8+(3V-C)\,x^2+2(V+C)\,x^4,\qquad x\in[-\tfrac12,\tfrac12].
    \label{eq:fquartic}
\end{equation}

The coordinate $x=0$ represents balanced weights, while $x=\pm1/2$ represents a single active branch. The quadratic coefficient decides local stability at balance, but comparison with the endpoints is still needed for global optimality.
The balanced point $x=0$ is a strict local maximum iff $3V-C<0$.

The stationary points solve $2(3V-C)x+8(V+C)x^3=0$: $x=0$ or $x^2=(C-3V)/(4(V+C))$, the latter interior local minima when $C>3V$ (positive quartic coefficient). Hence for $C>3V$ the global maximum is $\max\{f(0),f(\pm\tfrac12)\}=\max\{(V+C)/8,\;V\}$, which is the balanced point iff $C\geq7V$. For $C\leq3V$, $f$ increases monotonically in $|x|$ (both derivative terms have the same sign), so the maximum is at the boundary.

(iii) By Theorem~\ref{thm:grad_select}, the $(\ell,k)$ interference term of $\Var(\partial_\theta\calL)$ is $2\lambda^4w_\ell^2w_k^2C^{\nabla}_{\ell k}(\gamma)$ with $C^{\nabla}_{\ell k}(\gamma)=\mathbb E|\Gamma_{\ell k}|^2-\mathrm{Re}[e^{2i\Delta}\mathbb E(\Gamma_{\ell k}^2)]$, $\Delta=\gamma_\ell-\gamma_k$: a sinusoid in $2\Delta$ with mean $\mathbb E|\Gamma_{\ell k}|^2$ and amplitude $|\mathbb E(\Gamma_{\ell k}^2)|$, nonnegative by the Cauchy--Schwarz inequality. Its extremizers follow as in (i), with the sign of the sinusoid reversed, and the stated conflict with (i) is immediate by comparing the extremizing phases.

For the potential case, the maximizing condition $2(\gamma_\ell-\gamma_k)\equiv\pi-\arg\mathbb E(\Gamma_{\ell k}^2)$ reads $2(\gamma_\ell-\gamma_k)\equiv\psi_\ell-\psi_k$ when $\pi+\arg\mathbb E(\Gamma_{\ell k}^2)\equiv\psi_k-\psi_\ell$, and $\gamma=\psi/2$ satisfies it for every pair.

The shift by $\pi$ matters when a collection of gradient pairs forms a cycle. For example, if the arguments of nonzero gradient moments themselves have the unshifted potential form, their sum around a triangle is zero, whereas simultaneous gradient maximization would require it to equal $3\pi$ modulo $2\pi$. For one designated branch alone the relevant pairs form a star, so this cycle obstruction is absent; it can arise in an aggregate objective.

(iv) When $V_\nabla=0$, \eqref{eq:Fgrad} reduces to $2C_\nabla w^2(1-w)^2$, giving the stated boundary cases. Assume $V_\nabla>0$. Differentiation gives
\begin{equation}
 \mathcal G'(w)=4w[(V_\nabla+2C_\nabla)w^2-3C_\nabla w+C_\nabla],
\end{equation}
so $\mathcal G'(1/2)=V_\nabla/2$. The bracket has discriminant $C_\nabla(C_\nabla-4V_\nabla)$. For $C_\nabla\leq4V_\nabla$ it is nonnegative, and the maximum is $\mathcal G(1)=V_\nabla$. Otherwise its roots are
$w_\pm=(3C_\nabla\pm\sqrt{C_\nabla^2-4C_\nabla V_\nabla})/[2(V_\nabla+2C_\nabla)]$.
Both lie in $(0,1)$, with derivative signs $+,-,+$ on the three intervals. The global maximum is therefore $\max\{\mathcal G(w_-),\mathcal G(1)\}$.

The first candidate preserves interference between the branches; the second keeps only the differentiated branch. The remaining calculation finds the ratio of interference to branch signal at which the former becomes preferable.

At $w_-$, stationarity gives $V_\nabla w_-^2=C_\nabla(1-w_-)(2w_--1)$ and hence $\mathcal G(w_-)=C_\nabla w_-^2(1-w_-)$. As $C_\nabla/V_\nabla$ increases from $4$ to infinity, $w_-$ decreases from $2/3$ to $1/2$.

The comparison $\mathcal G(w_-)\geq V_\nabla$ becomes $w_-^4-2w_-+1\geq0$. Factoring as $(w_--1)(w_-^3+w_-^2+w_--1)$ gives the unique transition $w_*=0.543689\ldots$ in this interval. The stationarity relation then yields $C_\nabla/V_\nabla\geq r_*=7.413750\ldots$. Finally, $\mathcal G(w)+\mathcal G(1-w)=V_\nabla[w^4+(1-w)^4]+4C_\nabla w^2(1-w)^2$, so part (ii) applies with $(V,C)=(V_\nabla,2C_\nabla)$.
\end{proof}

\section{Quantum Fisher information and parameter directions}\label{app:qfim}

More circuit parameters need not provide more independent ways to change the output. The quantum Fisher information matrix (QFIM) measures locally distinguishable state changes, and its rank counts their independent directions. Here the output is the pure joint ancilla--system state before postselection. Its Fisher information need not equal that of the normalized successful output. Since a common phase is unobservable, the relevant bound is the dimension of the state orbit with that phase removed.

The \emph{state orbit} is the set of states obtained by applying the generated group to the input. Its \emph{stabilizer} consists of generator directions that leave that input unchanged up to a common phase. Subtracting the stabilizer dimension counts the directions that can actually change the physical state. All generators below are expressed in the same frame as $W$; a fixed change of preparation frame requires conjugating them as in Appendix~\ref{app:gradient}.

Theorem~1 of Larocca et al.~\cite{Larocca2023Overparam} gives the algebra-dimension bound on QFIM rank; their accompanying orbit bound further accounts for the input state. The following corollary expresses that orbit refinement as a stabilizer subtraction for the present joint circuit.

\begin{corollary}[QFIM rank bound and parameter budget]\label{cor:qfim}
Let $\ket{\Psi(\theta)}=W(\theta)\ket{\Psi_0}$, where $\ket{\Psi_0}\in\calH_a\otimes\calH_s$ is a fixed unit vector and every gate generator of the circuit $W(\theta)$ lies in $\gint$, and let $F(\theta)$ be the quantum Fisher information matrix of the pure state $\ket{\Psi(\theta)}$, with entries
\begin{equation}
    F_{pp'}=4\,\mathrm{Re}\big[\langle\partial_p\Psi|\partial_{p'}\Psi\rangle-\langle\partial_p\Psi|\Psi\rangle\langle\Psi|\partial_{p'}\Psi\rangle\big]
\end{equation}
for parameters $\theta_p,\theta_{p'}$, writing $\partial_p=\partial/\partial\theta_p$. For every $\theta$,
\begin{equation}
    \operatorname{rank}F(\theta)\;\leq\;\dim\gint-\dim\stab_{\gint}(\Psi_0),
    \label{eq:qfim_bound}
\end{equation}
where $\stab_{\gint}(\Psi_0)=\{X\in\gint:X\ket{\Psi_0}\in i\mathbb R\ket{\Psi_0}\}$.
\end{corollary}

\begin{proof}[Proof of Corollary~\ref{cor:qfim}]\phantomsection\label{proof:cor:qfim}

A tangent vector describes the first-order state change under a parameter variation. After projecting away the state vector, these tangents exclude changes of common phase. Their real inner products form the QFIM up to a factor of four, so its rank cannot exceed the dimension of their span.

For gate $p$ with anti-Hermitian generator $G_p\in\gint$, let $W_{>p}$ be the product of subsequent gates. Then $\partial_p\ket\Psi=\widehat G_p\ket\Psi$, where $\widehat G_p=W_{>p}G_pW_{>p}^{-1}\in\gint$: conjugating a generator by an element of its generated group preserves the Lie algebra. Let $\Pi^\perp$ project orthogonally to $\mathbb C\ket\Psi$. The QFIM is four times the real Gram matrix of $\Pi^\perp\widehat G_p\ket\Psi$. These vectors lie in the image of $T_\Psi:\gint\to\calH_a\otimes\calH_s$, $T_\Psi(X)=\Pi^\perp X\ket\Psi$. The kernel consists of directions mapped to zero. It is exactly $\stab_{\gint}(\Psi)$: anti-Hermiticity forces any eigenvalue on $\ket\Psi$ to be imaginary. Hence the rank is at most $\dim\gint-\dim\stab_{\gint}(\Psi)$.

Finally $W\in G_{\INT}$ gives $\stab_{\gint}(\Psi)=W\stab_{\gint}(\Psi_0)W^{-1}$, so this dimension equals that in \eqref{eq:qfim_bound}.
\end{proof}

\end{document}